\input pipi.sty
\input epsf.sty
\input psfig.sty
\magnification1040
\vsize=21cm
\def\otightboxit{\relax}
\raggedbottom
\nopagenumbers
\rightline\timestamp
\rightline{FTUAM 07-11} 
\rightline{arXiv:0710.1150}    

\bigskip
\hrule height .3mm  
\vskip.6cm
\centerline{{\bigfib The pion-pion scattering amplitude. III: Improving the analysis}} 
\centerline{{\bigfib  with forward dispersion relations
and Roy equations\vphantom{\Bigg|}}}
\medskip
\centerrule{.7cm}
\vskip1.7cm
\setbox1=\vbox{\hsize65mm {\noindent\fib R.~Kami\'nski} 
\vskip .1cm
\noindent{\addressfont Department of Theoretical Physics\hb
Henryk Niewodnicza\'nski Institute\hb of Nuclear Physics,\hb
Polish Academy of Sciences,\hb
31-242, 
Krak\'ow, Poland,}}
\medskip
\setbox8=\vbox{\hsize65mm {\noindent\fib J. R. Pel\'aez} 
\vskip .1cm 
\noindent{\addressfont Departamento de F\'{\i}sica Te\'orica,~II\hb
 (M\'etodos Matem\'aticos),\hb
Facultad de Ciencias F\'{\i}sicas,\hb
Universidad Complutense de Madrid,\hb
E-28040, Madrid, Spain}}
\line{\hfil\box1\hfil\box8\hfil}
\smallskip 
\setbox7=\vbox{\hsize65mm \fib and} 
\centerline{\box7}
\smallskip
\setbox9=\vbox{\hsize65mm {\noindent\fib F. J. 
Yndur\'ain} 
\vskip .1cm
\noindent{\addressfont Departamento de F\'{\i}sica Te\'orica, C-XI\hb
 Universidad Aut\'onoma de Madrid,\hb
 Canto Blanco,\hb
E-28049, Madrid, Spain.}\hb}
\smallskip
\centerline{\box9}
\bigskip
\setbox0=\vbox{\abstracttype{Abstract}
We complete and improve the fits to  experimental $\pi\pi$
scattering amplitudes, both at low and high energies, that we performed in the previous papers of this
series.  We then verify that the corresponding amplitudes satisfy analyticity requirements, 
in the form of partial wave analyticity at low energies, 
forward dispersion relations (FDR) at all 
energies, and Roy equations below$\bar{K}K$ threshold; the first by construction, the last two, inside
experimental errors. Then we repeat the fits including as constraints FDR and Roy equations. 
The ensuing central values of the various scattering amplitudes verify very accurately FDR and, 
especially, Roy equations, and change very little from what we found by just fitting data,
 with the exception of the D2 wave phase shift, for which one parameter  moves by
$1.5\;\sigma$. These improved parametrizations therefore provide a reliable representation of
pion-pion  amplitudes with which one can test various physical relations.
 We also present a list of low energy parameters and other observables.
  In particular, we find  
$a_0^{(0)}=0.223\pm0.009\,M^{-1}_\pi$,
  $a_0^{(2)}=-0.0444\pm0.0045\,M^{-1}_\pi$  and
$\delta_0^{(0)}(m^2_K)-\delta_0^{(2)}(m^2_K)=50.9\pm1.2^{\rm o}$. }
\centerline{\box0}
\brochureendcover{Typeset with \physmatex}
\brochureb{\smallsc r. kami\'nski,  j. r. pel\'aez and f. j.  yndur\'ain}{\smallsc 
the pion-pion scattering amplitude. iii: improving the analysis }{1}
\brochuresection{1. Introduction}
\noindent
In two recent papers,\ref{1,2} that we will 
consistently denote by PY05, KPY06, 
we have presented a set of fits to the data on 
$\pi\pi$ scattering phase shifts and inelasticities, and 
we also checked how well forward dispersion
relations (FDR, henceforth) are satisfied by  different 
 $\pi\pi$ scattering phase shift analyses
(including our own), finding in fact that some among the more widely used sets of phase shifts, 
 failed to pass this test by several standard deviations. 
 We then performed  (in PY05)
 a consistent 
energy-dependent phase shift analysis of  
$\pi\pi$ scattering amplitudes in which we  constrained the fits 
by requiring verification of FDR.

In the present paper we complete and improve on the results in PY05 and KPY06 
in various ways. First of all (\sect~2) we incorporate the  fit to data in ref.~3
(to be called GMPY07 henceforth)  
for the S0 wave which is very precise thanks to use of recent experimental results\ref{4}
and a more appropriate parametrization,  which 
provides an  accurate determination of $\delta_0^{(0)}(s)$ 
below $\bar{K}K$ threshold. 
The precision of this determination is such that we have to refine also the analysis 
of some other waves. 
Thus, we improve the S2 wave as given in PY05 by requiring smoother junction between the low energy 
($s^{1/2}\leq932\,\mev$) and intermediate energy  ($932\,\mev\leq s^{1/2}\leq1420\,\mev$) regions. 
We also (slightly) improve the fit to the inelasticity
 of the D0 wave, smoothing the onset of
inelasticity above the $\bar{K}K$ threshold, and the parametrization of the P and F 
waes removing their ghosts. Finally, we 
improve the error analysis of the D2 wave. 

We also make (\sect~3) some improvements in the high energy 
($s^{1/2}>1.42\,\gev$) input, especially in the momentum transfer dependence of the 
amplitudes, necessary to
evaluate  Roy equations.\ref{5} 
Indeed, for Roy equations, we  need Regge formulas away from the forward 
direction, so we extend our analysis 
of the Regge amplitudes there.
 For the imaginary part of a scattering amplitude, $\imag F(s,t)$, 
one only expects Regge theory to be valid for $s\gg\lambdav^2$, 
$|t|\ll s$ (with $\lambdav\simeq0.35\,\gev$ the QCD parameter); 
this will limit the validity of the Roy equations to energies 
 $\lsim1\,\gev$ (as a matter of fact, we here only evaluate them 
below $\bar{K}K$ threshold).
 The corresponding Regge parameters were obtained, 
in the forward direction, by relating the $\pi\pi$ cross section to 
the $\pi N$ and $NN$ cross sections, using factorization, and fitting all three processes. 
Away from the forward direction, we  fix other parameters  
using also sum rules, in \sects~5 and 6.

All of this (Regge parameters and partial wave amplitudes) constitute a set of pion-pion amplitudes,
obtained with {\sl unconstrained fits to data}, 
that we denote by ``UFD~Set", which we have collected in Appendix~A.

Next, in \sect~4, we test FDR and Roy equations for this UFD~Set. 
We get good verification in both cases, to a level of average agreement corresponding to 
$1.00$ standard deviations, for FDR, and to $0.97$  standard deviations 
for the Roy equations.
The fact that FDR and Roy equations are practically satisfied within errors, 
makes it reasonable to improve the fits to data 
including as a constraint the verification of forward dispersion
 relations {\sl and} Roy equations: 
this should provide a set of parametrizations to partial waves 
fully compatible with analyticity and, hence, 
with more reliable central values. 
This we do in \sect~5; here we also 
constrain the fits by requiring verification of two sum 
rules which relate higher ($s^{1/2}>1420\,\mev$) and lower energies, 
and which permit us to refine the values for Regge parameters 
away from the forward direction. 
This procedure provides a set of $\pi\pi$ amplitudes, 
that we call {\sl constrained fit to data}, or CFD~Set, 
that not only fit data, but also satisfy sum rules,  
FDR and Roy equations, well within the rather small errors we now have;
 in particular, the degree of
verification of the FDR below 932~\mev\ and, especially, 
 of the Roy equations, is spectacular.
 It turns out that 
all parameters for all waves remain inside the error bars we obtained by 
just fitting the data
(UFD~Set), 
except for the D2 wave where, in particular, one 
parameter changes 
by $1.5\;\sigma$. 
The result of these constrained fits is collected in Appendix~B for the 
partial wave amplitudes and Regge parameters. 
As it happened in PY05, we still find a marked hump in the S0 wave
both for the unconstrained and constrained  fits, between 400 and  900~\mev\ of energy, 
which structure is thus shown to be compatible with 
both FDR and Roy equations. 
In \sect~6 we present the values of the low energy parameters 
and other observables that follow  from both fits to data (UFD~Set) and, 
especially,
constrained fits (CFD~Set). The values of some parameters 
are improved using sum rule determinations. In particular, for the scattering lengths for
the S0, S2 waves we find the very accurate determination
$$a_0^{(0)}=0.223\pm0.009\;M_{\pi}^{-1},
\quad a_0^{(2)}=-0.0444\pm0.0045\;M_{\pi}^{-1};
\equn{(1.1)}$$
here, and in what follows, $\;M_{\pi}$ represents the mass of the charged pion, that we take
 $\;M_{\pi}=139.57\;\mev.$ 
The article is finished  in \sect~7 where we give the conclusions, and a short discussion of our
results;  particularly, comparing them with other independent evaluations 
(theoretical and experimental) of the 
$\pi\pi$ scattering amplitudes. A few words on isospin breaking
 corrections (which affect very little our results) for the S0, P waves are also said there.

\booksection{2. Fits to data}

\noindent
In the present Section we briefly summarize the 
methods and results in PY05, KPY06, together with a few improvements, obtained
 fitting experimental data for
$\pi\pi$  partial waves, up to the energy of 1.42~\gev. Above this energy we
assume the  scattering amplitudes to be given by 
Regge formulas, that we discuss in \sect~3 below.
The reason why we choose to use Regge theory above 1.42~\gev\ is that,
 whereas there are uncertainties
in the Regge expressions at the lower part of the  energy range where we use them, say between 
1.42~\gev\ and $\sim1.8\,\gev$, these are substantially smaller than the 
uncertainties in $\pi\pi$ phase shift analyses in the same energy region.  
Broadly speaking we distinguish two energy regions in the fits to 
experimental partial wave data: the low energy region, 
$s^{1/2}\leq s_{ i}^{1/2}$, $s_{ i}^{1/2}\sim1\,\gev$, and the intermediate energy region,
 $s_{ i}^{1/2}\lsim
s^{1/2}\leq1.42\,\gev$.  $s_{ i}^{1/2}$ is the energy above which one 
{\sl cannot} consider that {\sl inelastic} processes are negligible.
 The precise value $s_{ i}^{1/2}$ where we separate ``low energy" 
from ``intermediate energy" depends on each wave.
 
In the low
energy region
we write {\sl model-independent} parametrizations that take into account unitarity and analyticity; 
this last, by making a conformal mapping,
$$s\to w(s)=\dfrac{\sqrt{s}-\sqrt{s_i-s}}{\sqrt{s}+\sqrt{s_i-s}}$$
and expanding the function\fnote{Up to a kinematical factor, and up to  poles, 
that we separate explicitly; see the specific formulas for each wave.}
$\cot\delta_l^{(I)}(s)$ in powers of this variable 
$w$. The details may be found in Appendix~C,
 where we also explain how using this variable  does {\sl not} imply any model, while
accelerating the convergence: we, generally speaking,
 only need between two and four terms in this
expansion. We then fit experimental $\pi\pi$ phase shift data\ref{4,6} and, for the 
S0 wave at intermediate energy, also $\pi\pi\to\bar{K}K$ data.\ref{7} 
In four cases (the waves 
D0, D2, F, G0) we include in the fit the values of the 
scattering length and, for the D0 wave, also the effective range parameter. 
These are obtained from experiment via the Froissart--Gribov representation; 
see PY05, KPY06 for details.
In the case of the P wave, we do {\sl not} fit the data on $\pi\pi$ scattering,
 but the vector form factor
of the pion, which gives  much more precise results.\ref{8}
In the intermediate energy region we make {\sl phenomenological} fits to experimental 
phase shifts and elasticity parameters, basically polynomial fits. 
There is, however, an exception to this: the S0 wave. 
Here, most of the inelasticity is due to the $\bar{K}K$ channel, so 
we can make a two-channel calculation; see KPY06. 
The precision of the fits, particularly for the S0, P and D0 waves, is such 
that we require a (slight) improvement of the analyses of PY05, 
KPY06 in some cases.
\booksubsection{2.1. The S0 wave}
\noindent
The first case where we improve on the analysis of
PY05, KPY06 is the S0 wave, at low energy. 
Although this is discussed in detail in GMPY07, we say a few words here for 
completeness.
To write the corresponding expansion, it is necessary to separate off the 
 pole of the effective range function  that lies on the 
real axis, viz., the pole due to the so-called Adler zero of the 
partial wave amplitude. This lies near the beginning of the left hand cut, at
$s=\tfrac{1}{2}z_0^2,$ 
 $z_0\simeq M_\pi$.
We  fit using
the expression
$$\eqalign{
\cot\delta_0^{(0)}(s)=&{{s^{1/2}}\over{2k}}\,{{M^2_\pi}\over{s-\tfrac{1}{2}z_0^2}}\,
\Big\{\dfrac{z_0^2}{M_\pi\sqrt{s}}+{B}_0+{B}_1w(s)+{B}_2w(s)^2\Big\}.\cr
}
\equn{(2.1a)}$$
We here fix $z_0=M_\pi$; in \sect~5 we will allow $z_0$ to vary.
We find, taking the best fit\fnote{In ref.~3 we gave several different parametrizations 
for the S0 wave.
These different parameters do not correspond to
different physical scenarios, but just to the use of two or three
terms in the conformal expansion, to what data sets are fitted, or to wheter or not we  explicitly
factorize the  zeros  in the amplitudes.
Such parametrizations  correspond to the same 
physical scenario, as indeed 
their resulting phase shifts overlap within errors.
We here choose,  among those parametrizations,
the one with better analytic properties, and which fits best  the largest
sample of consistent and reliable data points in the elastic region.}
 in ref.~3,
$${B}_0=4.3\pm0.3,\quad {B}_1=-26.9\pm0.6,\quad
{B}_2=-14.1\pm1.4.
\equn{(2.1b)}$$

\topinsert{\bigskip
\setbox0=\vbox{{\psfig{figure=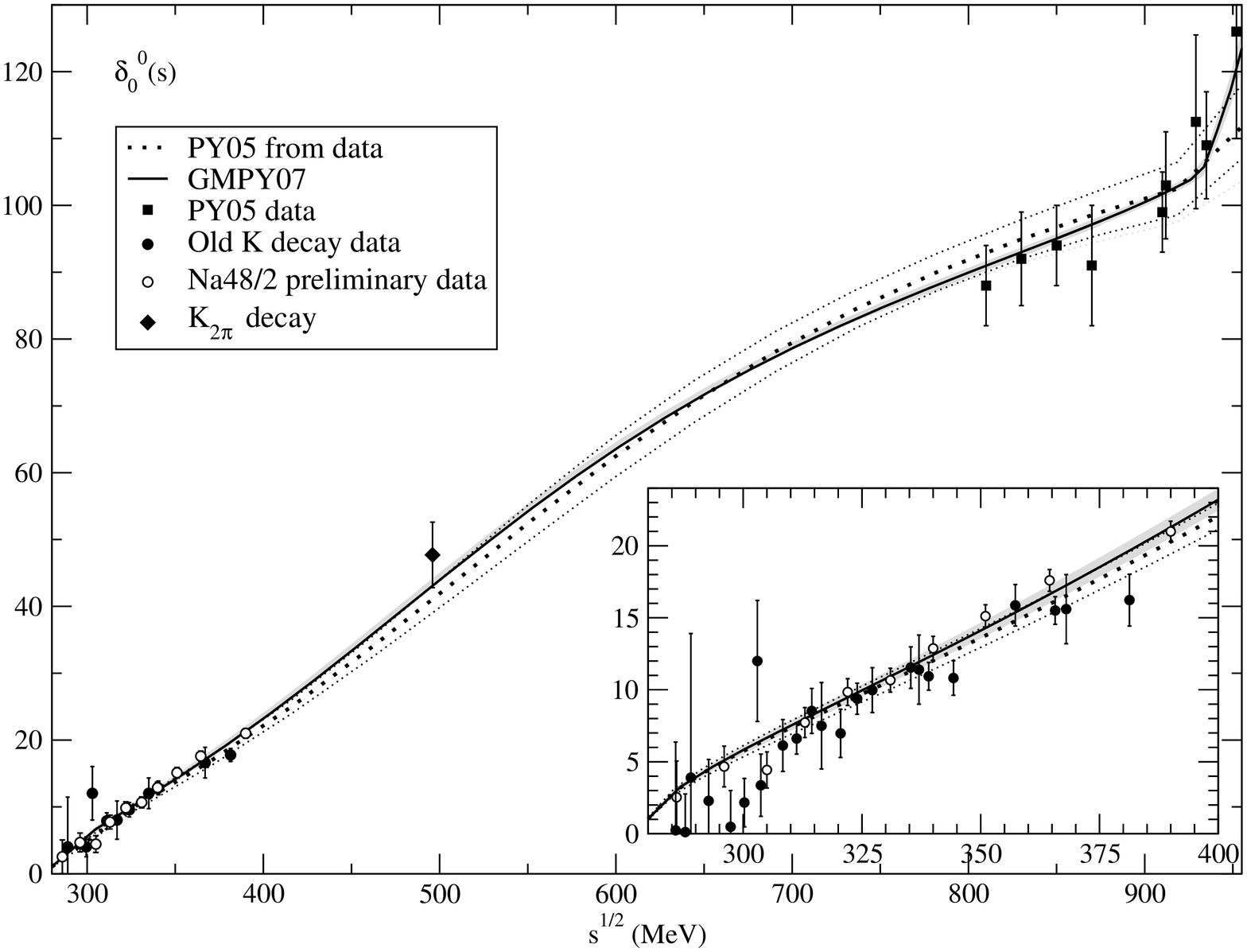,width=15.5truecm,angle=-90}}}
\centerline{\box0}
\setbox6=\vbox{\hsize 12truecm\captiontype\figurasc{Figure 2.1. }
{The fits to S0  from GMPY07 (solid line and gray band) and in PY05 (dotted lines). 
In the indent, a blow-up
of the low energy region. The fitted data  from refs.~4,~6
 are also shown. }}
\centerline{\box6}
}\endinsert\bigskip
The  ensuing numerical results for the phase shift
 are very similar to what was obtained
in  PY05, but have a sounder theoretical basis and, above all, are much more accurate.
The  curves corresponding to this, and that in PY05, are shown in \fig~2.1, 
together with some experimental data. 
Because at 932~\mev\ we match the low energy to the K-matrix 
fit at intermediate energy, and the 
low energy fit has changed (in particular with much smaller errors)
 from what we had in  KPY06, 
we have also to slightly modify the parameters of the K-matrix fit. 
The resulting values for the parameters are given in Appendix~A.
They do not change much from what we had in KPY06, but 
they are now determined more accurately.

The corrections due to isospin breaking have been computed for this wave, 
at very low energy; 
we discuss the (slight) modification they imply in \sect~7, and the corresponding 
values of the parameters are given also in Appendix~A.

\topinsert{
\bigskip
\setbox0=\vbox{{\psfig{figure=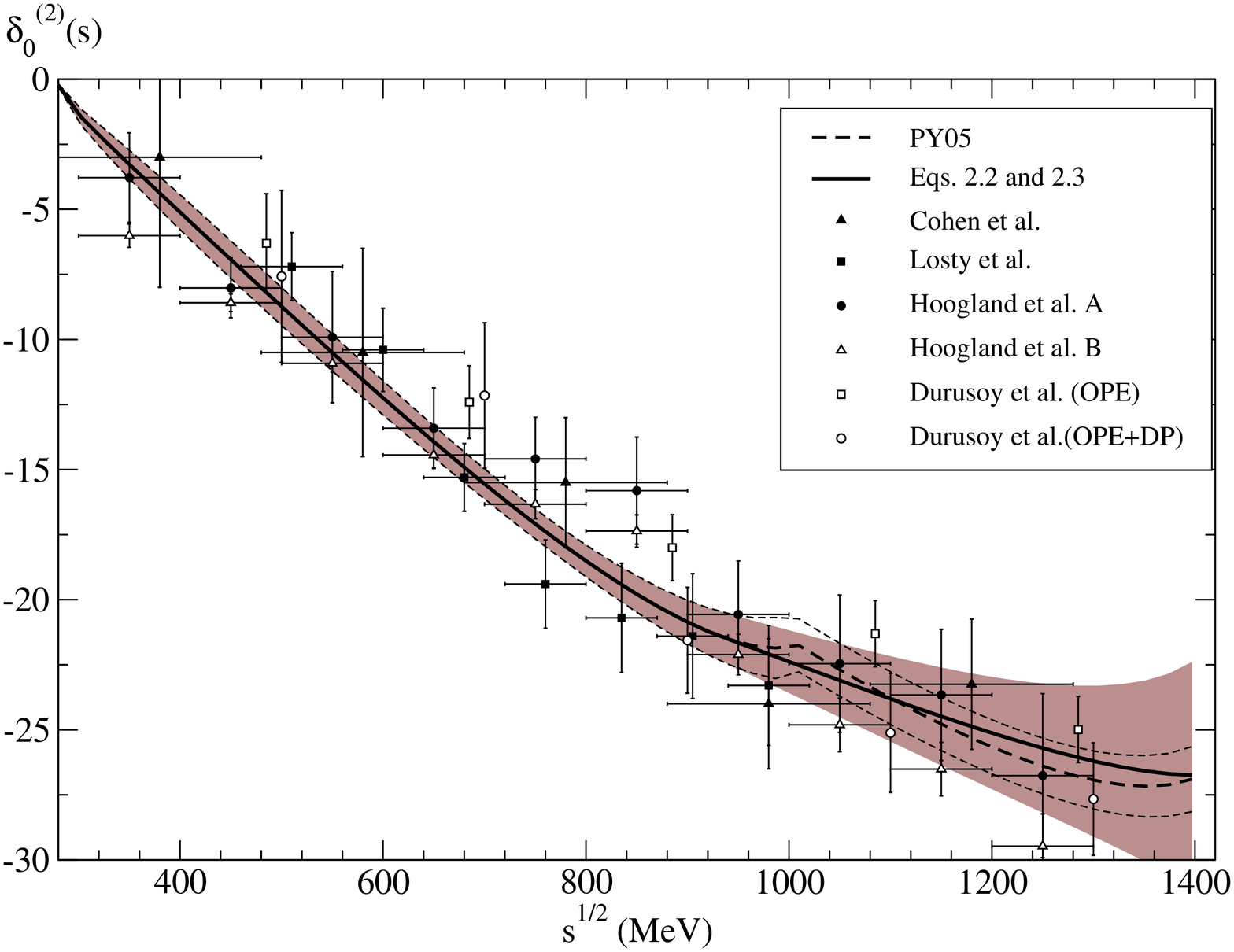,width=13.7truecm,angle=-90}}}
\centerline{\box0}
\setbox6=\vbox{\hsize 12truecm\captiontype\figurasc{Figure 2.2. }{The fits to the S2 phase shift 
here (solid line and gray band) and in PY05 (dashed lines). Some experimental data from ref.~6 
 are also shown. }}
\centerline{\box6}
}\endinsert
\bigskip

\booksubsection{2.2. The S2 wave}

\noindent
The second case where we improve on the analysis in PY05, KPY06 is for the 
S2 wave. The reason is that the precision of our calculations has improved so much 
that one is sensitive to the {\sl derivative} of the 
phase shift as one crosses into the region where 
inelasticity is not negligible. 
In the case of the S2 wave, the lowest inelastic process is 
$\pi\pi\to\pi\pi\rho$, which is a (quasi-)three body process. 
Therefore, we expect $\delta_0^{(2)}(s)$ to be continuous, and 
with a continuous derivative, until the first two-body channel 
($\rho\rho$) opens. We take this into account now by altering the PY05 fits to this wave, 
as follows. 
In the low energy region, $s^{1/2}\leq932\,\mev$, we maintain the fit in PY05. 
 We 
write
$$\cot\delta_0^{(2)}(s)=\dfrac{s^{1/2}}{2k}\,\dfrac{M_{\pi}^2}{s-2z_2^2}\,
\left\{B_0+B_1\dfrac{\sqrt{s}-\sqrt{{s_l}-s}}{\sqrt{s}+\sqrt{{s_l}-s}}\right\},
\quad z_2\equiv M_\pi;\quad 
s_l^{1/2}=1.05\;\gev; 
\equn{(2.2a)}$$
$s^{1/2}\leq 932\;\mev.$
 We  fix $z_2=M_\pi$; later, in \sect~5, we will allow $z_2$ to vary.
For the errors, and since the low and high energy pieces are 
very strongly correlated, we take this into account and get somewhat improved errors.
Altogether we find a $\chidof=11.2/(21-2)$ and the parameters  
$$B_0=-80.4\pm2.8,\quad B_1=-73.6\pm10.5.
\equn{(2.2b)}$$

For the high energy region we  neglect the inelasticity below $1.45$ \gev\  
for the fit to the  {\sl phase}.
We then fit  high energy data ($s^{1/2}\geq0.95\,\gev$), 
requiring agreement of the 
central value and derivative with the low energy determination 
given in (2.2) at  the energy $s^{1/2}_M=932\,\mev$.  
We write
$$\eqalign{
\cot\delta_0^{(2)}(s)=&\,\dfrac{s^{1/2}}{2k}\,\dfrac{M_{\pi}^2}{s-2M^2_\pi}\,
\left\{B_{h0}+B_{h1}\left[w_h(s)-w_h(s_M)\right]+
B_{h2}\left[w_h(s)-w_h(s_M)\right]^2\right\},\cr s^{1/2}\geq &\,932\;\mev;\cr
 B_{h0}=&\,B_0+B_1w_l(s_M),\quad B_{h1}=
B_1\left.\dfrac{\partial w_l(s)}{\partial w_h(s)}\right|_{s=s_M},\cr
}
\equn{(2.3a)}$$
and
$$\eqalign{w_l(s)=\dfrac{\sqrt{s}-\sqrt{s_l-s}}{\sqrt{s}+\sqrt{s_l-s}};\quad
s_l^{1/2}=1050\;\mev,\cr
w_h(s)=\dfrac{\sqrt{s}-\sqrt{s_h-s}}{\sqrt{s}+\sqrt{s_h-s}};\quad
 s_h^{1/2}=1450\;\mev.\cr
}
$$
$B_{h2}$  is a free parameter. We get a reasonable $\chidof=13.8/(13-1)$ and
$$B_{h2}=112\pm38.
\equn{(2.3b)}$$
Both the present fit and that in PY05 may be found in \fig~2.2.

The present fit has slightly smaller errors than the old fit
 in PY05 in the low energy region, 
and larger ones above 1~\gev.

The inelasticity  we still describe by the 
empirical fit in PY05:
$$\eta_0^{(2)}(s)=\cases{
1-\epsilon(1-s_l
/s)^{3/2},\quad 
\epsilon=0.18\pm0.12;\quad s>{s_l}=(1.05\;\gev)^2;\cr
1,\qquad s<{s_l}=(1.05\;\gev)^2.\cr}
\equn{(2.4)}$$

\booksubsection{2.3. The inelasticity for the D0 wave}

\noindent
We now consider the elasticity parameter for the D0 wave. In PY05, KPY06 we made a {\sl
phenomenological}  fit to $\eta_2^{(0)}$:

\topinsert{
\setbox0=\vbox{\psfig{figure=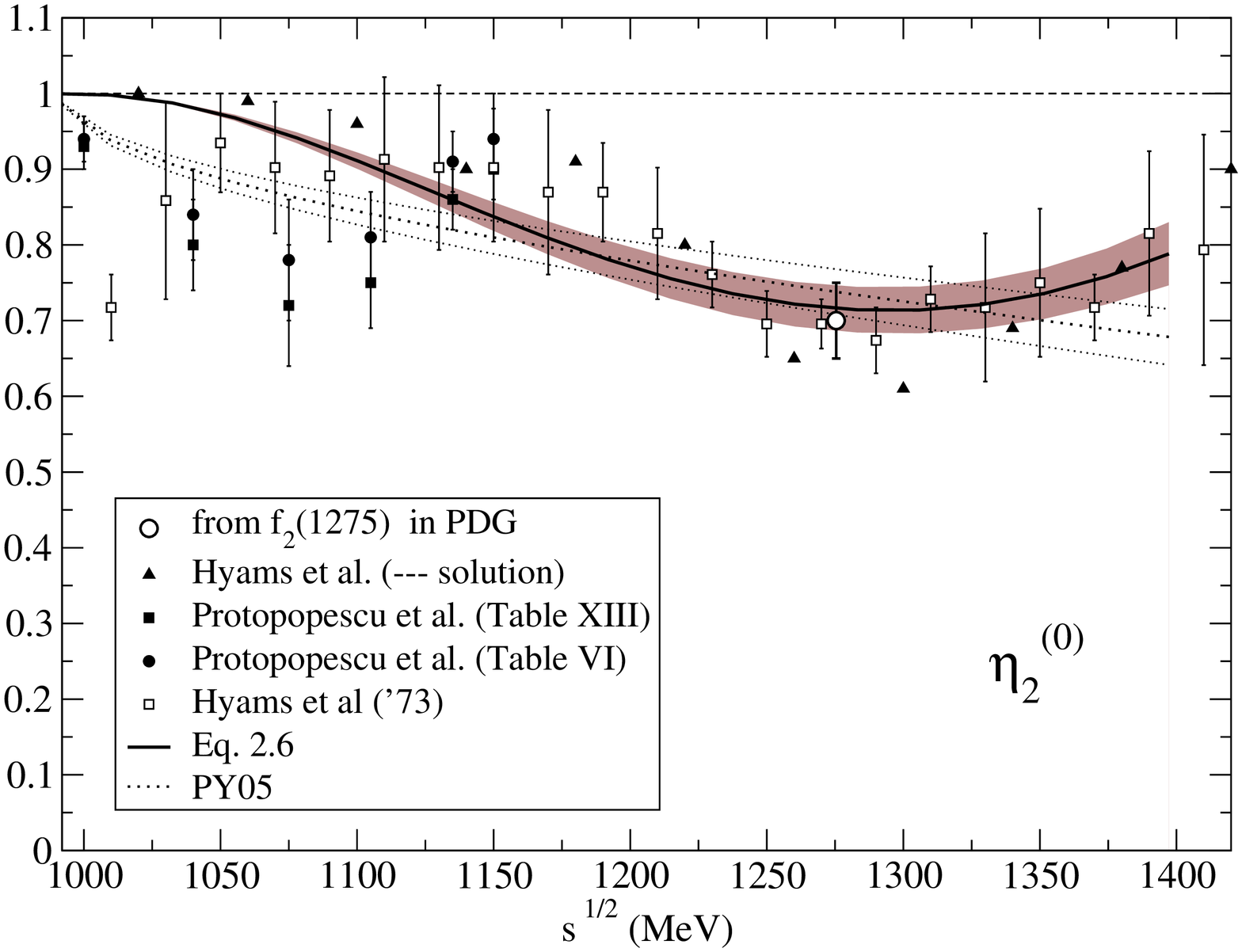,width=12.2truecm,angle=-90}}  
\setbox6=\vbox{\hsize 11.truecm\noindent\petit\figurasc{Figure 2.3. }{Fit
 to $\eta_2^{(0)}$ (continuous line and dark area
that covers the uncertainty), from Eq.~(2.6) here. 
The dotted lines follow from KPY06 [Eq. (2.5) here].
The  elasticity on the $f_2(1270)$, from 
 the Particle Data Tables, is also shown as a large white dot; 
other data are from ref.~6.
}} 
\centerline{\tightboxit{\box0}}
\medskip
\centerline{\box6}
\bigskip}\endinsert

$$\eqalign{
\eta_2^{(0)}(s)=&\,\cases{1,\qquad  s< 4m_K^2,\;
\cr
1-\epsilon\,\dfrac{k_2(s)}{k_2(M^2_{f_2})},\quad
 s> 4m_K^2;\quad\epsilon=0.262\pm0.030,\quad
 k_2=\sqrt{s/4-m^2_K}.\cr} \cr
}
\equn{(2.5)}$$
This  provides a fit to the elasticity parameter on the average.

The problem with (2.5) is that
 it rises too brusquely at $\bar{K}K$
threshold,\fnote{We thank H.~Leutwyler 
for this remark. This type of  problem is important for the D0 wave, because  $\bar{K}K$
threshold is not far  from the very strong $f_0(1270)$ resonance; but it also exists for 
P and D2 waves. We have checked that, in the case of these waves, the influence of 
the incorrect threshold behaviour is much smaller than our errors, 
hence negligible.}  proportional to $k_2$,
which causes distortions for
$s^{1/2}$ near
$2m_K$; 
 a behaviour $k_2^5$ near this threshold would be  indicated. 
For this reason, we here also try the following parametrization:
$$\eqalign{
\eta_2^{(0)}(s)=&\,\cases{1,\qquad  s< 4m_K^2,\phantom{\Big|}
\cr
1-\epsilon\,\left(1-\dfrac{4m^2_K}{s}\right)^{5/2}
\left(1-\dfrac{4m^2_K}{M^2_{f_2}}\right)^{-5/2}
\left\{1+r\left[1-\dfrac{k_2(s)}{k_2(M^2_{f_2})}\right]\right\},\cr
 s> 4m_K^2;\quad
 k_2=\sqrt{\dfrac{s}{4}-m^2_K}.\cr} \cr
\epsilon=&0.284\pm0.030;\quad r=2.54\pm0.31;\quad M_{f_2}=1275.4\;\mev.\cr
}
\equn{(2.6)}$$
This will underestimate the inelasticity near $2m_K$, since, in fact, most of the inelasticity of the 
D0 wave is {\sl not} due to  $\bar{K}K$ but to the four pion states;
 but will have a smooth  
threshold behaviour. Moreover, (2.6) is more flexible than (2.5) and it will, by construction,
 have the correct inelasticity 
around the $f_2(1270)$ resonance, which is the most important region. 
If evaluating the dispersion relations with  (2.5), (2.6) 
 we find that they are almost equivalent. 
We will 
use (2.6) in our calculations in the present paper because, as stated, it reproduces best 
the inelasticity around the  $f_2(1270)$ resonance.

The  elasticity parameter here and that from PY05 are shown in \fig~2.3.

\booksubsection{2.4. The D2 wave   over the whole range}

\noindent
Finally, we discuss the D2 wave. The data for the corresponding phase shift 
are very poor (and nonexistent for the inelasticity). 
For this reason,  
we made, in PY05, a {\sl phenomenological} fit in the whole energy range from 
threshold to 1.42~\gev. 
We wrote
$$\cot\delta_2^{(2)}(s)=
\dfrac{s^{1/2}}{2k^5}\,\Big\{B_0+B_1 w(s)+B_2 w(s)^2\Big\}\,
\dfrac{{M^4_\pi} s}{4({M^2_\pi}+\deltav^2)-s}
\equn{(2.7a)}$$
with $\deltav$ a free parameter fixing the zero of the phase shift and
$$w(s)=\dfrac{\sqrt{s}-\sqrt{s_0-s}}{\sqrt{s}+\sqrt{s_0-s}},\quad
 s_0^{1/2}=1450\,\mev,$$
and we  got a mediocre fit, $\chidof=71/(25-4)$;  the values of the parameters
 were
$$B_0=(2.4\pm0.3)\times10^3,\quad B_1=(7.8\pm0.8)\times10^3,\quad
 B_2=(23.7\pm3.8)\times10^3,\quad
\deltav=196\pm20\,\mev.
\equn{(2.7b)}$$
When we fitted requiring fulfillment of dispersion relations, 
 the parameters moved to (PY05, Appendix~A)
$$B_0=(2.9\pm0.2)\times10^3,\quad B_1=(7.3\pm0.8)\times10^3,
\quad B_2=(25.4\pm3.6)\times10^3,\quad
\deltav=212\pm19\,\mev.
\equn{(2.7c)}$$
The corresponding phase shift moves by a little more than one standard 
deviation.
This shows that the errors in (2.7b) were  underestimated. 
We improve this by keeping the central values of the parameters 
we found in the fit to data, but enlarging the errors 
by adding quadratically the difference between the central values in (2.7a) and (2.7b), 
which may be considered as a ``systematic" error in the fit.\fnote{It 
should be remarked that the present method, as indeed any method
 for estimating ``systematic" errors, is arbitrary to a large extent. In PY05 
we took into account the poor quality of the fit (which reflects the incompatibility 
of the various sets of experimental data) by scaling the 
purely statistical errors by the square root of the \chidof,
$\sqrt{71/(25-4)}\simeq1.8$. This produced (2.7b). We could 
have scaled instead by the \chidof\ itself; this would have given errors 
like in (2.7d) for the  parameter, $B_0$, and
a bit larger than those in (2.7d) for the other 
parameters. Likewise, we could have 
taken as central values the averages of the central values in (2.7b) and (2.7c).
We consider our method to be reasonable and, as we find in the present paper, 
the increase in the errors  quite justified.} 
Thus we find the parameters
$$B_0=(2.4\pm0.5)\times10^3,\quad B_1=(7.8\pm1.0)\times10^3,\quad
 B_2=(23.7\pm4.2)\times10^3,\quad
\deltav=196\pm25\,\mev.
\equn{(2.7d)}$$
It should be noted that the errors in $\delta_2^{(2)}$ are still small, 
compared to the experimental errors; 
(2.7d) gives errors for $\delta_2^{(2)}$  below the level of 0.8\degrees.

For the elasticity parameter we  write the same formula as in KPY06: 
 above $1.05\,\gev$,
$$\eta_2^{(2)}(s)=1-\epsilon (1-\hat{s}/s)^3,\quad \hat{s}^{1/2}=1.05\;\gev,\quad
\epsilon=0.2\pm0.2.
\equn{(2.7e)}$$

\booksubsection{2.5. The G0 wave}

\noindent
With respect to the G0 wave, we gave in PY05 a parametrization for its imaginary part 
based on dominance by the resonance $f_4$,
$$\eqalign{
\imag \hat{f}_4^{(0)}(s)=&\,\left(\dfrac{k(s)}{k(M^2_{f_4})}\right)^{18}
{\rm BR}\dfrac{M^2_{f_4}\gammav^2}{(s-M^2_{f_4})^2+M^2_{f_4}
\gammav^2[k(s)/k(M^2_{f_4})]^{18}};\cr
s^{1/2}\geq 1\;\gev;\qquad{\rm BR}=&\,0.17\pm0.02,\quad
 M_{f_4}=2025\pm8\;\mev,\quad \gammav=194\pm13\;\mev.\cr}
\equn{(2.8)}$$
Unfortunately, this much {\sl underestimates} the value at low energy. 
In fact, from the Froissart-Gribov representation we can evaluate 
rather accurately the scattering length, finding
$ a_4^{(0)}=(8.0\pm0.4)\times10^{-6}\,M_{\pi}^{-9}$. 
At low energy one can write
$$\imag \hat{f}_4^{(0)}(s)\simeq [k(s)]^{18}\,[a_4^{(0)}]^2,
\equn{(2.9)}$$
which disagrees with what one would find from (2.8a) by many orders of magnitude. 
A simple formula that interpolates between low and high energy is 
$$\eqalign{
\imag \hat{f}_4^{(0)}(s)=&\,\left(\dfrac{k(s)}{k(M^2_{f_4})}\right)^{18}
{\rm BR}\dfrac{M^2_{f_4}\gammav^2\ee^{2c(1-s/M^2_{f_4})^2}}{(s-M^2_{f_4})^2+M^2_{f_4}
\gammav^2[k(s)/k(M^2_{f_4})]^{18}};\cr
{\rm BR}=&\,0.17\pm0.02,\quad
 M_{f_4}=2025\pm8\;\mev,\quad \gammav=194\pm13\;\mev;\quad c=9.23\pm0.46.\cr}
\equn{(2.10)}$$
This gives the correct values near the $f_4$ resonance and at low energy, the only 
regions where we have experimental information.
Note that this interpolation is rather arbitrary, but there is no point in trying to 
improve it as there are no data to which one can fit more realistic formulas. 
Any reasonable interpolation would give the same {\sl order of magnitude} 
estimate for the contribution of this wave to dispersion relations and sum rules.

We have, in our
calculations, not taken into account the contributions of G0 or G2 waves other than  to
check that they are considerably smaller than the  experimental errors: this is the only interest of
the parametrization.

\booksection{2.6. Low energy S0, P and F waves: ghost removal}

\noindent
When cutting the low energy expansions 
$$\cot\delta(s)=K(s)\{B_0+B_1w+\cdots\}$$
(with $K$ an appropriate kinematical factor) at a finite order, a ghost, i.e.,  a spurious 
pole in the partial wave amplitude $\hat{f}(s)$ appears in the vicinity of the point $s=0$ 
for the S0, P and F waves. 
As remarked in
GMPY07 (see especially Appendix~A there), such ghost poles are rather harmless, 
their effect being at the percent level: removing the ghosts is 
 little more than an aesthetical 
 requirement. Nevertheless, we will here improve our formulas by 
writing them in such a manner that the
ghosts disappear. This was already done for the S0 wave in
GMPY07, and the S0 parametrization used in the present paper takes this into account; now
we remove also the ghosts for P and F
waves.  As a matter of fact, this improves the consistency of our results, slightly but
systematically.

In PY05 and KPY06 we used the following  formula for the P wave,
$$\cot\delta_1(s)=\dfrac{s^{1/2}}{2k^3}
(M^2_\rho-s)\left\{B_0+B_1\dfrac{\sqrt{s}-\sqrt{s_0-s}}{\sqrt{s}+\sqrt{s_0-s}}
\right\};\quad s_0^{1/2}=1.05\;\gev.
\equn{(2.11a)}$$
The best result, from ref.~8 (see also PY05), is 
\smallskip
$$\eqalign{
B_0=&\,1.069\pm0.011,\quad B_1=0.13\pm0.05,\quad M_{\rho}=773.6\pm0.9\,{\rm MeV}.
\cr }
\equn{(2.11b)}$$

Instead of this we now write  a parametrization where the ghost is absent,
$$\cot\delta_1(s)=\dfrac{s^{1/2}}{2k^3}
(M^2_\rho-s)\left\{\dfrac{2M^3_\pi}{M^2_\rho
\sqrt{s}}+B_0+B_1\dfrac{\sqrt{s}-\sqrt{s_0-s}}{\sqrt{s}+\sqrt{s_0-s}}
\right\};\quad s_0^{1/2}=1.05\;\gev,
\equn{(2.12a)}$$
and find
$$B_0=1.055\pm0.011,\quad B_1=0.15\pm0.05;
\equn{(2.12b)}$$ 
the difference with what follows from (2.11) is less than 0.7\% at $\pi\pi$ 
threshold, decreasing to
0.05\%  at $\bar{K}K$ threshold.

For the F wave, one can remove the ghost without changing the parameters we found in 
PY05 (within the significant digits), so we have
$$\eqalign{
\cot\delta_3(s)=&\,\dfrac{s^{1/2}}{2k^7}\,M^6_\pi\,
\left\{\dfrac{2\lambda
M_\pi}{\sqrt{s}}+B_0+B_1\dfrac{\sqrt{s}-\sqrt{s_0-s}}{\sqrt{s}+\sqrt{s_0-s}}\right\},\quad
s_0^{1/2}=1.45\;\gev;
\cr
 B_0=&\,(1.09\pm0.03)\times 10^5,\quad
B_1=(1.41\pm0.04)\times 10^5,\quad \lambda=0.051\times 10^5.
\cr}
\equn{(2.13)}$$

\booksubsection{2.7. Other waves}

\noindent
The results of the remaining fits for this UFD Set may be found in PY05 and KPY06,
 with details of the fitting procedure and
the 
far from trivial matter of the selection of data.
The results of all the fits are collected in Appendix~A, for ease of reference.

\booksubsection{2.8. Matching}

\noindent
Before turning to the calculations of forward dispersion relations, and Roy equations, a few words 
have to be said about the matching points between our low energy and intermediate energy
regimes,  and between 
intermediate energy and high energy regime, that we discuss later. Such matchings are, of course,
artificial: for example, and as discussed about the D0 wave, the inelasticity is small, but not
zero, below 
$\bar{K}K$ threshold. A fully satisfactory matching is however not possible; it would require a
multichannel evaluation, and hence introducing a number of parameters impossible to fix 
with the existing experimental data. In KPY06, and here, we have requested 
matching of the {\sl central values} of the  phase shifts,
 $\delta_l^{(I)}(s)$, at the 
matching points themselves, usually (but not always) the $\bar{K}K$ threshold.
However, and because the errors in the low energy and intermediate energy ranges are independent, 
this produces jumps (when varying the parameters inside their error bars) 
which increase the errors of the dispersive integrals artificially. 
This is, unfortunately, an unavoidable feature of our analysis: the 
results deteriorate somewhat when one is very near the matching points.

In fact, the situation is even less clear near the $\bar{K}K$ threshold. 
In our analysis we neglect isospin breaking effects, and therefore
we have taken it at an average between the $K^+K^-$ and $\bar{K}^0K^0$ 
thresholds,  $s^{1/2}=992\,\mev$. 
Since the  $K^+K^-$ and $\bar{K}^0K^0$ 
thresholds differ by some 8~\mev, the threshold itself is thus
not  well defined to this extent: $992\,\pm4\,\mev$. 
All in all, the net result is that our dispersive (or even direct calculations) of 
the $\pi\pi$ amplitudes suffer from uncontrollable errors in a, fortunately narrow, 
band of less than or around 6~\mev\ (to be on the safe side)
 around the matching points. We have
avoided these matching regions  when calculating the fulfillment of dispersive relations.

With respect to the matching between intermediate and high energy regions, the situation is different. 
It is clear that, near $s^{1/2}=1420\,\mev$, which is the corresponding 
matching point, the Regge expression can only agree with the real amplitudes in the mean
(as can be seen in the cases where we have precise data, as  for  $\pi N$ scattering).
We expect that, since these Regge amplitudes only appear in integrals, 
the fluctuations will be averaged out to lie within the 
errors.
 However, we here have a problem similar
to  that of the low energy matching with intermediate energy: 
this lack of correlation of the errors  causes artificially enhanced error bars near the 
matching points. Therefore, our calculation should be used  
excluding a (narrow) band below the matching point. 
Here we have refrained to compare calculations above
1400~\mev, which is  sufficient to render most of the  fluctuations 
smaller than the experimental errors.

\booksection{3. Regge formulas}

\noindent
Regge formulas have been obtained for $\pi\pi$ scattering,
 {\sl in the forward direction}, 
by fitting experimental data for the various $\pi\pi$ total 
cross sections. 
This provides expressions that are not very precise.
One  improves this 
by  use of factorization. It is then 
 possible to include information on total cross sections 
for $\pi N$ and $NN$ scattering\ref{9} ($NN$
 includes antinucleon-nucleon
scattering), which furnishes us with 
precise results for the contributions to 
$\pi\pi$ scattering of the three Regge poles\fnote{The $P'$
is in  fact a combination of two Regge trajectories, associated with 
the $f_2(1270)$ and $f'_2(1525)$ resonances.} 
$P$, $P'$ and $\rho$. Here, we will use these Regge 
expressions above 1.42~\gev. 

These results are sufficient to calculate {\sl forward} 
dispersion relations. For Roy equations, however, we require also 
the imaginary parts of the scattering amplitudes, $\imag F(s,t)$, for rather large values of 
$|t|$; in our calculation, 
up to $t=-0.43\,\gev^2$. In fact, these values are so large that 
one does not expect Regge theory to hold in the extreme range. 
What we do to circumvent this problem is to {\sl enlarge} 
the errors in the $t$ dependence of the parametrizations so that they cover, 
in the whole $t$ range, all fits to experimental data. 
This will provide, at large $t$, a {\sl phenomenological} representation of the 
corresponding scattering amplitude.
For isospin zero exchange, we take the expressions given in PY05, in the forward
direction, and enlarge the errors away from the forward direction. For
$\rho$ exchange  we take, at $t=0$, the parameters described in KPY06  and, for 
$t\neq0$, an uncertainty that covers the extreme
fits in ref.~10 for $\pi N$ scattering, and assume that $\pi\pi$ scattering will vary in a
similar manner.

We 
 write
$$\eqalign{
\imag F^{(I_t=1)}(s,t)\simeqsub_{{s\to\infty}\atop{t\,{\rm fixed}}}&\,
\beta_\rho\,\dfrac{1+\alpha_\rho(t)}{1+\alpha_\rho(0)}\,
\phiv(t)\ee^{bt}(s/\hat{s})^{\alpha_\rho(t)};\quad
\alpha_\rho(t)=\,\alpha_\rho(0)+t\alpha'_\rho+\tfrac{1}{2}t^2\alpha''_\rho;\cr
\beta_\rho=&\,1.22\pm0.14,\quad\alpha_\rho(0)=0.46\pm0.02;\quad
\alpha'_\rho=0.90\;{\gev}^{-2};\quad \alpha''_\rho=-0.3\;{\gev}^{-4}\cr
\phiv(t)=&\,1+d_\rho t+e_\rho t^2;
\quad b=2.4\pm0.2\;{\gev}^{-2}.\cr}
 \equn{(3.1a)}$$
$\hat{s}$ is a scale parameter, that we consistently take $\hat{s}\equiv1\,\gev^2$. 
We set
$$d_\rho=2.4\pm0.5,\quad e_\rho=0\pm2.5\;{\gev}^{-4}.
\equn{(3.1b)}$$

For the Pomeron and $P'$, one can write, also for $ s^{1/2}>1.42\;\gev$,
$$\eqalign{ 
\imag F^{(I_t=0)}_{\pi\pi}(s,t)&\,\simeqsub_{{s\to\infty}\atop{t\,{\rm fixed}}}
P(s,t)+P'(s,t),\cr
P(s,t)=&\,\beta_P\psiv_P(t)\,\alpha_P(t)\,\dfrac{1+\alpha_P(t)}{2}\,
\ee^{bt}(s/\hat{s})^{\alpha_P(t)},\quad
\alpha_P(t)=1+t\alpha'_P;\cr
P'(s,t)=&\,\beta_{P'}\psiv_{P'}(t)\,\dfrac{\alpha_{P'}(t)
[1+\alpha_{P'}(t)]}{\alpha_{P'}(0)[1+\alpha_{P'}(0)]}\,
\ee^{bt}(s/\hat{s})^{\alpha_{P'}(t)},\quad
\alpha_{P'}(t)=\alpha_{P'}(0)+t\alpha'_{P'};\cr
\beta_P=&\,2.54\pm0.04,\quad  \alpha'_P=0.20\pm0.10\;{\gev}^{-2},\quad \psiv_P(t)= 1+c_P t;\cr 
\beta_{P'}=&\,0.83\pm0.05,\quad\alpha_{P'}(0)=0.54\pm0.02,\quad
 \alpha'_{P'}=0.90\;{\gev}^{-2};\cr
\psiv_{P'}(t)=&\,1+c_{P'} t;\quad b=2.4\pm0.2\,{\gev}^{-2}.\cr
}
\equn{(3.2a)}$$
We may fix 
$$c_P=0.0\pm1.0\;\gev^{-2};\quad c_{P'}=-0.4\pm0.4\;\gev^{-2}.
\equn{(3.2b)}$$
If we do so, we cover the fits of Rarita et al.\ref{10} and of Froggatt and
Petersen.\ref{11}

Note that we do not give  errors for the slopes $\alpha'_\rho$ and $\alpha'_{P'}$
because the variation of $\phiv$, $\psiv_{P'}$  
covers  possible variations of the Regge slopes:
$\phiv$ varies a lot at large $t$. In fact,  we have 
a range of variation
$$-0.56\lsim\phiv(-0.4\;{\gev}^2)\lsim0.64,$$
and something similar for $\psiv_{P'}$.
 Fortunately, however, the evaluations for the Roy equations below 1~\gev\
 depend very little on the 
 scattering amplitudes for large $s$ and large $|t|$.

Finally, for exchange of isospin two we write 
$$\imag F^{(I_t=2)}(s,t)\simeqsub_{s\to\infty}
\beta_2\ee^{bt}(s/\hat{s})^{\alpha_\rho(t)+\alpha_\rho(0)-1},\quad\beta_2=0.2\pm0.2; \quad
s\geq(1.42\;{\gev})^2. 
\equn{(3.3)}$$ 

These fits are expected to represent {\sl experimental} data 
for energies between 1.42~\gev\ and $\sim20\,\gev$ and for 
$4M^2_\pi\geq t\gsim -0.4\,\gev^2$, with less reliability at the 
more negative values of $t$.
At  values of $s^{1/2}$  larger than 20~\gev, one would have to use more complicated
formulas, taking into account in particular the logarithmic growth of the total cross
sections.\ref{9} 
For our purposes the formulas given above are sufficiently accurate, since the influence
of the energy region 
much above 20~\gev\ for forward dispersion relations or Roy equations is 
negligible.

The values of the Regge parameters can be improved by requiring verification of dispersion relations,
and of two 
sum rules that relate directly the Regge behaviour, for nonzero $t$, to low energy 
amplitudes [see below Eqs.~(5.1) and (5.2)]. 
The resulting numbers are collected in Appendix~B.

\booksection{4. Dispersion relations}
\vskip-0.5truecm
\booksubsection{4.1. Forward dispersion relations}

\noindent
In this Section we will evaluate forward dispersion relations for the three
independent $\pi\pi$ scattering amplitudes. 
For these calculations we will take the parameters for all partial waves from the fits to
data described in the previous Sections (and collected in Appendix~A).

Although the form of the dispersion relations has been given before, we repeat them here.
For 
 $\pi^0\pi^0$ scattering
we write
$$\real F_{00}(s)-F_{00}(4M_{\pi}^2)=
\dfrac{s(s-4M_{\pi}^2)}{\pi}\pepe\int_{4M_{\pi}^2}^\infty\dd s'\,
\dfrac{(2s'-4M^2_\pi)\imag F_{00}(s')}{s'(s'-s)(s'-4M_{\pi}^2)(s'+s-4M_{\pi}^2)}.
\equn{(4.1a)}$$
The result of the calculation is shown in \fig~4.1{\sc a}, where the continuous curve is
the  real part evaluated from the parametrizations, 
and the dashed line  is the result of
the dispersive integral, i.e., the right hand side of 
(4.1a).

The dispersion relation  for  $\pi^0\pi^+$ 
scattering reads, with $F_{0+}(s)$ the forward $\pi^0\pi^+$ amplitude,
$$\real F_{0+}(s)-F_{0+}(4M_{\pi}^2)=
\dfrac{s(s-4M^2_\pi)}{\pi}\pepe\int_{4M_{\pi}^2}^\infty\dd s'\,
\dfrac{(2s'-4M^2_\pi)\imag F_{0+}(s')}{s'(s'-s)(s'-4M_{\pi}^2)(s'+s-4M_{\pi}^2)}.
\equn{(4.1b)}$$
In \fig~4.1{\sc b} we show the fulfillment of (4.1b).

Finally, the dispersion relation for the $I_t=1$ scattering 
amplitude does not require subtractions, and  reads 
$$\real F^{(I_t=1)}(s,0)=\dfrac{2s-4M^2_\pi}{\pi}\,\pepe\int_{4M^2_\pi}^\infty\dd s'\,
\dfrac{\imag F^{(I_t=1)}(s',0)}{(s'-s)(s'+s-4M^2_\pi)}. 
\equn{(4.1c)}$$
The result  is shown graphically in \fig~4.1{\sc c}.

{
\setbox3=\vbox{
{\psfig{figure=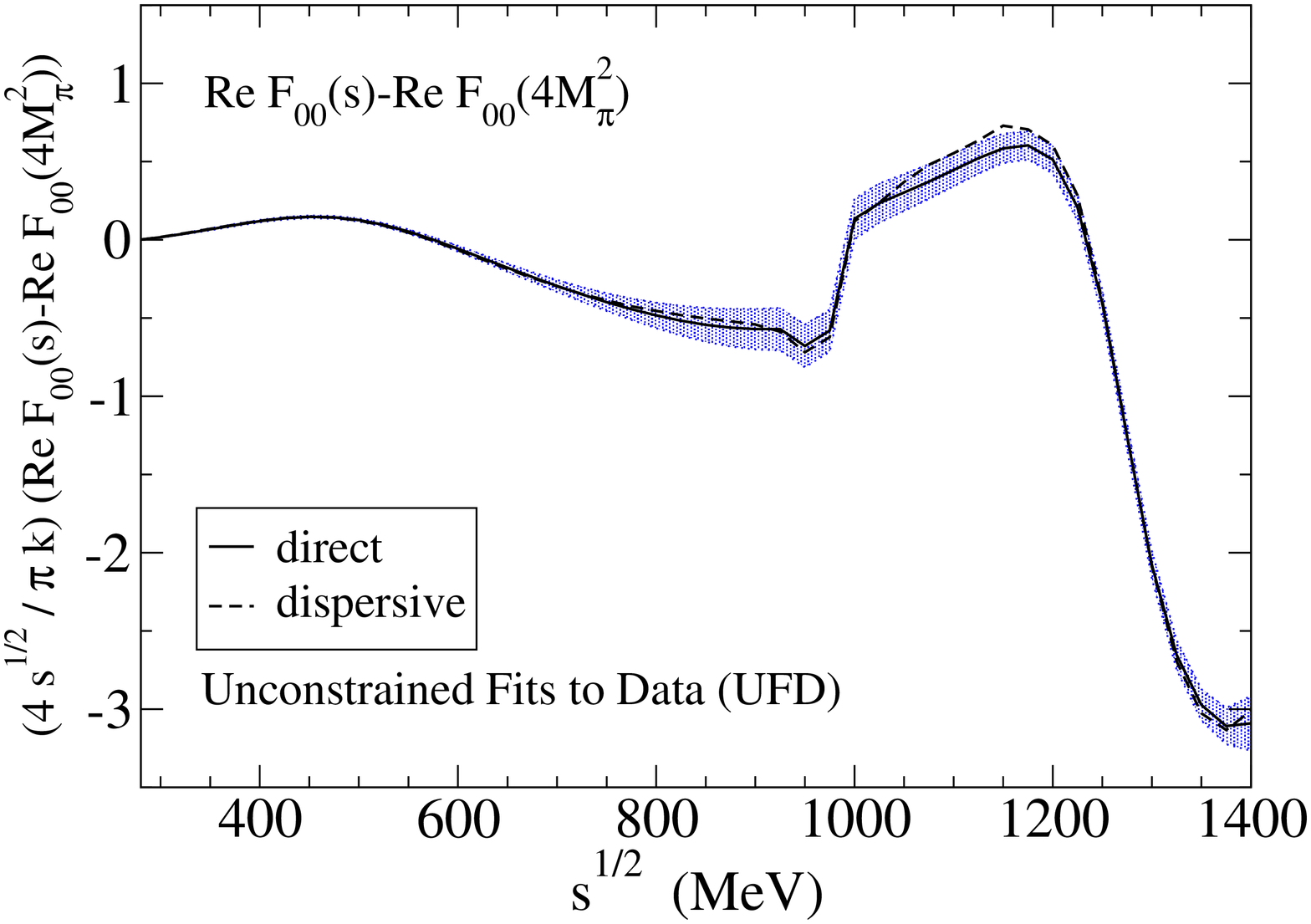,width=9.0truecm,angle=-90}}} 
\setbox6=\vbox{\hsize 5.4truecm\noindent\petit\figurasc{Figure 4.1a. }{\hfuzz1.truecm
The $\pi^0\pi^0$ dispersion relation.
 Continuous line: left hand side of (4.1a), evaluated
directly
 with the
parametrizations (the gray band covers the error).
 Dashed line: the result of the dispersive integral.
\hb\phantom{x}\hb\phantom{x}\hb\phantom{x}
}}
\line{\otightboxit{\box3}\hfil\box6}
\setbox3=\vbox{\hfuzz1.truecm
{\psfig{figure=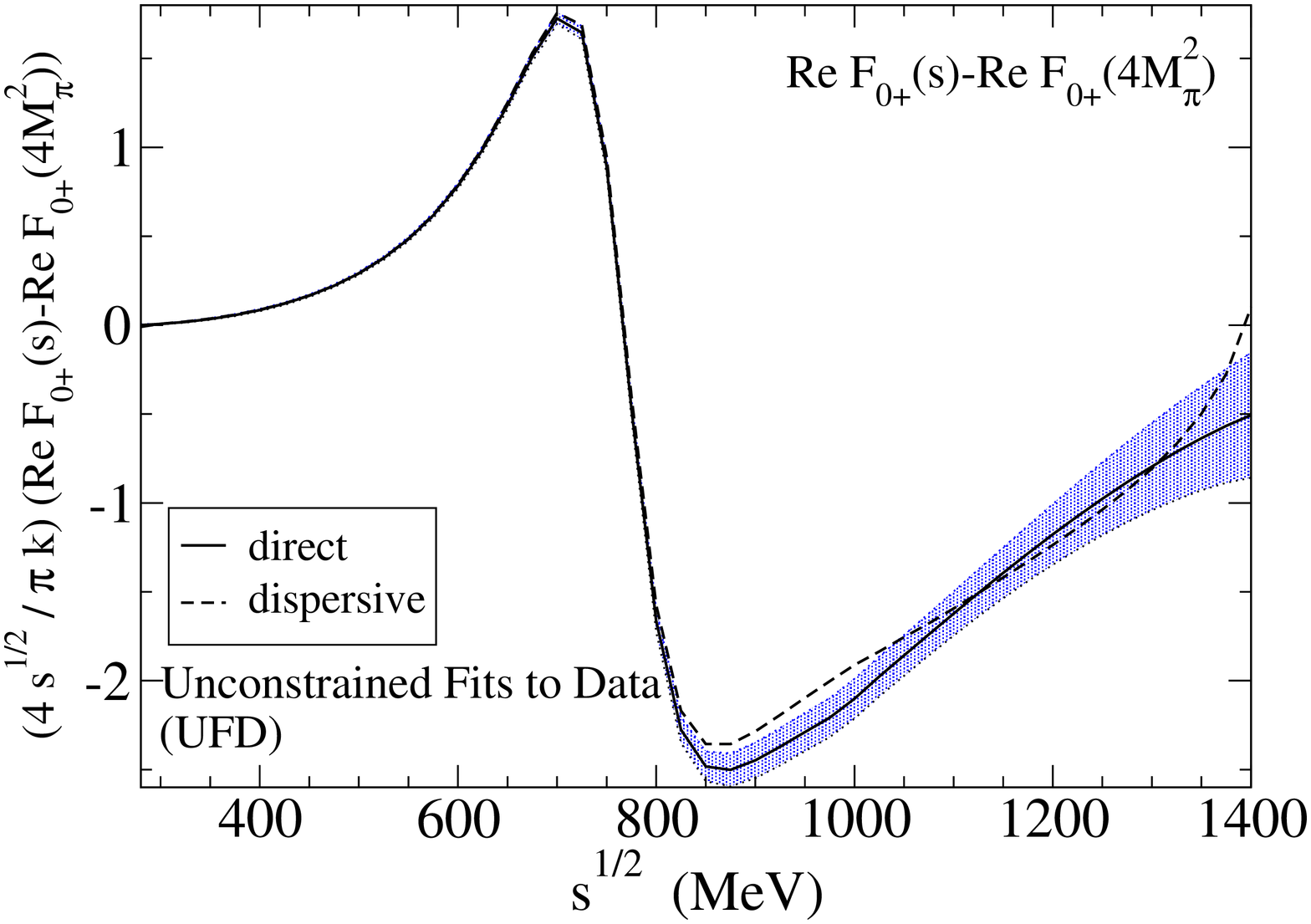,width=9.0truecm,angle=-90}}} 
\setbox6=\vbox{\hsize 5.4truecm\noindent\petit\figurasc{Figure 4.1.b }{
The $\pi^0\pi^+$ dispersion relation with the
 new P and D2 waves. Continuous line:  left hand side of (4.1b), evaluated 
directly with the
parametrizations. The gray band covers the errors.\hb
 Dashed line: the result of the dispersive integral.\hb\phantom{x}
\hb\phantom{x}\hb\phantom{x}
}}
\line{\otightboxit{\box3}\hfil\box6}
\setbox3=\vbox{\hfuzz1truecm
{\psfig{figure=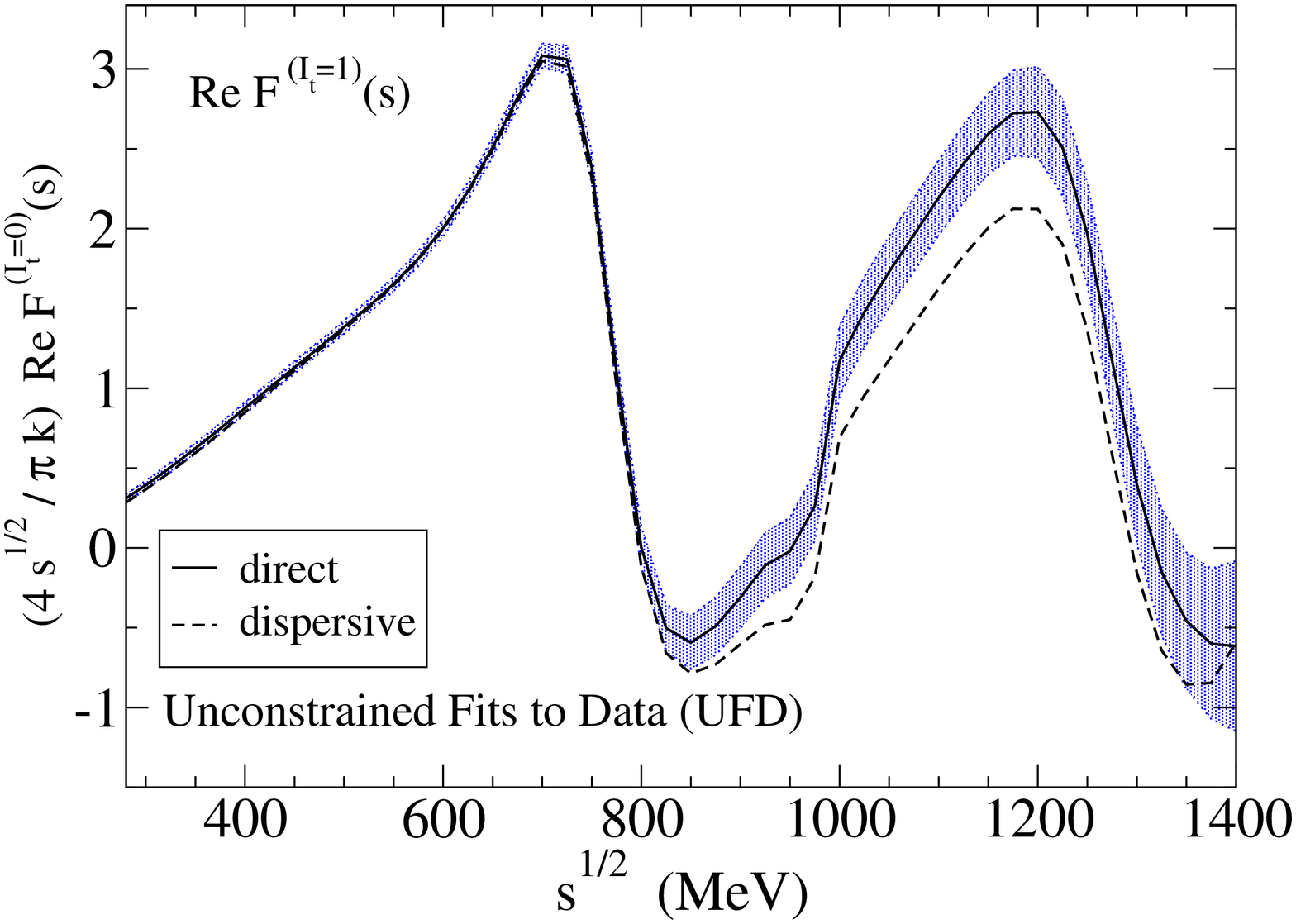,width=9.truecm,angle=-90}
}} 
\setbox6=\vbox{\hsize 5.4truecm\noindent\petit\hfuzz0.5truecm\figurasc{Figure 4.1c. }{
The  dispersion relation for the $I_t=1$ amplitude.
Continuous line: real part and error (shaded
area)
 evaluated directly with the para\-metrizations.
 Dashed line: the result of the dispersive integral.\hb\phantom{x}\hb\phantom{x}
\hb\phantom{x}
}}
\line{\otightboxit{\box3}\hfil\box6}
}

\goodbreak

To measure {\sl quantitatively} the fulfillment of the dispersion relations we evaluate 
the average (squared) distance between the real parts, calculated 
with our parametrizations, and the same real parts but now 
calculated with the aid of the dispersive integrals, a quantity that 
we denote by $\bar{d}^2$. This quantity is
defined as follows. First, we rewrite the dispersion relations as
the discrepancies $\deltav_i$ (shown graphically in \fig~4.2), 

$$\deltav_{00}(s)\equiv\real F_{00}(s)-F_{00}(4M_{\pi}^2)-
\dfrac{s(s-4M_{\pi}^2)}{\pi}\pepe\int_{4M_{\pi}^2}^\infty\dd s'\,
\dfrac{(2s'-4M^2_\pi)\imag F_{00}(s')}{s'(s'-s)(s'-4M_{\pi}^2)(s'+s-4M_{\pi}^2)},
\equn{(4.2a)}$$

$$\deltav_{0+}(s)\equiv\real F_{0+}(s)-F_{0+}(4M_{\pi}^2)-
\dfrac{s(s-4M^2_\pi)}{\pi}\pepe\int_{4M_{\pi}^2}^\infty\dd s'\,
\dfrac{(2s'-4M^2_\pi)\imag F_{0+}(s')}{s'(s'-s)(s'-4M_{\pi}^2)(s'+s-4M_{\pi}^2)},
\equn{(4.2b)}$$
and
$$\deltav_{1}(s)\equiv\real
F^{(I_t=1)}(s,0)-\dfrac{2s-4M^2_\pi}{\pi}\pepe\int_{4M^2_\pi}^\infty\dd s'\,
\dfrac{\imag F^{(I_t=1)}(s',0)}{(s'-s)(s'+s-4M^2_\pi)}.
\equn{(4.2c)}$$
These quantities would vanish,  $\deltav_i=0$, 
if the dispersion relations were exactly satisfied.
Because our fits have errors, we can only require vanishing within the uncertainties that 
said  errors induce in the $\deltav_i$, that we call
  $\delta\deltav_i$. Therefore, we define the quantities
({\sl average discrepancies})
$$\bar{d}_i^2\equiv\dfrac{1}{\hbox{number of points}}
\sum_n\left(\dfrac{\deltav_i(s_n)}{\delta\deltav_i(s_n)}\right)^2.
\equn{(4.3)}$$
This we do for all three relations (4.2). 
The values of the $s_n$ are taken   at energy intervals of 25~\mev. 
For the dispersion relation for $I_t=1$, we also include the value at threshold, 
known at times as the (first) Olsson sum rule. 
For the other two dispersion relations, since 
the $\deltav$ vanish identically at threshold, 
we include a point {\sl below} threshold, at $s=2M^2_\pi$.
This is useful, among other things, to fix the location of the Adler zeros for the S0, S2 waves.

If we had a fit to FDRs (which we {\sl do not}) instead of an {\sl evaluation}, 
$\bar{d}^2$ would be the average chi-squared of the fit: in our case, 
$\bar{d}^2$ 
is simply a measure of how well the forward dispersion 
relations are satisfied by the data fits, which are
{\sl independent} for each wave, and independent of dispersion relations.
When calculating this $\bar{d}^2$, we  use, in the present Section, the parameters for 
phase shifts and inelasticities discussed in the previous Sections, 
and collected in Appendix~A.

The  average discrepancies in the various cases are given in Eq.~(4.4) below:
$$\matrix{&s^{1/2}\leq
932\;\mev&s^{1/2}\leq1420\;\mev\vphantom{\Big|}\cr
\pi^0\pi^0\;\;\hbox{FDR}&\bar{d}^2=0.12&\bar{d}^2=0.29\cr
\pi^0\pi^+\;\;\hbox{FDR}&\bar{d}^2=0.84
&\bar{d}^2=0.86\cr
I_t=1\;\;\hbox{FDR}&\bar{d}^2=0.66&\bar{d}^2=1.87.\cr
}\equn{(4.4)}$$ 
Below 932 MeV, the average $\bar{d}^2$ shows a remarkably good fulfillment of
FDRs. Still, the situation
over the whole range is such that the  $\bar{d}^2$  for the $I_t=1$ FDR is well
above unity, indicating
that there is room for improvement.

\topinsert{
\setbox0=\vbox{\hsize8truecm

\setbox1=\vbox{{\psfig{figure=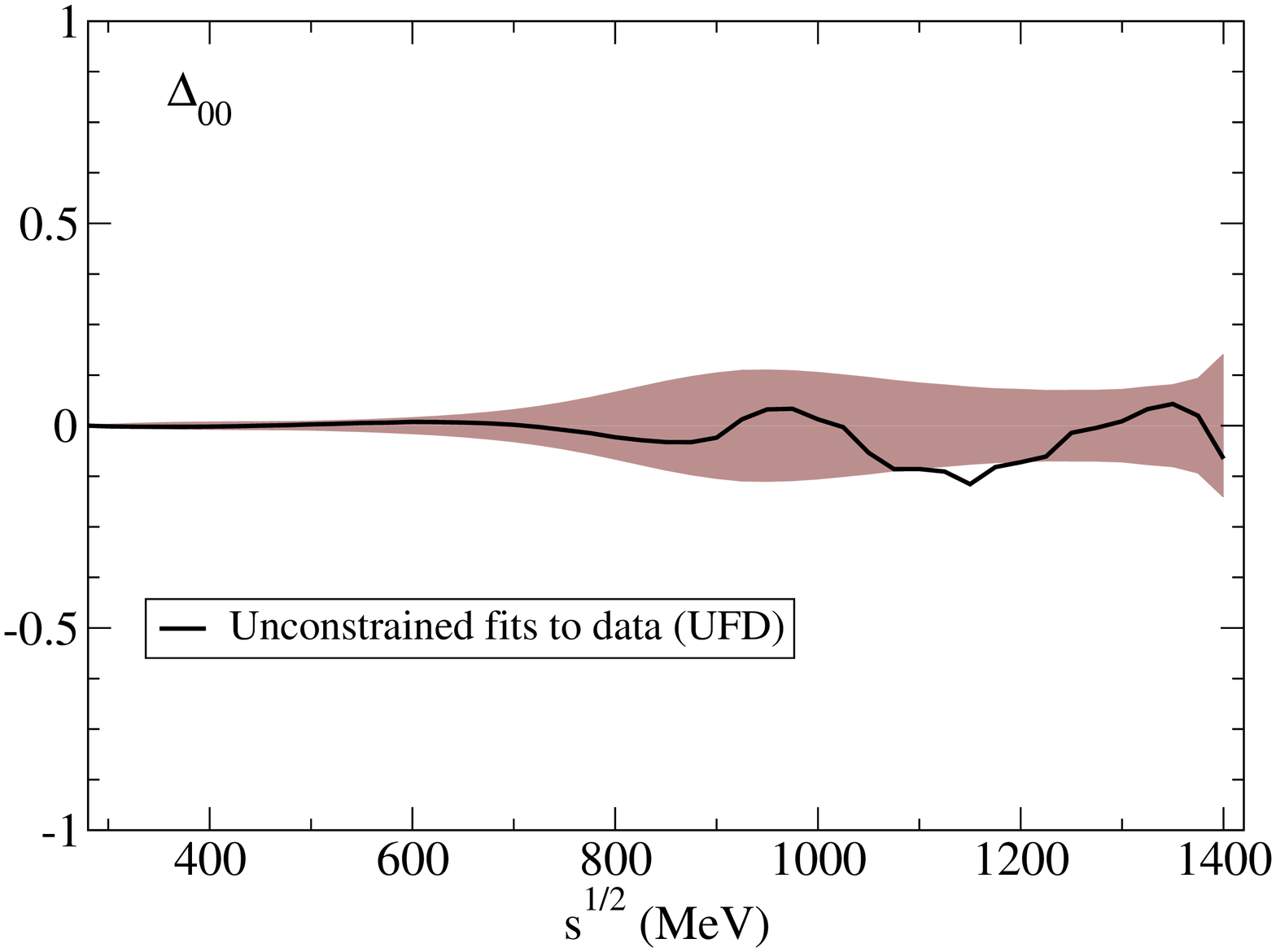,width=8.5truecm,angle=-90}}} 
\setbox2=\vbox{{\psfig{figure=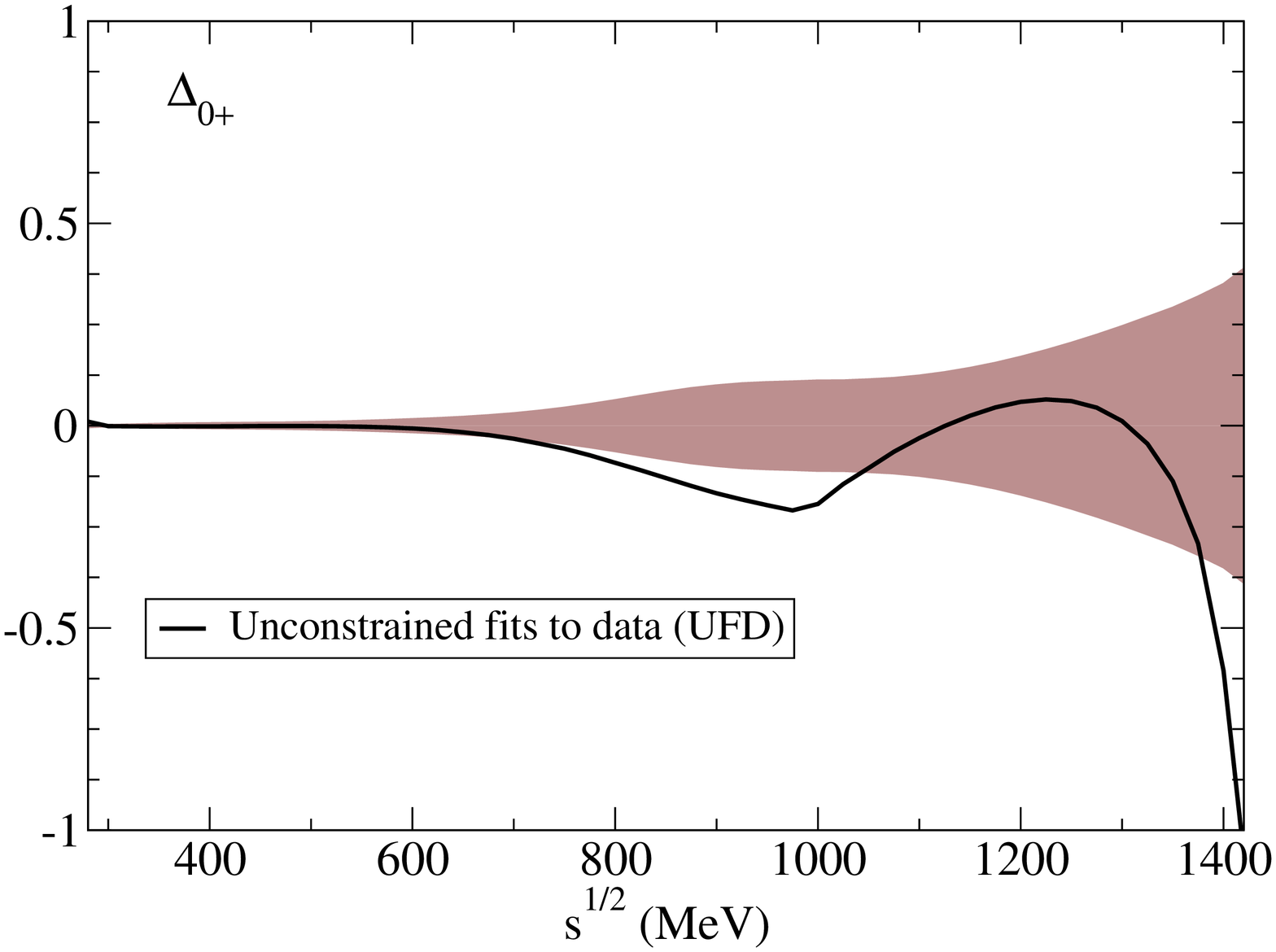,width=8.5truecm,angle=-90}}} 
\setbox3=\vbox{{\psfig{figure=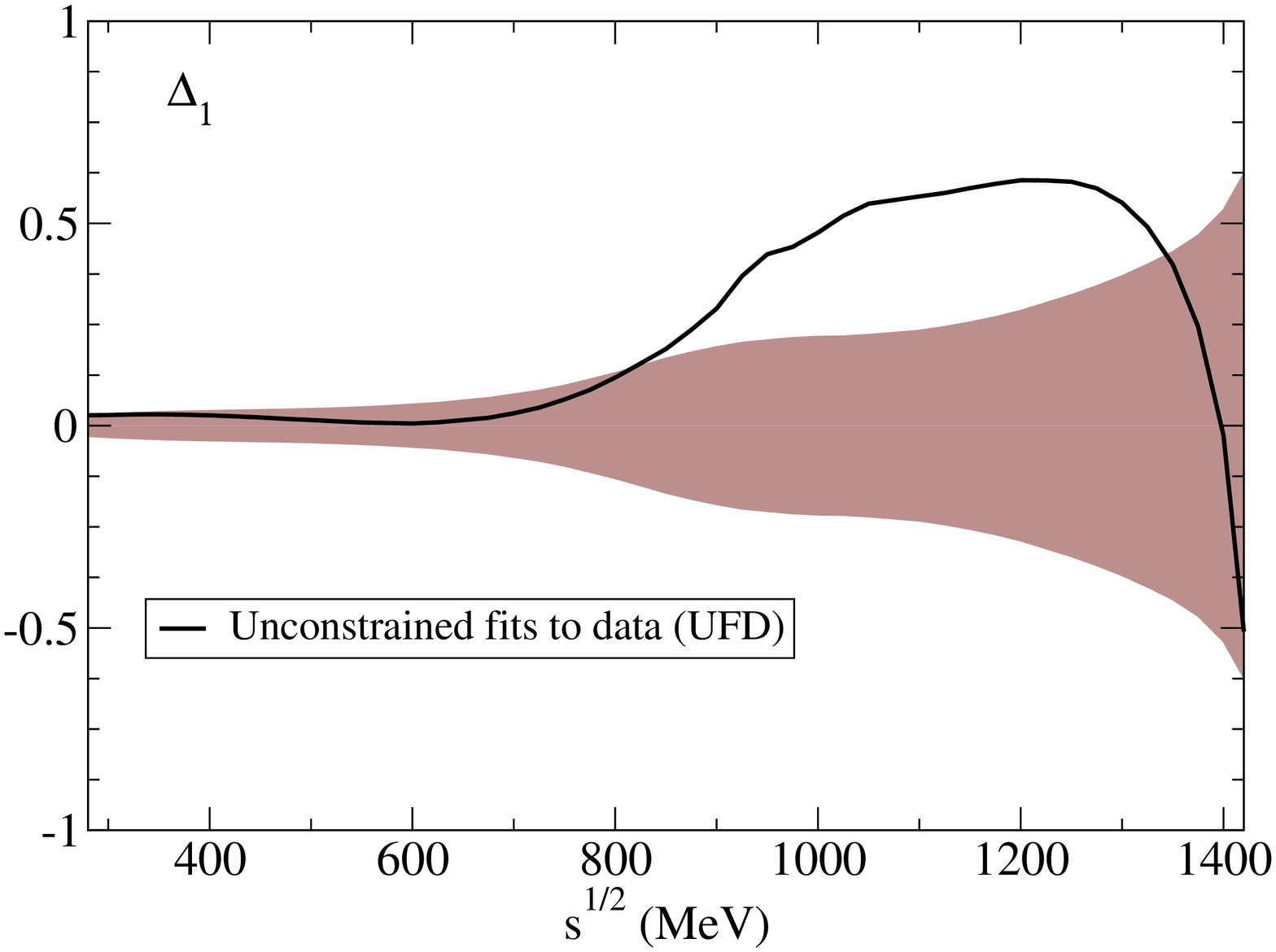,width=8.5truecm,angle=-90}}}
\centerline{\tightboxit{\box1}}
\centerline{\tightboxit{\box2}}
\centerline{\tightboxit{\box3}}
}
\setbox6=\vbox{\hsize 15truecm\captiontype\figurasc{Figure 4.2. }
{Fulfillment of 
dispersion
relations,  with the  central parameters  in (4.1a).  The error bands are also shown.}\hb}
\centerline{{\box0}}
\bigskip
\centerline{\box6}
}\endinsert

\booksubsection{4.2. Roy equations}

\noindent
 Roy equations are fully equivalent to nonforward 
dispersion relations, plus some $t-s$ crossing symmetry, 
 projected on the various partial waves. They can be written as
$$\real f^{(I)}_l(s)=C_l^{(I)}a_0^{(0)}+{C'_l}^{(I)}a_0^{(2)}
+\sum_{l',I'}
\pepe\int_{4M^2_\pi}^\infty\dd s' 
K_{l,l';I,I'}(s',s)\imag f^{(I')}_{l'}(s').
\equn{(4.5a)}$$
$C_l^{(I)}$, ${C'_l}^{(I)}$ are known constants, and the kernels $K_{l,l';I,I'}$ 
are also known.

We  also define the quantities
$$\deltav^{(I)}_l(s)\equiv \real f^{(I)}_l(s)-C_l^{(I)}a_0^{(0)}-{C'_l}^{(I)}a_0^{(2)}
-\sum_{l',I'}
\pepe\int_{4M^2_\pi}^\infty\dd s' 
K_{l,l';I,I'}(s',s)\imag f^{(I')}_{l'}(s');
\equn{(4.5b)}$$
they would vanish if Roy equations were exactly fulfilled.

 Roy equations are only valid 
up to $s=64\,M^2_\pi\simeq1\;\gev^2$ because, for larger values, the integrand receives
new contributions from the double spectral functions, not contained in (4.5).
Moreover, when the value of
$t$ over which one integrates  to project the partial waves is $|t|\gg\lambdav^2_{\rm QCD}$, for 
$s\simeq 2\;\gev^2$, the Regge expressions are not valid.
This is a further limitation of the validity of the Roy equations 
  to energies below $\sim1~\gev$.

We here calculate up to  $\bar{K}K$ threshold, and only test the waves S0, S2, P.\fnote{A 
preliminary review of these results was presented at the 
4th Int'l Conf. 
on Quarks and Nuclear Physics, Madrid, June 2006.\ref{13}} 
We define the equivalent of the average discrepancies $\bar{d}^2$ we used for the 
forward dispersion relations,
$$\bar{d}_{l,I}^2\equiv\dfrac{1}{\hbox{number of points}}
\sum_n\left(\dfrac{\deltav_l^{(I)}(s_n)}{\delta\deltav_l^{(I)}(s_n)}\right)^2,
\equn{(4.6)}$$
 and find the results shown in \figs~4.3, where we plot what our parametrizations give for 
$\real f_l^{(I)}$ (denoted by ``in") and what follows from the integrals in the right hand side of 
(4.5a), denoted by ``out".

Numerically, we have the results
$$\bar{d}_{S0}^2=0.54,\quad \bar{d}_{S2}^2=1.63,\quad \bar{d}_{P}^2=0.74:
\equn{(4.7)}$$
a reasonable fulfillment, but with it is clear that this can be
improved, particularly for the S2 wave for which 
the discrepancy is larger than unity.

\break

{
\setbox2=\vbox{\hfuzz1truecm
{\psfig{figure=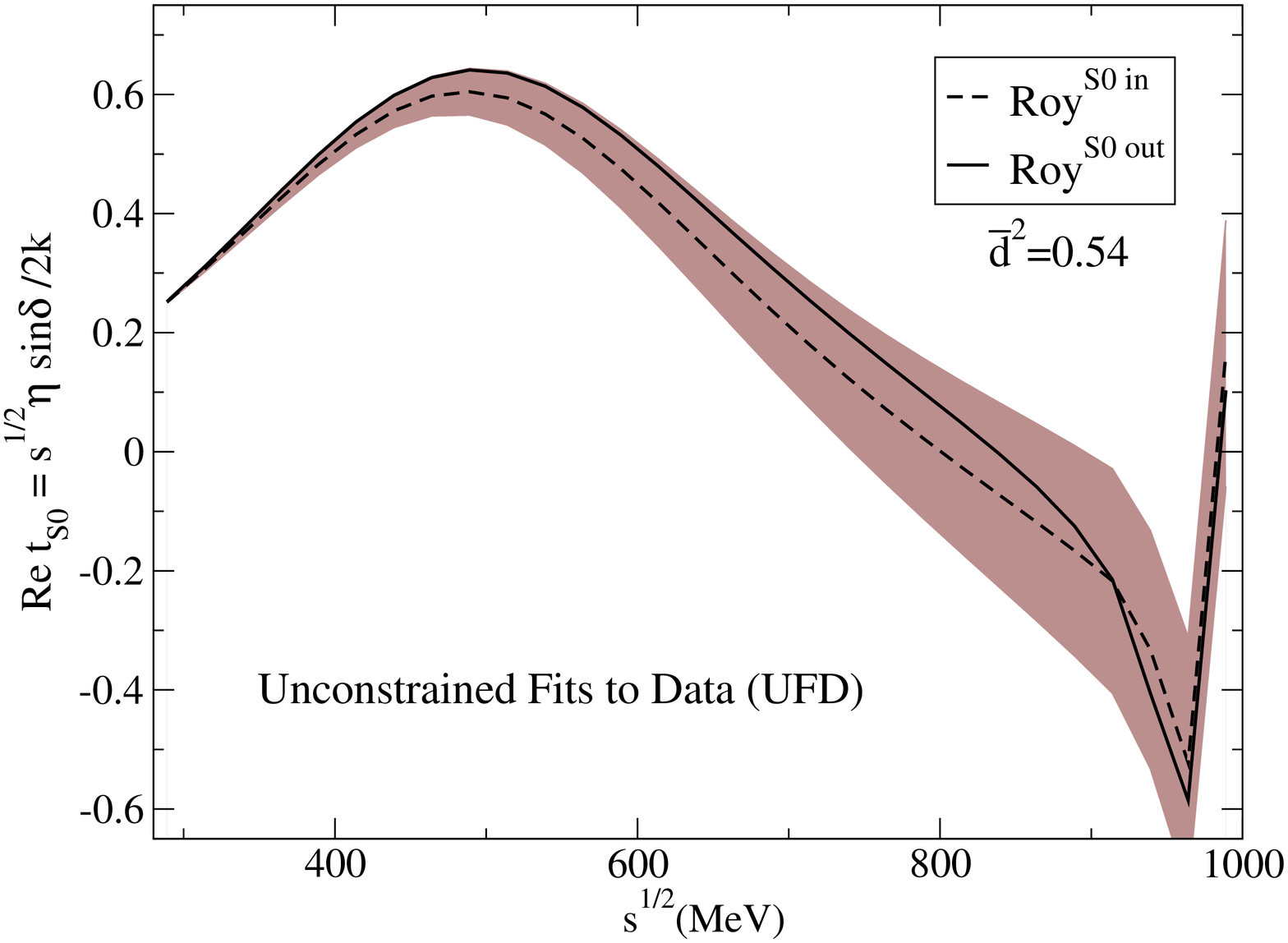,width=8.8truecm,angle=-90}}}   
\setbox6=\vbox{\hsize 4.3truecm\noindent\petit\figurasc{Figure 4.3a. }{
Fulfillment of the  Roy equation for the S0 wave.\hb Continuous line:  the result of the dispersive
integral.\hb
 Dashed line: real part, with the 
dark band the error band.\hb\phantom{x}\hb\phantom{x}
}}
\line{\tightboxit{\box2}\hfil\box6}
\setbox2=\vbox{\hfuzz1truecm
{\psfig{figure=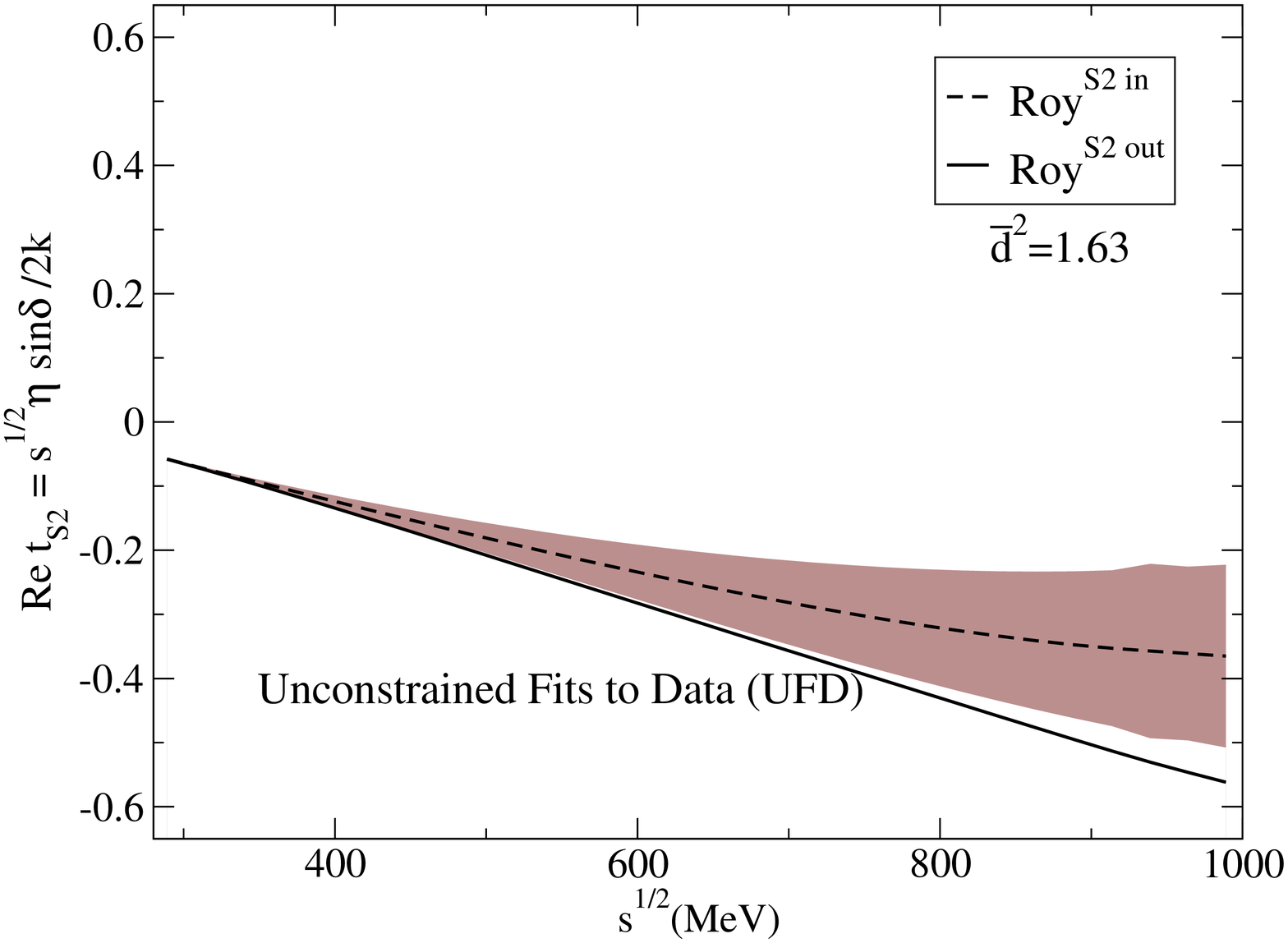,width=8.8truecm,angle=-90}}}   
\setbox6=\vbox{\hsize 4.3truecm\noindent\petit\figurasc{Figure 4.3b.  }{
Fulfillment of the  Roy equation for the S2 wave.\hb Continuous line:  the result of the dispersive
integral.\hb
 Dashed line: real part, with the 
dark band the error band.\hb\phantom{x}\hb\phantom{x}
}}
\line{\tightboxit{\box2}\hfil\box6}
\setbox2=\vbox{\hfuzz1truecm
{\psfig{figure=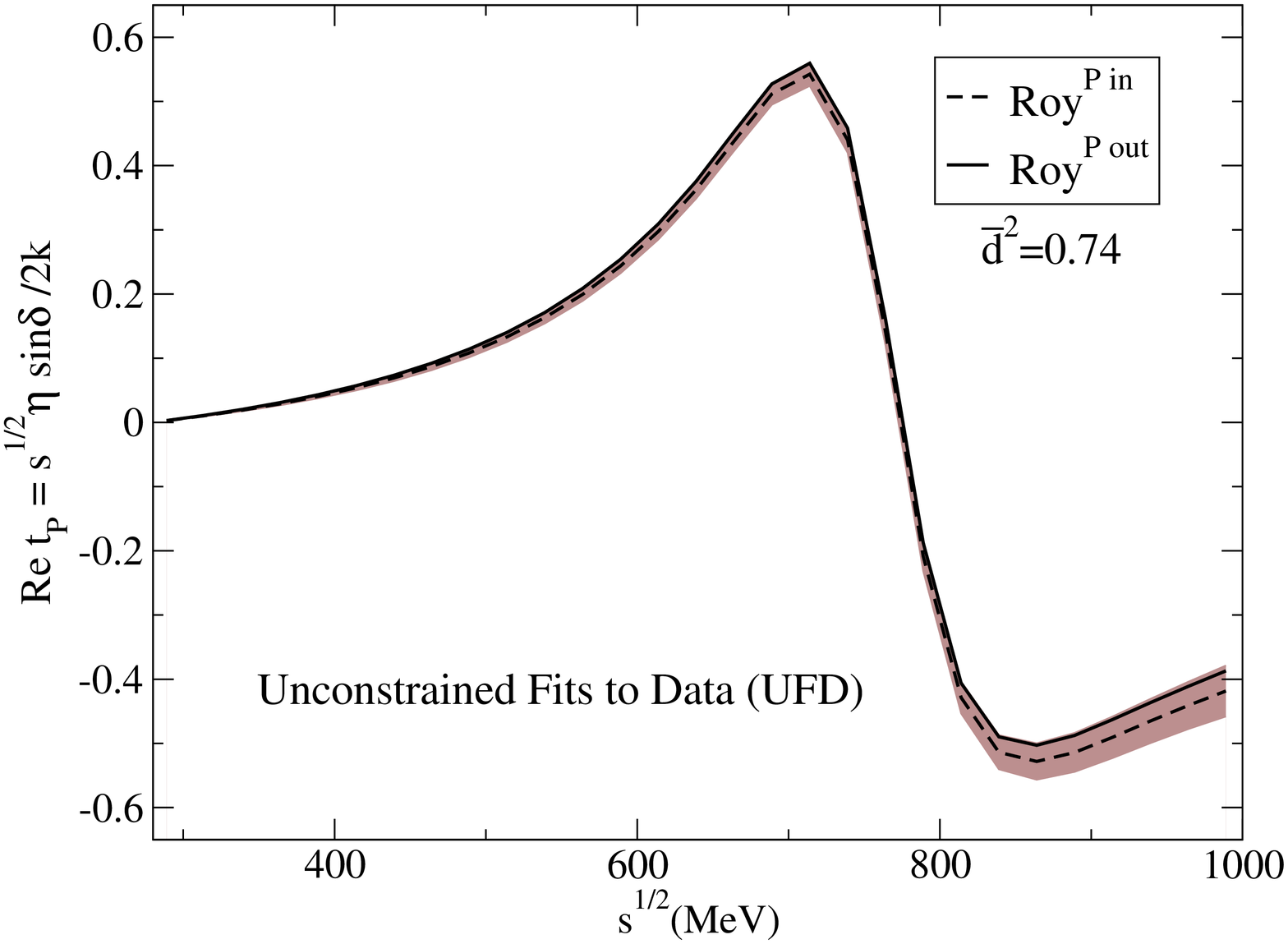,width=8.8truecm,angle=-90}}}   
\setbox6=\vbox{\hsize 4.3truecm\noindent\petit\figurasc{Figure 4.3c.  }{
Fulfillment of the  Roy equation for the P wave.\hb Continuous line: 
 the result of the dispersive integral.\hb
 Dashed line: real part, with the 
dark band the error band.\hb\phantom{x}\hb\phantom{x}
}}
\line{\tightboxit{\box2}\hfil\box6}
}
\booksection{5. Improvement of the parametrizations of the partial waves:\hb
 fits to data constrained
  requiring
fulfillment of dispersion relations}
\vskip-0.5truecm
\booksubsection{5.1. Two sum rules}

\noindent
Apart from forward dispersion relations and Roy equations, we will also require 
fulfillment, within errors, of two sum rules that 
relate high energy (Regge) parameters for $t\neq0$ to low energy P and D waves.

The first sum rule is  (PY05)
$$\eqalign{
I\equiv&\, \int_{4M^2_\pi}^\infty\dd s\,
\dfrac{\imag F^{(I_t=1)}(s,4M^2_\pi)-\imag F^{(I_t=1)}(s,0)}{s^2}-
 \int_{4M^2_\pi}^\infty\dd s\,\dfrac{8M^2_\pi[s-2M^2_\pi]}{s^2(s-4M^2_\pi)^2}
\imag F^{(I_s=1)}(s,0)=0.\cr
}
\equn{(5.1)}$$
The contributions of the S waves cancel in (5.1), so only the P, D,  F and G waves 
contribute. 
At high energy the integrals are dominated by rho exchange.

The second sum rule we consider is that given in Eqs.~{(B.6), (B.7)} of ref.~14.
 It reads,
$$J\equiv\int_{4M^2_\pi}^\infty\dd s\,\Bigg\{
\dfrac{4\imag F'^{(0)}(s,0)-10\imag F'^{(2)}(s,0)}{s^2(s-4M^2_\pi)^2}
-6(3s-4m^2_\pi)\,\dfrac{\imag F'^{(1)}(s,0)-\imag F^{(1)}(s,0)}{s^2(s-4M^2_\pi)^3}
\Bigg\}=0.
\equn{(5.2)}
$$
Here $F'^{(I)}(s,t)\equiv\partial F^{(I)}(s,t)/\partial\cos\theta$.
At high energy, the integral is dominated by isospin zero Regge trajectories.
We also define a discrepancy for these sum rules:
$$\bar{d}^2_I=\left(\dfrac{I}{\delta I}\right)^2,
\quad
\bar{d}^2_J=\left(\dfrac{J}{\delta J}\right)^2.
\equn{(5.3)}
$$
These two sum rules are reasonably satisfied, if using the partial wave
parameters  obtained from data (\sect~2), and 
the Regge parameters determined  from factorization and fits to data (\sect~3).
\booksubsection{5.2. Minimization procedure}
\noindent
Because forward dispersion relations and Roy equations are satisfied almost 
within the fluctuations induced by the  experimental errors, it makes sense
 to repeat the fits to experiment 
requiring verification within errors of forward dispersion relations and Roy equations, 
to which we add the sum rules (5.1), (5.2) to control Regge parameters away from the forward direction.
We do this by minimizing the  quantity $\chi^2$ defined, with self-evident notation, by
$$\chi^2\equiv
\left\{\bar{d}^2_{00}+\bar{d}^2_{0+}+\bar{d}^2_{I_t=1}+\bar{d}^2_{S0}+\bar{d}^2_{S2}+\bar{d}^2_P\right\}
W
+\bar{d}^2_I+\bar{d}^2_J+\sum_i\left(\dfrac{p_i-p_i^{\rm exp}}{\delta p_i}\right)^2.
\equn{(5.4)}
$$
Here, $p_i^{\rm exp}$ are the parameters that we have found in the unconstrained fits 
to experimental data, 
and $\delta p_i$ are their errors. 
Thus the sum over $p_i$ runs over $B_n$s, zeros ($z_0,\,z_2,\,\deltav$), inelasticity parameters 
$\epsilon_n$, $r$, etc, and over the K-matrix parameters for the 
S0 wave. 
The presence of the sum $\sum_i[(p_i-p_i^{\rm exp})/\delta p_i]^2$,
 of course, ensures fit to experimental
data. 

The quantity $W$ in (5.4) is a weight, which can be estimated in two different manners. 
First, it will serve to give each of the dispersion relations (FDR or Roy) 
a weight appropriate to the information that they carry. 
For example, for the FDR for $\pi^0\pi^0$ scattering, 
the quantity $\real F_{00}(s)- F_{00}(4M^2_\pi)$ can be fixed giving 
the slope at $s=4M^2_\pi$, the value of  $\real F_{00}(s)- F_{00}(4M^2_\pi)$
 and its derivative at each of the points $s$ where 
its changes  direction; and the same at the end point, $s=1.42\;\gev^2$ (see \fig~4.1.{\sc a}): 
altogether, 13 values. Any reasonably smooth function that fits these 13 
values is sure to follow $\real F_{00}(s)- F_{00}(4M^2_\pi)$ in all the range:
putting extra weight would be imposing redundant constraints.
For other dispersion relations the number is a bit smaller; 
in general, a number $6\lsim W\lsim 13$ is obtained.
An alternate method to find $W$ is to increase it so that all dispersion relations 
are satisfied within errors, that is to say, all the corresponding $\bar{d}^2$ 
are less than or equal to unity. This occurs for $W\sim9$. 
In our calculation  we have taken $W=9$, although we have verified 
that results practically indistinguishable are obtained for 
$7\leq W\leq12$.

\booksubsection{5.3. Results for the constrained fits (CFD)}

\noindent
The results of the fits to data, constrained by requiring fulfillment of the 
FDR and Roy equations plus the sum rules (5.1) and (5.2), 
that we call Set~CFD, are summarized in 
Appendix~B. This CFD~Set is obtained by minimizing $\chi^2$ as defined 
in Eq.~(5.4). In general, the parameters hardly
change with respect to what  we had from fits to data, Appendix~A; but there are a few waves for which 
there are noteworthy alterations. First of all, we have
 the S0 and S2 waves. Because now we are requiring 
verification of the FDR {\sl below} threshold, we can leave the location 
of the Adler zeros $z_0$, $z_2$ free. 
This produces some changes in the parameters $B_n$, since there is a strong 
correlation between them and the location of the Adler zeros. 
However, the {\sl phase shifts} themselves are practically identical 
to what we had in fits to data, Appendix~A.
Then we have  the D2 wave, the only one that changes appreciably. It 
 moves by a bit more than one standard deviation. 
The S2 wave also moves appreciably after constraining its fit; 
we discuss the two later on (\sects~5.4 and 5.5). 
All other waves change so little that the difference between Sets UDF and CFD
 is almost inappreciable.

For the present CFD~Set, FDRs and Roy equations are, of course, 
 better satisfied than before. 
For the FDR we find
\smallskip
$$\matrix{&s^{1/2}\leq 932\;\mev&s^{1/2}\leq1420\;\mev\vphantom{\Big|}\cr
\pi^0\pi^0\quad\hbox{FDR}&\bar{d}^2=0.13&\bar{d}^2=0.31\cr
\pi^0\pi^+\quad\hbox{FDR}&\bar{d}^2=0.83&\bar{d}^2=0.85\cr
I_t=1\quad\hbox{FDR}&\bar{d}^2=0.13&\bar{d}^2=0.70,\cr
}\equn{(5.5)}$$ 
and, for the Roy equations,
$$\bar{d}_{S0}^2=0.23,\quad \bar{d}_{S2}^2=0.25,\quad \bar{d}_{P}^2=2\times10^{-3}.
\equn{(5.6)}$$
Furthermore, for the sum rules (5.1) and (5.2) we have $\bar{d}^2_I=0.02$ and 
 $\bar{d}^2_J=0.55$. 
The verification of FDR and Roy equations are 
shown graphically in Figs.~5.1, 5.2 and 5.3.

The overall average $\bar{d}^2$ is now substantially smaller than unity; 
particularly for the Roy equations. However, 
for the FDR for $\pi^0\pi^+$ scattering  
is near unity. 
This is probably due to the D2 wave, which is likely still a bit
 away from its correct 
location, and to the P wave at energies above 1380~\mev, where our parametrization 
fails to take account of the 
$\rho(1450)$ resonance. 
This  indicates that one is at the limit of accuracy 
for experimentally-based parametrizations of 
scattering amplitudes.

Besides sum rules and dispersion relations, another independent test of our amplitudes is the 
Adler sum rule that relates the pion decay constant to 
pion-pion scattering amplitudes with one pion off its mass sell. This has been recently
evaluated\ref{15}
with our scattering UFD amplitudes, and a very satisfactory fulfillment of the sum rule is found; 
the discrepancy $\deltav_\pi$ that measures the accuracy with which the sum rule is 
fulfilled (and which should vanish if it was satisfied exactly) is found to be
$\deltav_\pi=0.021\pm0.053$.

\vfill\eject

{
\setbox3=\vbox{
{\psfig{figure=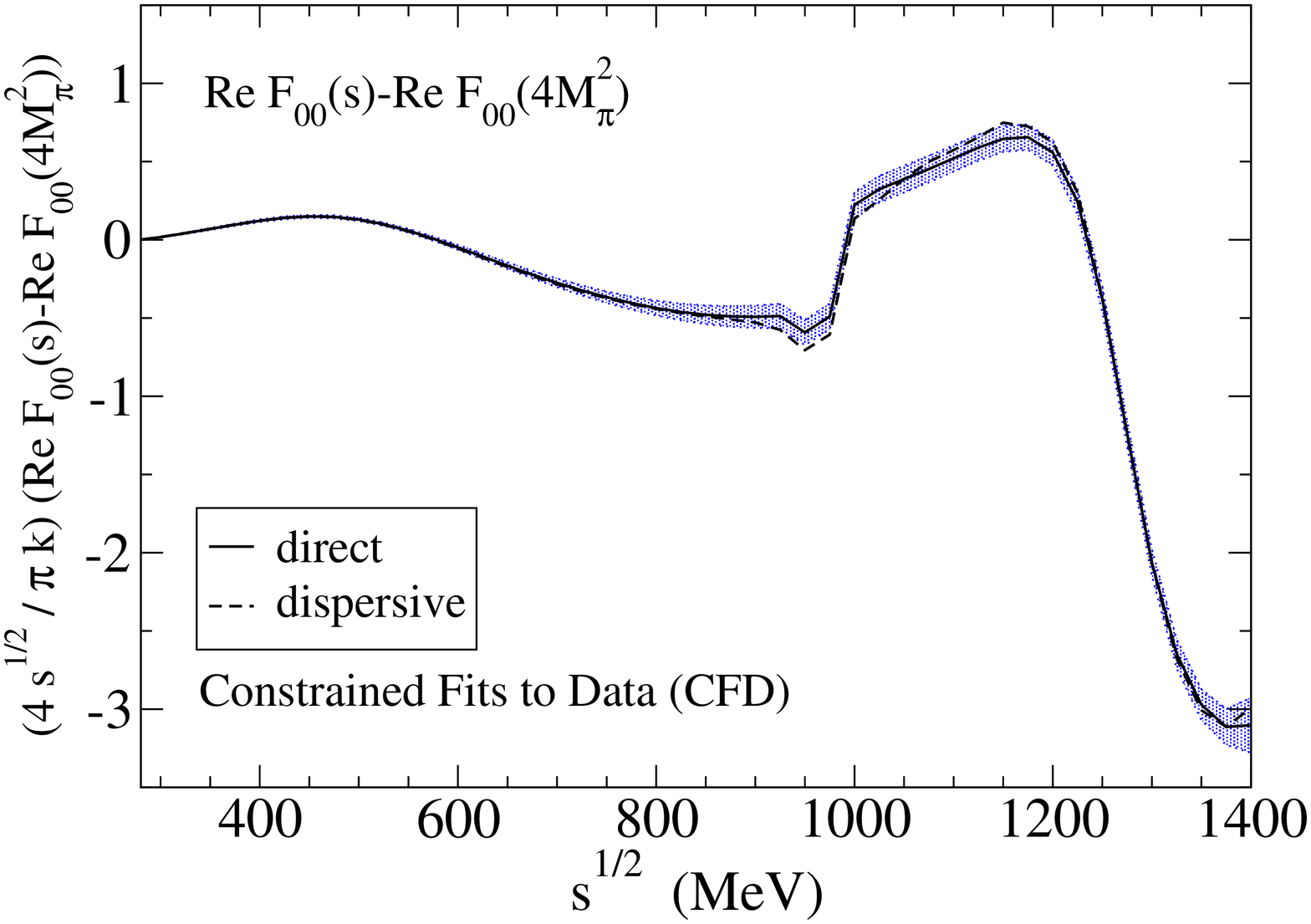,width=9.truecm,angle=-90}}} 
\setbox6=\vbox{\hsize 5.4truecm\noindent\petit\figurasc{Figure 5.1a. }{\hfuzz1.truecm
The $\pi^0\pi^0$ dispersion relation with the CFD amplitudes.\hb
 Continuous line: real part, evaluated
directly
 with the
parametrizations (the gray band covers the error).\hb
 Dashed line: the result of the dispersive integral.
\hb\phantom{x}\hb\phantom{x}\hb\phantom{x}
}}
\line{\otightboxit{\box3}\hfil\box6}
\setbox3=\vbox{\hfuzz1.truecm
{\psfig{figure=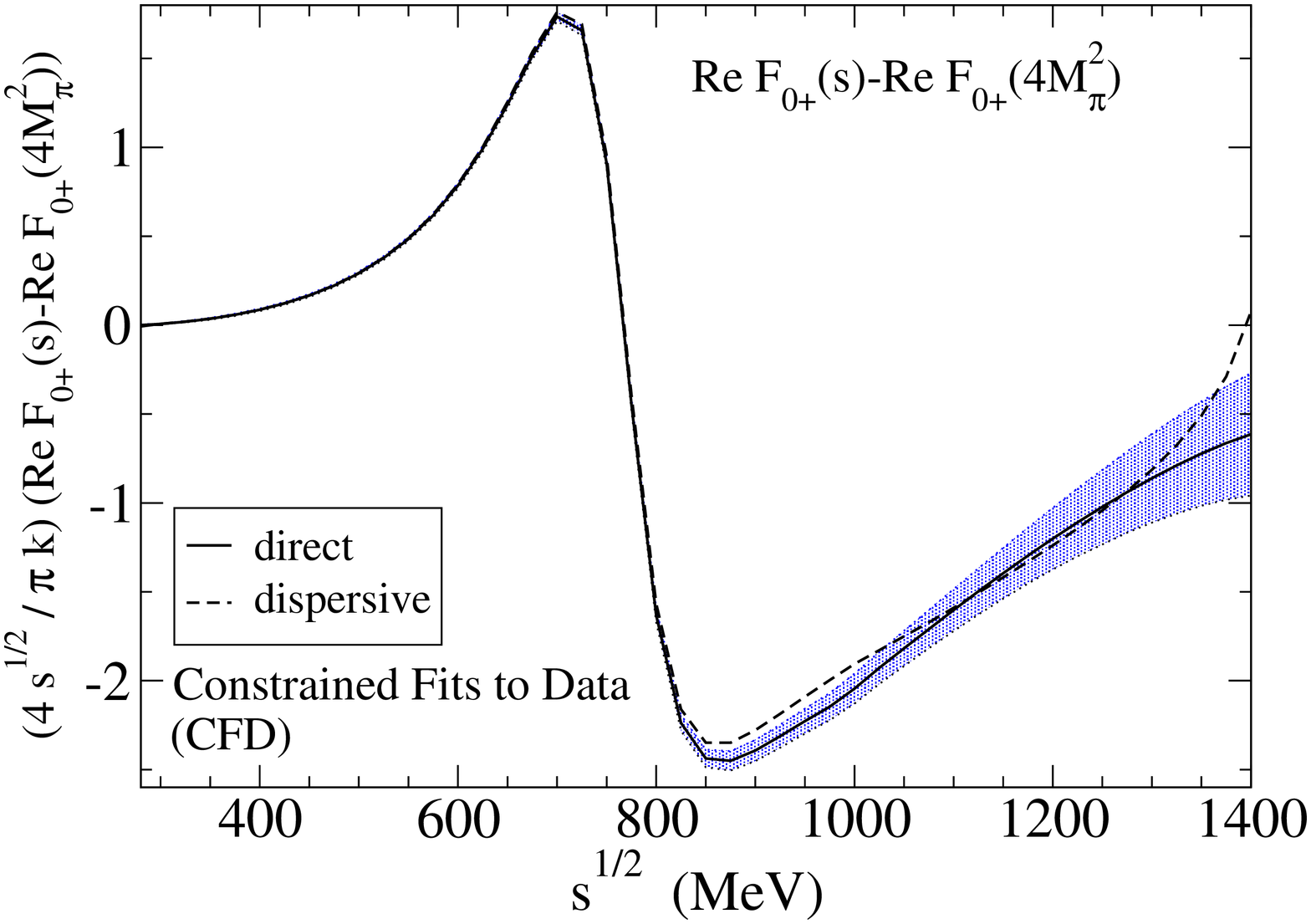,width=9.truecm,angle=-90}}} 
\setbox6=\vbox{\hsize 5.4truecm\noindent\petit\figurasc{Figure 5.1b }{
The $\pi^0\pi^+$ dispersion relation  with the CFD amplitudes.\hb
 Continuous line: real part, evaluated 
directly with the
parametrizations.\hb The gray band covers the error.\hb
 Dashed line: the result of the dispersive integral.\hb\phantom{x}
\hb\phantom{x}\hb\phantom{x}
}}
\line{\otightboxit{\box3}\hfil\box6}
\setbox3=\vbox{\hfuzz1truecm
{\psfig{figure=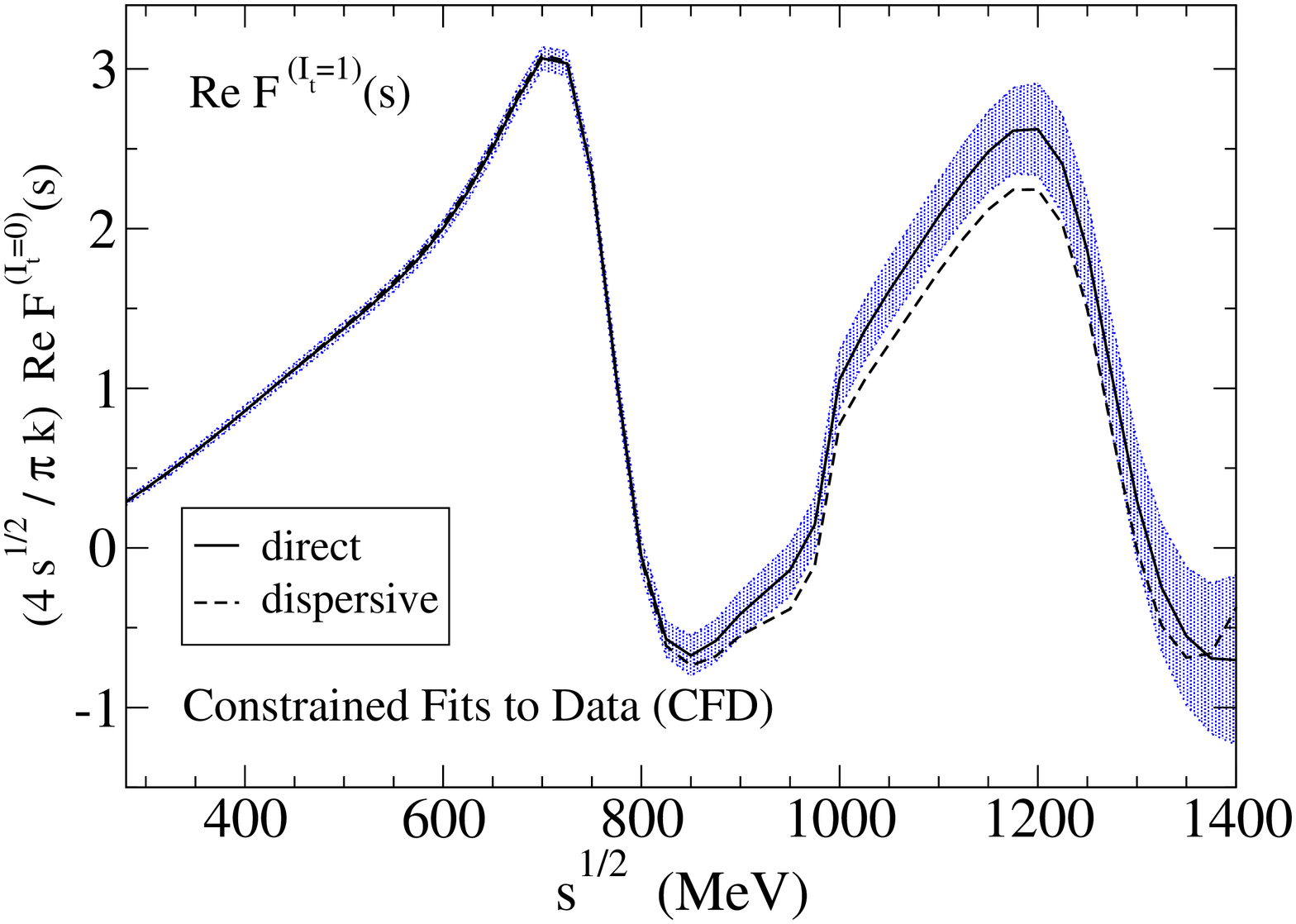,width=9.truecm,angle=-90}
}} 
\setbox6=\vbox{\hsize 5.4truecm\noindent\petit\hfuzz0.5truecm\figurasc{Figure 5.1c. }{
The  dispersion relation for  $I_t=1$ scattering with the CFD amplitudes.\hb
Continuous line: real part and error (shaded
area)
 evaluated directly with the parametrizat\-ions.\hb
 Dashed line: the result of the dispersive integral.\hb\phantom{x}\hb\phantom{x}
\hb\phantom{x}
}}
\line{\otightboxit{\box3}\hfil\box6}
}
 
\midinsert{
\setbox0=\vbox{\hsize8truecm 
\setbox1=\vbox{{\psfig{figure=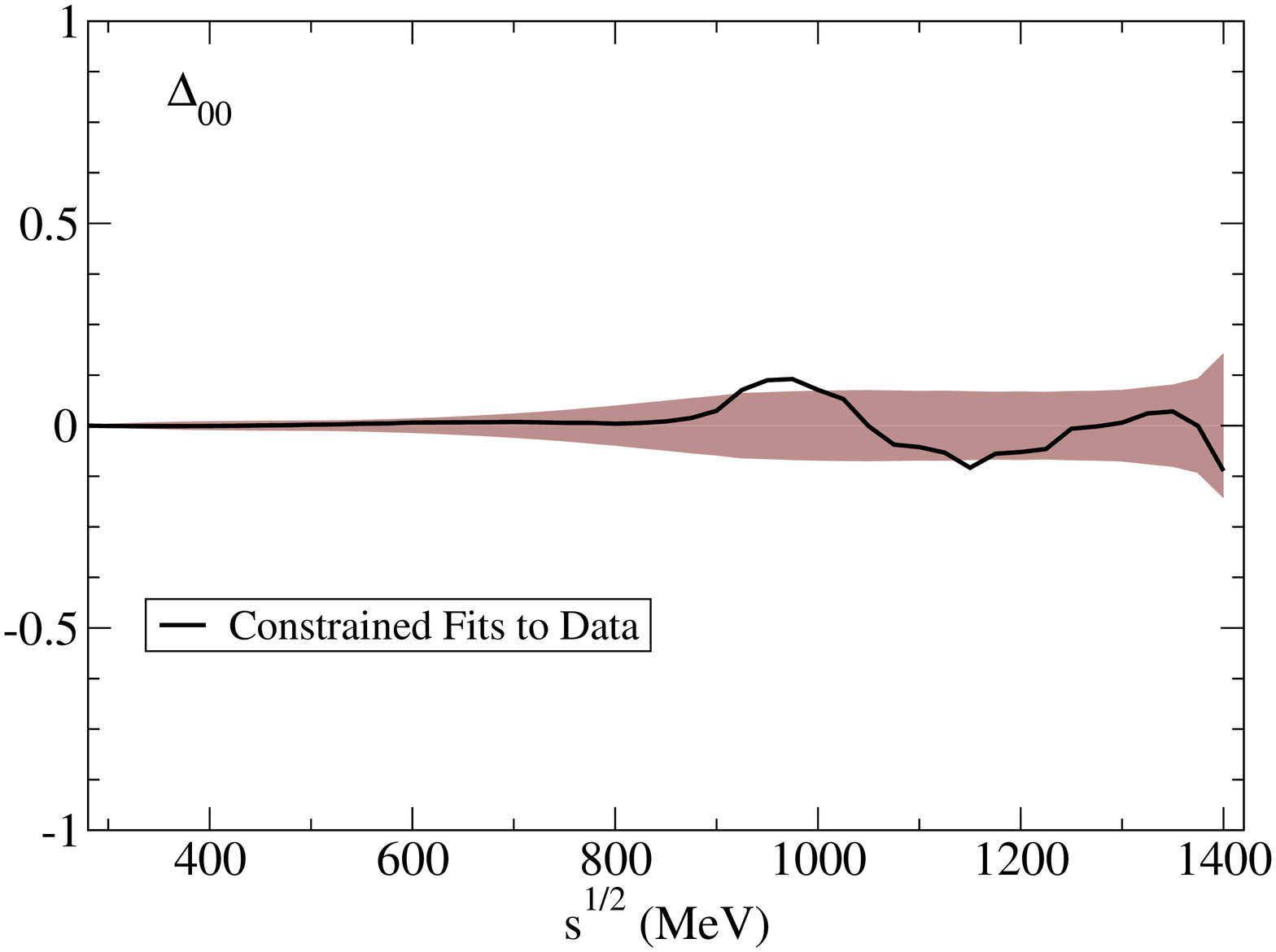,width=8.5truecm,angle=-90}}} 
\setbox2=\vbox{{\psfig{figure=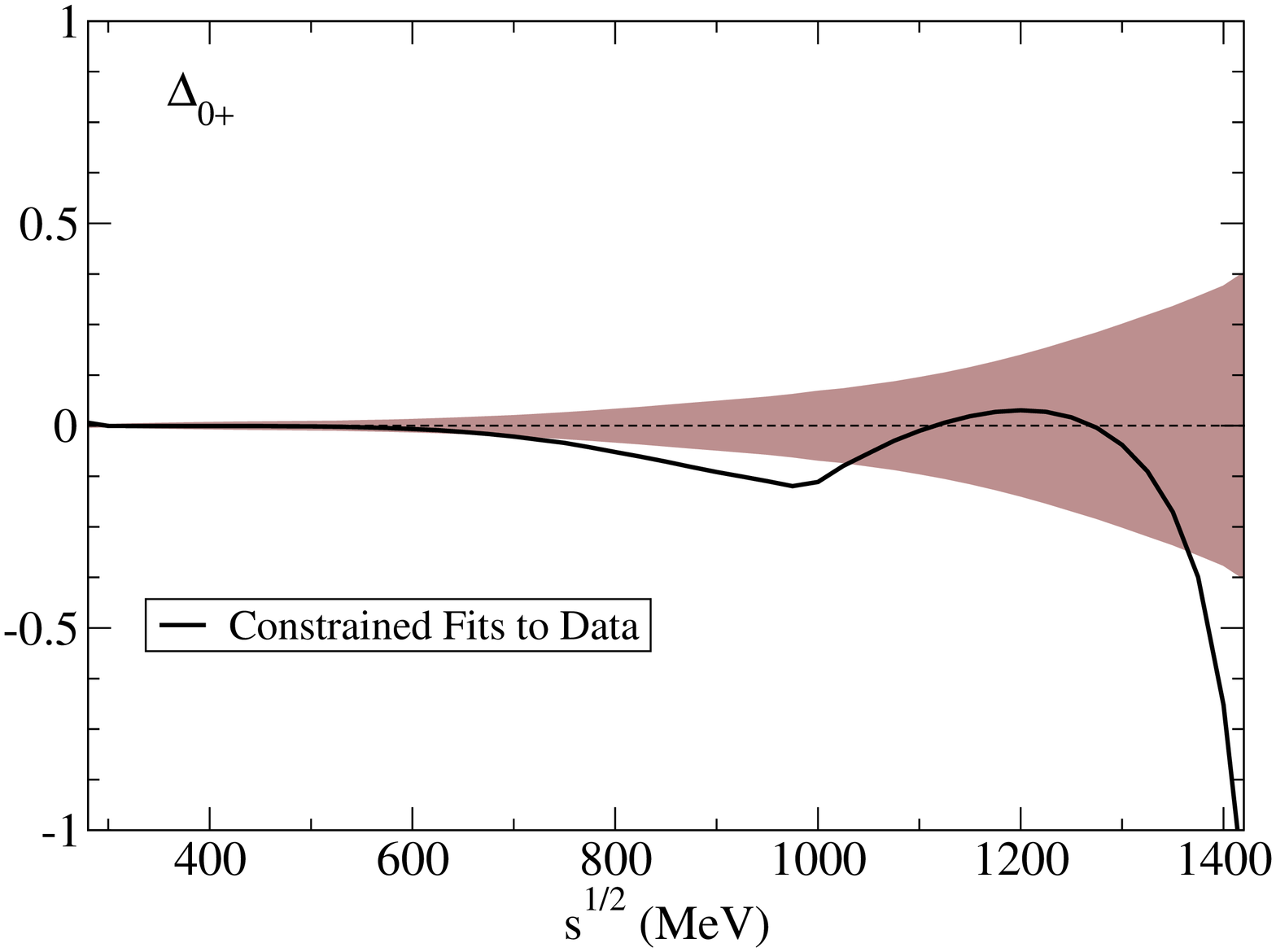,width=8.5truecm,angle=-90}}}
\setbox3=\vbox{{\psfig{figure=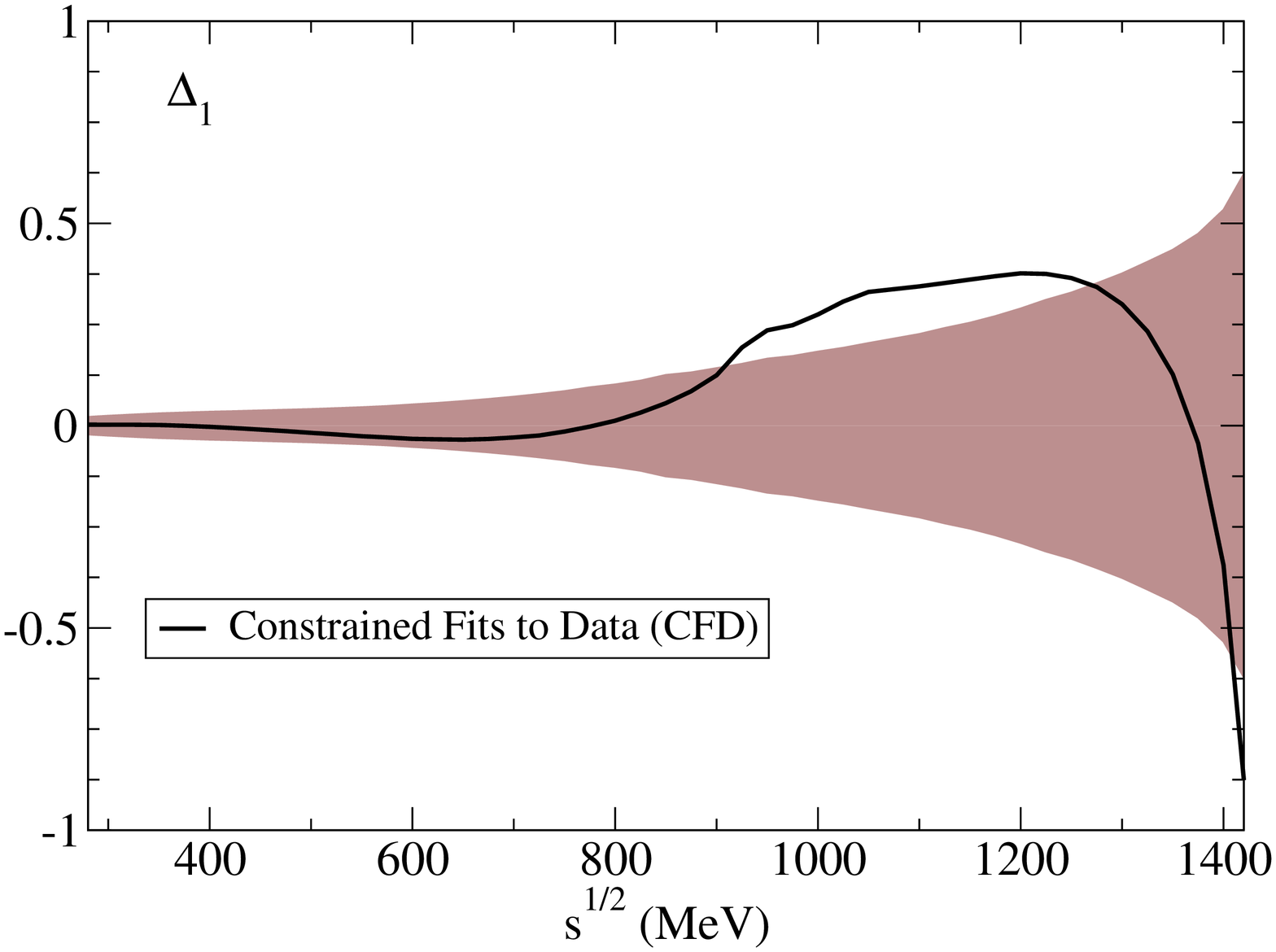,width=8.5truecm,angle=-90}}}
\centerline{\tightboxit{\box1}}
\centerline{\tightboxit{\box2}}
\centerline{\tightboxit{\box3}}}
\setbox6=\vbox{\hsize 15truecm\captiontype\figurasc{Figure 5.2. }{Fulfillment of 
dispersion
relations,  with the  central parameters  in (4.1a). 
 The error bands are also
shown.}\hb}
\centerline{{\box0}}
\bigskip
\centerline{\box6}
}\endinsert

{
\setbox2=\vbox{\hfuzz1truecm
{\psfig{figure=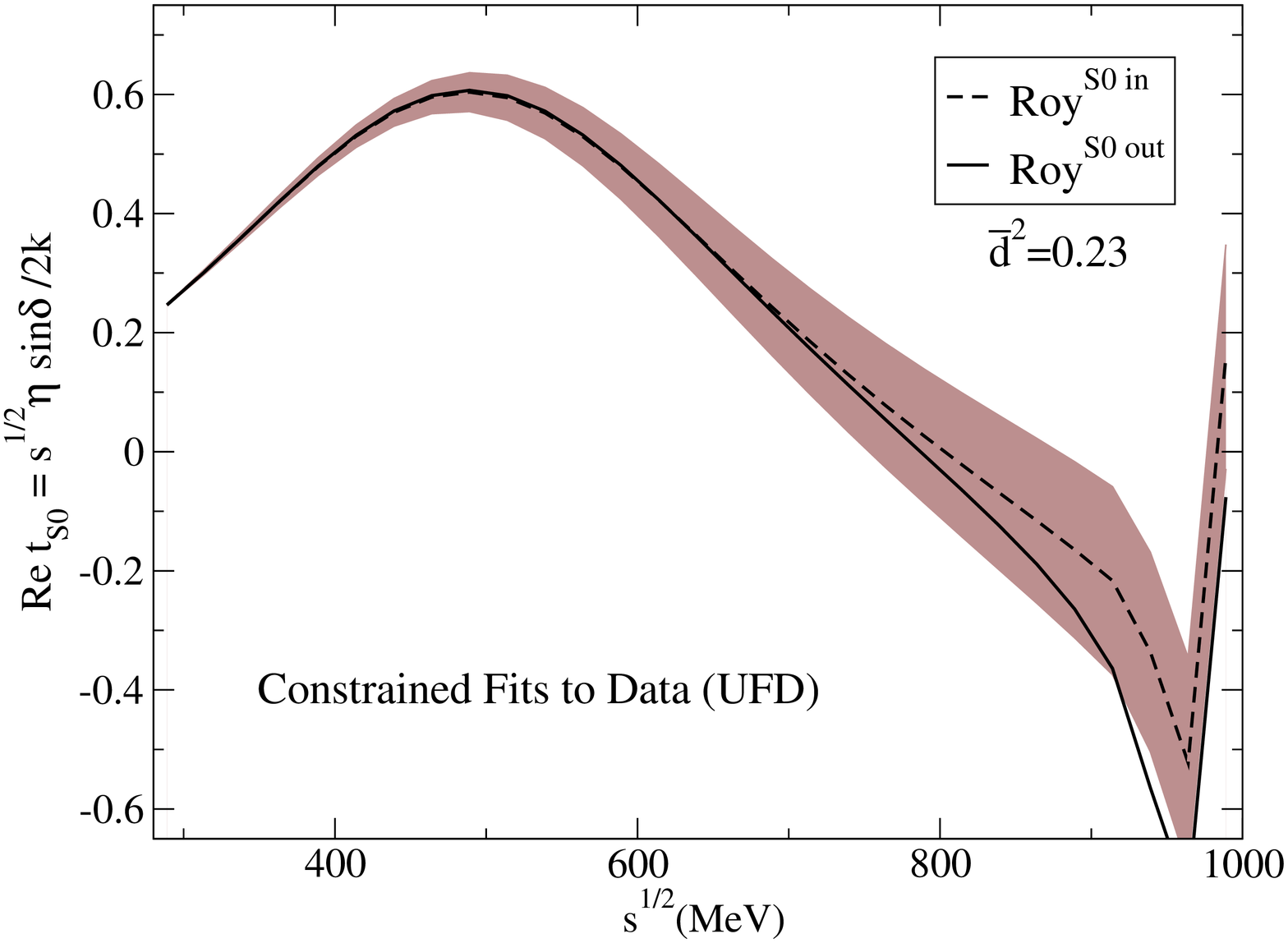,width=8.8truecm,angle=-90}}}   
\setbox6=\vbox{\hsize 4.3truecm\noindent\petit\figurasc{Figure 5.3a. }{
Fulfillment of the  Roy equation for the S0 wave.\hb Continuous line: 
 the result of the dispersive
integral.\hb
 Dashed line: real part, with the 
dark band the error band.\hb\phantom{x}\hb\phantom{x}
}}
\line{\tightboxit{\box2}\hfil\box6}
\setbox2=\vbox{\hfuzz1truecm
{\psfig{figure=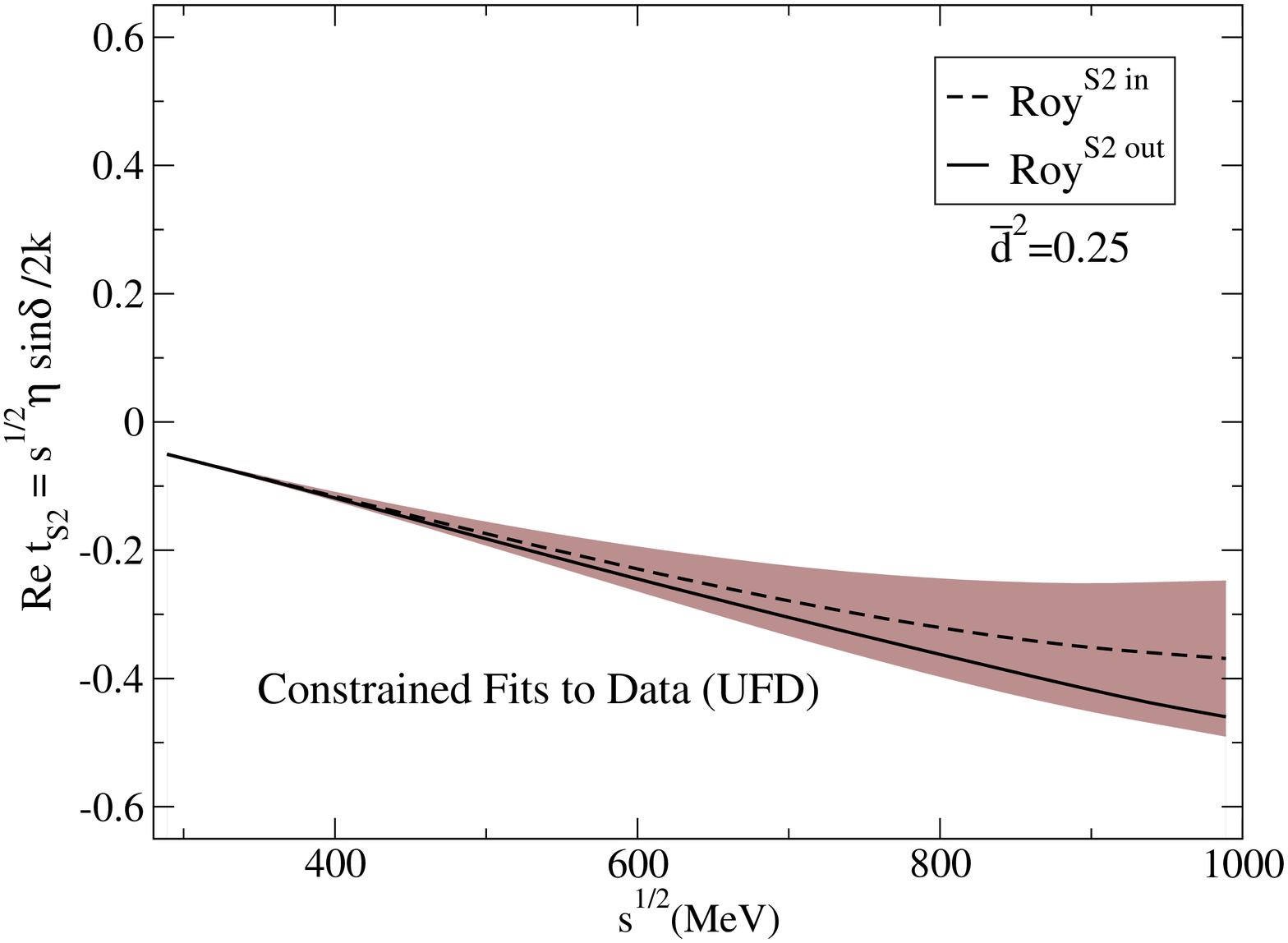,width=8.8truecm,angle=-90}}}   
\setbox6=\vbox{\hsize 4.3truecm\noindent\petit\figurasc{Figure 5.3b.  }{
Fulfillment of the  Roy equation for the S2 wave.\hb Continuous line:
  the result of the dispersive
integral.\hb
 Dashed line: real part, with the 
dark band the error band.\hb\phantom{x}\hb\phantom{x}
}}
\line{\tightboxit{\box2}\hfil\box6}
\setbox2=\vbox{\hfuzz1truecm
{\psfig{figure=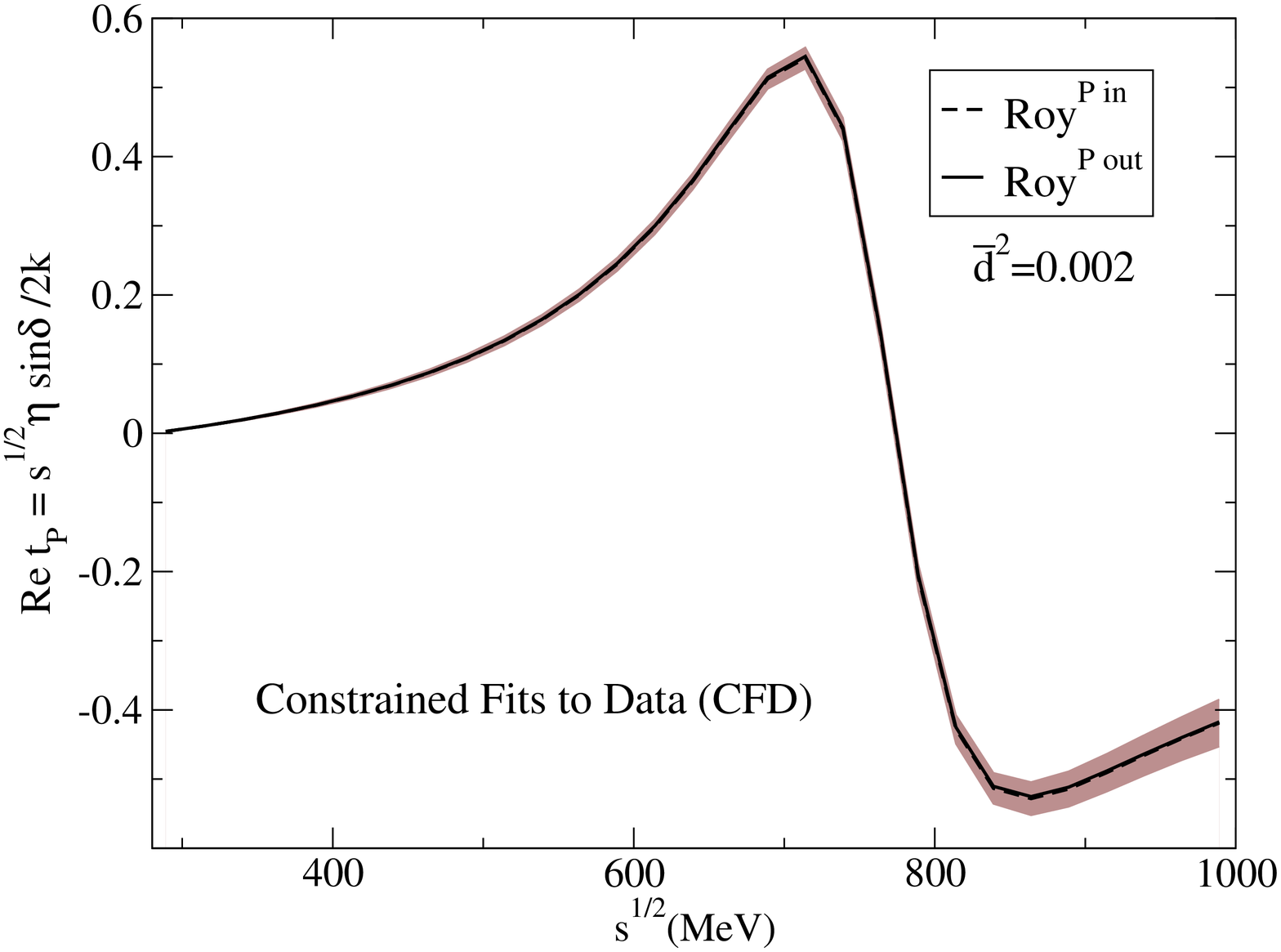,width=8.8truecm,angle=-90}}}   
\setbox6=\vbox{\hsize 4.3truecm\noindent\petit\figurasc{Figure 5.3c.  }{
Fulfillment of the  Roy equation for the P wave.\hb Continuous line: 
 the result of the dispersive integral.\hb
 Dashed line: real part, with the 
dark band the error band.\hb\phantom{x}\hb\phantom{x}
}}
\line{\tightboxit{\box2}\hfil\box6}
}

\booksubsection{5.4. Results: comparison of UFD and CFD Sets}

\noindent
We here present the comparison of our Sets~UFD and CFD, i.e., what we have
by directly fitting data, and what is obtained constraining 
the fits by imposing also FDRs and Roy equations.\fnote{For the S0 wave, 
we give here only the values 
obtained neglecting isospin breaking; the values of the 
parameters obtained taking isospin breaking into account may 
be found in Appendix~B.  
} 
Note that in the CFD~Set we have here only altered the {\sl central} values;  
 we leave the errors that follow from fits to data, i.e., 
we assume errors as in Set~UFD.
Note also that, in the following formulas,
 the parameters are as defined in Appendices~A and B.
\medskip
\noindent{\sl {\rm S0} wave}.\quad  
We have, for S0 below 932~\mev,
$$\matrix{&\hbox{UFD},\;z_0\equiv M_\pi&
\hbox{CFD},\;z_0\;{\rm free}\vphantom{\Big|}\cr
 B_0&4.3\pm0.3&4.41\pm0.3\cr
B_1&-26.7\pm0.6&-26.25\pm0.6\cr
B_2&-14.1\pm1.4&-15.8\pm1.4\cr
z_0&M_\pi&166.1\pm4.2\,\mev.}
\equn{(5.7a)}$$
Above 932~\mev,
$$\matrix{&\hbox{UFD}&\hbox{CFD}\vphantom{\Big|}\cr
\alpha_1&0.843\pm0.017&0.843\pm0.017\cr
\alpha_2&0.20\pm0.06&0.20\pm0.06\cr
\beta_1&1.02\pm0.02&1.02\pm0.02\cr
\beta_2&1.33\pm0.013&1.33\pm0.013\cr
\gamma_{11}&3.10\pm0.11&3.10\pm0.11\cr
\gamma_{12}&1.82\pm0.05&1.81\pm0.05\cr
\gamma_{22}&-7.00\pm0.04&-7.00\pm0.04\cr
M_1&888\pm4\;\mev&888\pm4\cr
M_2&1327\pm4\;\mev&1327\pm4.}
\equn{(5.7b)}$$

\medskip
\noindent{\sl {\rm S2} wave}.\quad We now find,
$$\matrix{&\hbox{UFD},\;z_0\equiv M_\pi
&\hbox{CFD},\;
z_2\;{\rm free}\vphantom{\Big|}\cr
 B_0&-80.4\pm2.8&-80.2\pm2.8\cr
B_1&-73.6\pm10.5&-69.4\pm10.5\cr
B_{h2}&109\pm38&120\pm38\cr
z_2&M_\pi&145.0\pm3.6\,\mev.}
\equn{(5.8a)}$$
For the inelasticity,
$$\matrix{&\hbox{UFD}&\hbox{CFD}\vphantom{\Big|}\cr
\epsilon&0.17\pm0.12&0.18\pm0.12\cr
}
\equn{(5.8b)}$$
\medskip
\noindent{\sl {\rm P} wave}.\quad In this case we have kept the value of the 
$\rho$ resonance mass fixed when imposing dispersion relations; thus, for both 
Sets~UFD and CFD, $M_\rho=773.6\pm0.9\,\mev$.
 For the remaining parameters below 992~\mev\ we find,
\medskip
$$\matrix{&\hbox{UFD}&
\hbox{CFD}\vphantom{\Big|}\cr 
B_0&1.055\pm0.011&1.052\pm0.011\cr
B_1&0.15\pm0.05&0.17\pm0.05.\cr}
\equn{(5.9a)}$$
Above 992~\mev,
\medskip
$$\matrix{&\hbox{UFD}&\hbox{CFD}\vphantom{\Big|}\cr
\lambda_1&1.57\pm0.18&1.50\pm0.18\cr
\lambda_2&-1.96\pm0.49&-1.97\pm0.49\cr
\epsilon_1&0.10\pm0.06&0.09\pm0.06\cr
\epsilon_2&0.11\pm0.11&0.12\pm0.11\cr}
\equn{(5.9b)}$$

\medskip
\noindent{\sl {\rm D0} wave}.\quad We here keep the mass of the resonance fixed
at $M_{f_2}=1275.4\,\mev$ for 
both Sets~UFD and CFD.
 We have, below 992~\mev, 
\medskip
$$\matrix{&\hbox{UFD}&\hbox{CFD}\vphantom{\Big|}\cr
B_0&12.47\pm0.12&12.48\pm0.12\cr
B_1&10.12\pm0.16&10.12\pm0.16.}
\equn{(5.10a)}$$
Above 992~\mev,
\medskip
$$\matrix{&\hbox{UFD}&\hbox{CFD}\vphantom{\Big|}\cr
B_{h1}&43.7\pm1.8&43.5\pm1.8\cr
\epsilon_1&0.284\pm0.030&0.283\pm0.030\cr
r&2.54\pm0.31&2.53\pm0.31.}
\equn{(5.10b)}$$

\medskip
\noindent{\sl {\rm D2} wave}.\quad This is the only wave that changes substantially;
see \fig~5.4.  
We find now,
\medskip
$$\matrix{&\hbox{UFD}&\hbox{CFD}\vphantom{\Big|}\cr
B_0&(2.4\pm0.5)\times10^3&(3.1\pm0.5)\times10^3\cr
B_1&(7.8\pm1.0)\times10^3&(7.9\pm1.0)\times10^3\cr
B_2&(23.7\pm4.2)\times10^3&(24.7\pm4.2)\times10^3\cr
\deltav&196\pm25\;\mev&205\pm25\;\mev.\cr}
\equn{(5.11a)}$$
For the inelasticity parameter, 
$$\matrix{&\hbox{UFD}&\hbox{CFD}\vphantom{\Big|}\cr
\epsilon&0.2\pm0.2&0.15\pm0.2.\cr}
\equn{(5.11b)}$$
\medskip
\noindent{\sl {\rm F} wave}.\quad This wave is  unchanged within our precision:
\medskip
$$\matrix{&\hbox{UFD}&\hbox{CFD}\vphantom{\Big|}\cr
B_0&(1.09\pm0.03)\times10^5&(1.09\pm0.03)\times10^5\cr
B_1&(1.41\pm0.04)\times10^5&(1.41\pm0.04)\times10^5.}
\equn{(5.12)}$$ 
\medskip
\noindent{\sl Regge parameters}.\quad 
We only give the values of the Regge parameters that we have allowed to vary. 
The parameters correspond to the formulas in \sect~3.
\medskip
$$\matrix{&\hbox{UFD}&\hbox{CFD}\vphantom{\Big|}\cr
c_P&(0.0\pm1.0)\;\gev^{-2}&(0.53\pm1.0)\;\gev^{-2}\cr
c_{P'}&-0.4\pm0.4\;\gev^{-2}&-0.38\pm0.4\;\gev^{-2}\cr
\beta_{P'}&0.83\pm0.05&0.83\pm0.05\cr
\alpha_{P'}(0)&0.54\pm0.02&0.54\pm0.02\cr
\beta_\rho&1.22\pm0.14&1.30\pm0.14\cr
\alpha_\rho(0)&0.46\pm0.02&0.46\pm0.02\cr
\beta_2&0.20\pm0.2&0.22\pm0.2.\cr
}
\equn{(5.13)}$$
The only parameters that change appreciably (but both by only $\sim0.5\;\sigma$) 
are $c_P$, which was to be expected, and $\beta_\rho$.
\booksubsection{5.5. Comments}
\noindent
From Eqs.~(5.7) through (5.13) we see that the changes in most waves 
induced by the constraints given by FDR and Roy equations are very small; 
in many cases, minute or even nonexistent within the accuracy of our formulas. 
There are a few exceptions. 
First of all, the $B_n$s for the S0 and S2 waves change because now the location of the Adler zeros 
is left free (although the phase shifts themselves move very 
little). Secondly, the $B_0$ and $B_1$ parameters of the P wave vary by $0.3$
 and $0.4\;\sigma$ respectively, and, at high energy, 
the parameter $\lambda_1$ changes by $0.3\;\sigma$. And
thirdly, the only wave that suffers changes by more than one sigma is the
D2 wave, as was to be expected; the  parameter, $B_0$, moves  by $1.5\;\sigma$.
All  the other parameters of the S0, S2 and P waves,
 as well
as all the parameters of  
D0, F waves, change below the limit of relevance.
\topinsert{
\setbox0=\vbox{{\psfig{figure=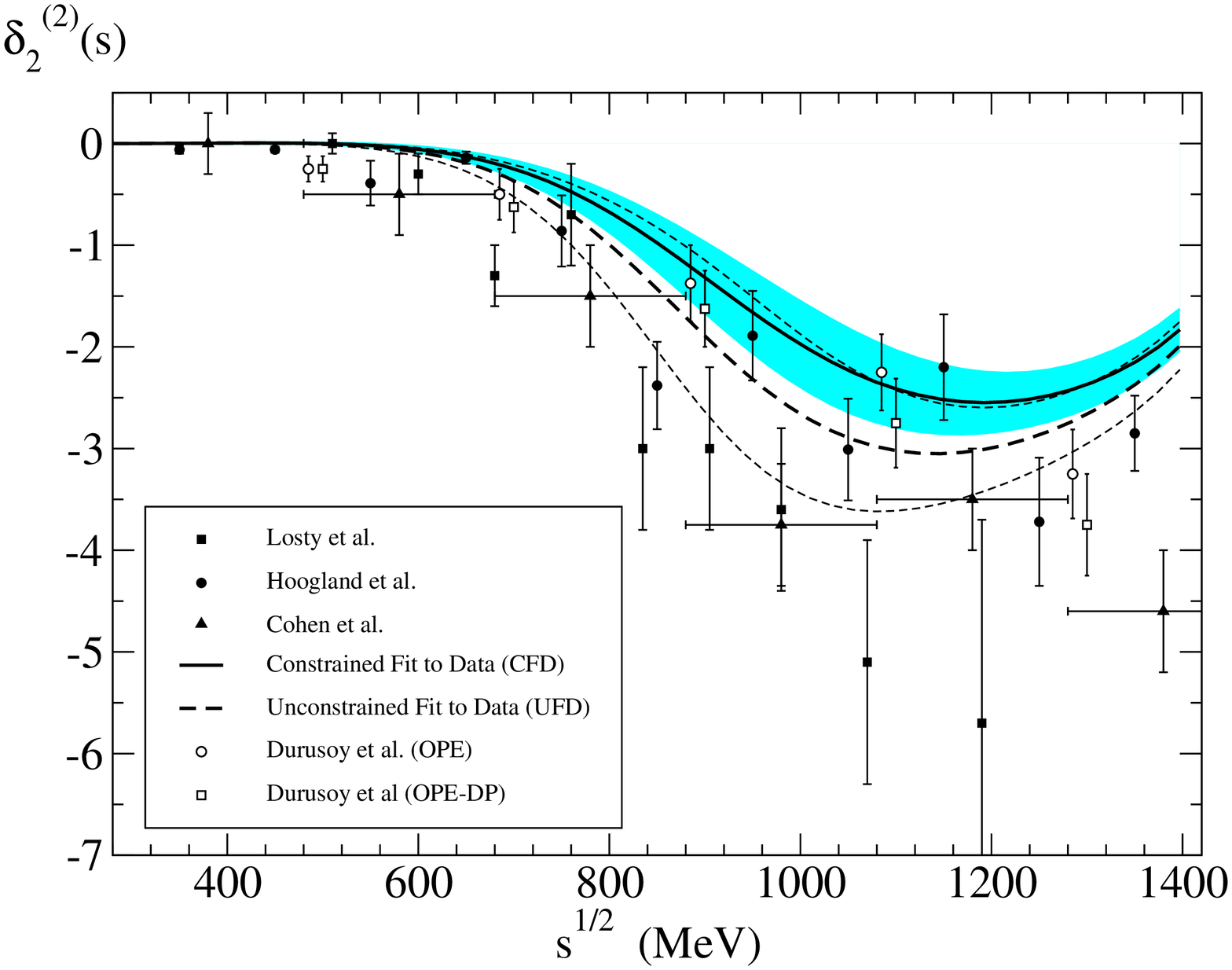,width=10.5truecm,angle=-90}}} 
\setbox6=\vbox{\hsize 15truecm\captiontype\figurasc{Figure 5.4. }{The  
$I=2$, D wave phase shift. Dashed line:  fit to data with Eq.~(2.7d).
Continuous line and shaded area:  after improving with dispersion relations.
Also shown are  data points from ref.~6.
}} 
\centerline{\tightboxit{\box0}}
\bigskip
\centerline{\box6}
\medskip
}\endinsert

\midinsert{
\setbox0=\vbox{{\psfig{figure=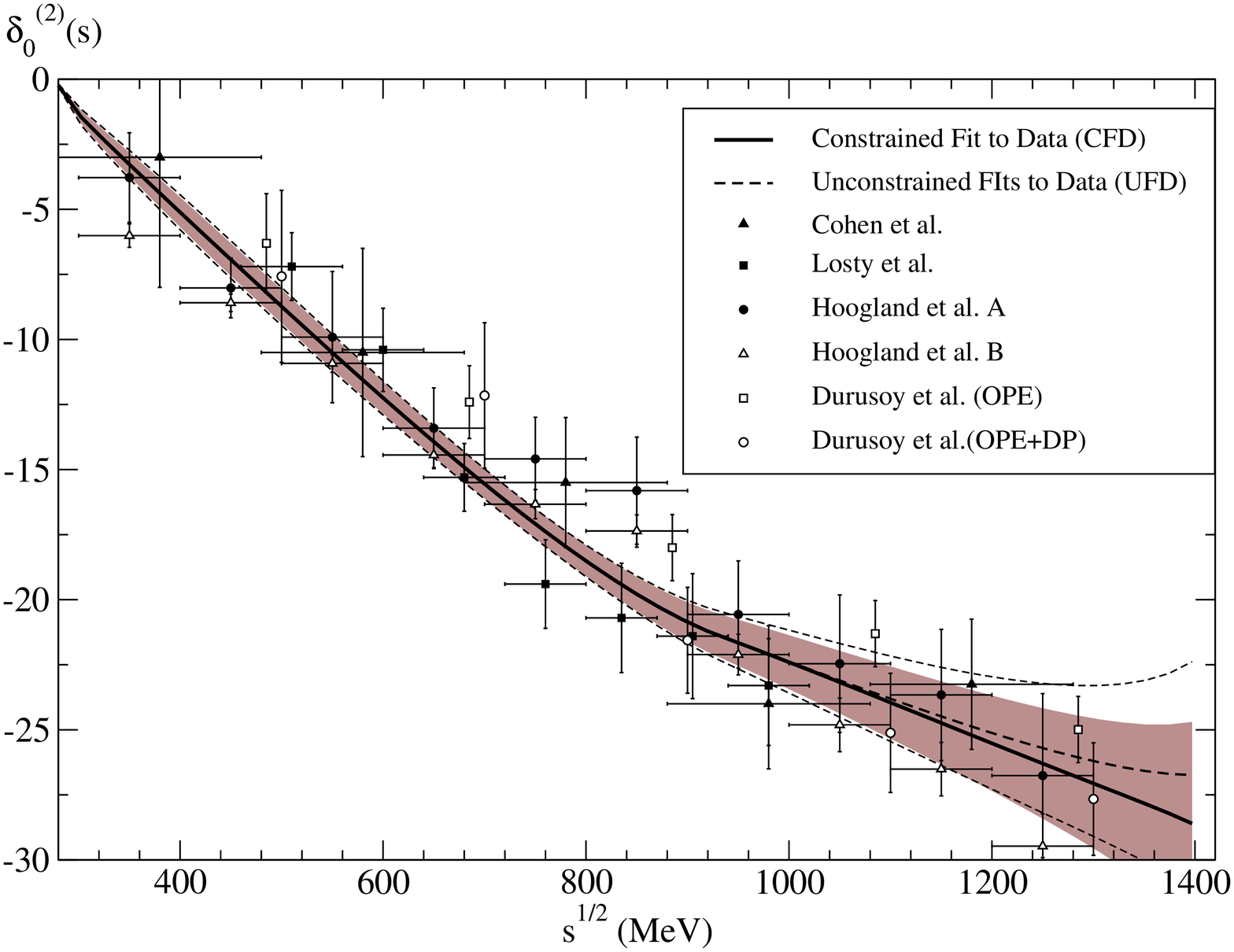,width=11.5truecm,angle=-90}}} 
\setbox6=\vbox{\hsize 15truecm\captiontype\figurasc{Figure 5.5. }{The  
$I=2$, $S$-wave phase shift. Dashed line: fit to data with 
Eqs.~(2.2), (2.3).
Continuous line and shaded area:  after improving with dispersion relations.
Also shown are data points from ref.~6.
}} 
\centerline{\tightboxit{\box0}}
\medskip
\centerline{\box6}
\medskip
}\endinsert
 
Altogether, the stability of the fit against imposing FDR and Roy equations is 
remarkable, showing its robustness. 
The stability is not obtained at the cost of large errors; quite the 
contrary. Except for the S2 wave at intermediate energies and
for the D2 wave, where the errors are 
larger, the errors in the other waves are as small (for the P, F waves) or much smaller 
(by a factor $\sim3$) than what we had in previous fits, in PY05, KPY06, 
{\sl even after improving} with FDR. 

Finally, a few words may be said on the Regge parameters. 
The parameters for exchange of isospin zero are almost unchanged 
when requiring fit to FDR, Roy equations and sum rules. 
This was to be expected; they are very well determined from 
$\pi N$ and $NN$ amplitudes using factorization. 
For exchange of isospin unity, 
only the parameter $\beta_\rho$ changes appreciably, and this by 
 $\sim0.5\;\sigma$. This shows the high 
degree of compatibility between our amplitudes above and below 1420~\mev.
\booksection{6. Low energy parameters and other observables}
\vskip-0.5truecm
\booksubsection{6.1. General}
\noindent
We   
present in the following Table~1  the low energy parameters 
(scattering lengths and effective range 
parameters, in units of the charged pion mass)
that follow from our calculations. Besides scattering lengths and effective
 range parameters,  defined as 
$$\dfrac{s^{1/2}}{2 M_\pi k^{2l+1}}\real \hat{f}_l^{(I)}(s)\simeqsub_{k\to0}
a_l^{(I)}+b_l^{(I)} k^2+\cdots\,;\quad \hat{f}_l^{(I)}=\sin\delta_l^{(I)}\ee^{\ii\delta_l^{(I)}},
 \quad
k=\sqrt{s/4-M^2_\pi},
$$  
where $\hat f_l^{(I)}$ is the partial wave of
 definite isospin $I$ and angular momentum $l$ (in the elastic region) and
$\delta_l^{(I)}$ stands for its corresponding phase shift, 
we give the quantities
 $a_0^{(0)}-a_0^{(2)}$ and $\delta_0^{(0)}(m^2_K)-\delta_0^{(2)}(m^2_K)$.
These quantities are 
relevant for pionic atom decays and  kaon decays. We also give the combination
 $2a_0^{(0)}-5a_0^{(2)}$  that 
appears in the Olsson sum rule, 
$$2a_0^{(0)}-5a_0^{(2)}= 3M_\pi\int_{4M_\pi^2}^\infty \dd s\,
\dfrac{\imag F^{(I_t=1)}(s,0)}{s(s-4M_\pi^2)},
\equn{(6.1)}$$
which corresponds to the $I_t=1$ amplitude
 forward dispersion relation evaluated at threshold.
This sum rule is very useful in determining a precise value for $a_0^{(2)}$.

Besides the Olsson sum rule, the results from direct fits may be improved with the help of   
 the Froissart--Gribov representation (whose explicit form may be found in PY05), 
 and the following sum rules: 
 the sum rule  [Eq.~(6.8)] in PY05, which we repeat here in Eq.~(6.2) for ease of reference, 

$$\eqalign{b_1=&\,
\dfrac{2}{3M_\pi }\int_{4M^2_\pi}^\infty\dd s\,\Bigg\{
\tfrac{1}{3}\left[\dfrac{1}{(s-4M^2_\pi)^3}-\dfrac{1}{s^3}\right]\imag F^{(I_t=0)}(s,0)
+\tfrac{1}{2}\left[\dfrac{1}{(s-4M^2_\pi)^3}+\dfrac{1}{s^3}\right]\imag F^{(I_t=1)}(s,0)\cr
-&\,\tfrac{5}{6}\left[\dfrac{1}{(s-4M^2_\pi)^3}-
\dfrac{1}{s^3}\right]\imag F^{(I_t=2)}(s,0) \Bigg\}
\cr}
\equn{(6.2)}$$
and two sum rules involving the effective range parameters for the S0, S2 waves. 
These are obtained  evaluating the limit as $s\to 4M^2_\pi$ of the ratio
$$\dfrac{F(s,0)-F(4M^2_\pi,0)}{s(s-4M^2_\pi)},$$
in the forward dispersion relations for $\pi^0\pi^0$ and $\pi^0\pi^+$ scattering.
We find,
$$\eqalign{
b_0^{(0)}+2b_0^{(2)}=6M_\pi\lim_{{s\to4M^2_\pi}\atop{s>4M^2_\pi}}\pepe
\int_{4M^2_\pi}^\infty\dd s'\,\dfrac{(2s'-4M^2_{\pi})\imag F_{00}(s')}
{s'(s'+s-4M^2_\pi)(s'-4M^2_\pi)(s'-s)},\cr
3a_1^{(1)}+b_0^{(2)}=4M_\pi\lim_{{s\to4M^2_\pi}\atop{s>4M^2_\pi}}\pepe
\int_{4M^2_\pi}^\infty\dd s'\,\dfrac{(2s'-4M^2_{\pi})\imag F_{0+}(s')}
{s'(s'+s-4M^2_\pi)(s'-4M^2_\pi)(s'-s)}.\cr
}
\equn{(6.3)}$$ 
Note that the limit has to be taken for $s$ {\sl larger} than $4M^2_\pi$; for 
$s<4M^2_\pi$, the derivative of $\real f_0(s)$ diverges like 
$({\rm Constant})/\ii k$. 
The principal part of the integrals is essential; 
the r.h.s. in (6.3) is convergent at the lower limit of integration only because 
it is a principal part, and one thus has
$$\pepe\int_0^\infty\dd x\,\dfrac{1}{(x-y)\sqrt{x}}=0,\quad{\rm for}\;y>0.$$

Taking the  value  $a_1=(38.2\pm1.3)\times10^{-3}\,M_{\pi}^{-3}$ from the 
Froissart--Gribov representation,
with CFD waves (cf.
 Table~1 below) this gives 
$$b_0^{(0)}=0.289\pm0.008\;M_{\pi}^{-3},\quad b_0^{(2)}=-0.081\pm0.0035\;M_{\pi}^{-3}.
\equn{(6.4)}$$

\topinsert{
\bigskip
\setbox0=\vbox{\petit
\setbox1=\vbox{ \offinterlineskip\hrule
\halign{
&\vrule#&\strut\hfil\ #\ \hfil&\vrule#&\strut\hfil\ #\ \hfil&
\vrule#&\strut\hfil\ #\ \hfil&
\vrule#&\strut\hfil\ #\ \hfil&\vrule#&\strut\hfil\ #\ \hfil\cr
 height2mm&\omit&&\omit&&\omit&&\omit&&\omit&\cr 
&\hfil \hfil&&\hfil Unc. fits (UFD) \hfil&&Constrained (CFD)&
&\hfil Sum rules, with CFD\hfil&
&\hfil Best values\hfil& \cr
 height1mm&\omit&&\omit&&\omit&&\omit&&\omit&\cr
\noalign{\hrule} 
height1mm&\omit&&\omit&&\omit&&\omit&&\omit&\cr
&$a_0^{(0)}$&&\vphantom{\Big|}$0.231\pm0.009$&&$0.223\pm0.010$&
&\hfil \hfil&&
$0.223\pm0.009 $& \cr 
\noalign{\hrule}
height1mm&\omit&&\omit&&\omit&&\omit&&\omit&\cr
&$a_0^{(2)}$&&\vphantom{\Big|}$-0.052\pm0.010$&&$-0.0451\pm0.0088$&
&\hfil \hfil&&
$-0.0444\pm0.0045$ $^{(e)}\vphantom{|}$& \cr 
\noalign{\hrule}
height1mm&\omit&&\omit&&\omit&&\omit&&\omit&\cr
&$a_0^{(0)}-a_0^{(2)}$&\vphantom{\Big|}&$0.282\pm0.014$&&$0.268\pm0.014 $&
&\hfil \hfil&&
$0.267\pm0.009^{(f)}$& \cr 
\noalign{\hrule}
height1mm&\omit&&\omit&&\omit&&\omit&&\omit&\cr
&$2a_0^{(0)}-5a_0^{(2)}$&\vphantom{\Big|}&$0.720\pm0.055$&&$0.672\pm0.048$&
&\hfil$0.667\pm0.018^{(a)}$ \hfil&&
$0.668\pm0.017$& \cr 
\noalign{\hrule}
height1mm&\omit&&\omit&&\omit&&\omit&&\omit&\cr
&$\delta_0^{(0)}(m^2_K)-\delta_0^{(2)}(m^2_K)$&\vphantom{\Big|}&$51.7\pm1.2\degrees$&
&$50.9\pm1.2\degrees$&
&\hfil \hfil&&
$50.9\pm1.2\degrees$& \cr 
\noalign{\hrule}
height1mm&\omit&&\omit&&\omit&&\omit&&\omit&\cr
&$b_0^{(0)}$&&\vphantom{\Big|}$0.288\pm0.009$&
&$0.291\pm0.009$&
&\hfil$0.289\pm0.008^{(d)}$ \hfil&&
$0.290\pm0.006$& \cr 
\noalign{\hrule}
height1mm&\omit&&\omit&&\omit&&\omit&&\omit&\cr
&$b_0^{(2)}$&&\vphantom{\Big|}$-0.085\pm0.010$&
&$-0.084\pm0.010$&
&\hfil$-0.081\pm0.0035^{(d)}$ \hfil&&
$-0.081\pm0.003$& \cr 
\noalign{\hrule} 
height1mm&\omit&&\omit&&\omit&&\omit&&\omit&\cr
&$a_1\quad(\times\,10^{3})$&&\vphantom{\Big|}$37.3\pm1.2$&&$38.0\pm1.2$&
&\hfil$ 38.2\pm1.3^{(b)}$ \hfil&&
$38.1\pm0.9$& \cr 
\noalign{\hrule}
height1mm&\omit&&\omit&&\omit&&\omit&&\omit&\cr
&\vphantom{\Big|}$b_1\quad(\times\,10^{3})$ 
\phantom{\big|}&&\phantom{\Bigg|}$5.18\pm0.23$&&$5.09\pm0.25$&
&\hfil ${{\displaystyle5.42\pm0.91^{(b)}
\atop{\displaystyle5.13\pm0.19^{(c)}}}}$  
\hfil&&
$5.12\pm0.15$& \cr
\noalign{\hrule} 
height1mm&\omit&&\omit&&\omit&&\omit&&\omit&\cr
&$a_2^{(0)}\quad(\times\,10^{4})$&\vphantom{\Big|}&$18.7\pm0.4$&
&\hfil $18.7\pm0.4$ \hfil&&$18.33\pm0.36^{(b)}$&&
$18.33\pm0.36$& \cr
\noalign{\hrule} 
height1mm&\omit&&\omit&&\omit&&\omit&&\omit&\cr
&\vphantom{\Big|}$a_2^{(2)}\quad(\times\,10^{4})$&& $2.5\pm1.1$&
&\hfil $2.4\pm0.9$ \hfil&&$2.46\pm0.25^{(b)}$&&
$2.46\pm0.25$& \cr
\noalign{\hrule}
height1mm&\omit&&\omit&&\omit&&\omit&&\omit&\cr
&$b_2^{(0)}\quad(\times\,10^{4})$&\vphantom{\Big|}&$-4.2\pm0.3$&
&\hfil$-4.2\pm0.3$  \hfil&&$-3.82\pm0.25^{(b)}$&&
$-3.82\pm0.25$ & \cr
\noalign{\hrule} 
height1mm&\omit&&\omit&&\omit&&\omit&&\omit&\cr
&\vphantom{\Big|}$b_2^{(2)}\quad(\times\,10^{4})$&&$-2.7\pm1.0$&
&\hfil $-2.5\pm0.8$ \hfil&&$-3.64\pm0.18^{(b)}$&&
$-3.59\pm0.18$ & \cr
\noalign{\hrule}
height1mm&\omit&&\omit&&\omit&&\omit&&\omit&\cr
&\vphantom{\Big|}$a_3\quad(\times\,10^{5})$&&$5.2\pm1.3$ &
&\hfil $5.2\pm1.3$ \hfil&
&$6.05\pm0.29^{(b)}$&&
$6.05\pm0.29$& \cr
\noalign{\hrule}
height1mm&\omit&&\omit&&\omit&&\omit&&\omit&\cr
&\vphantom{\Big|}$b_3\quad(\times\,10^{5})$&&$-4.7\pm2.6$&
&\hfil $-4.8\pm2.7$ \hfil&
&$-4.40\pm0.36^{(b)}$&&
$-4.41\pm0.36$& \cr
\noalign{\hrule}}
\vskip.05cm}
\centerline{\box1}
\bigskip
\medskip
\morenarrow{\noindent\petit
Units of $M_\pi$. The numbers  in the ``Sum rules" column  are from the 
Olsson sum rule$^{(a)}$, the  
Froissart--Gribov represen\-tation,$^{(b)}$ the sum rule in Eq.~(6.2),$^{(c)}$ 
and the sum rules in Eq.~(6.3).$^{(d)}$\hb
$^{(e)}$ This number is obtained composing the CFD values for 
$a_0^{(0)},\,a_0^{(2)}$ with the best value for $2a_0^{(0)}-5a_0^{(2)}$; which 
best value is obtained  from the Olsson sum rule. $^{(f)}$ This number takes into account the 
``best values" for $a_0^{(0)}$, $a_0^{(2)}$ given above.
 \hb 
For the best values  for $a_1$,
$b_1$, $b_0^{(0)}$ and $b_0^{(2)}$ 
  we have averaged what comes from constrained fits (CFD), with what
one finds from the sum rules 
(since they are practically independent).  The best
values for the other parameters are as follows: from the constrained fits (CFD),  for
$a_0^{(0)}$. For D0, D2 and F waves, the best 
values are those coming from
the Froissart--Gribov representation; because 
 our fits to data  {\sl impose} these values, it 
would not make sense to average them. However, we {\sl have} averaged 
the results for $b_2^{(2)}$ 
and $b_3$ 
since we did not impose their values when fitting the data.}
\bigskip
\centerline{\sc Table~1}
\bigskip
\hrule
}
\box0
\bigskip
}\endinsert

\booksubsection{6.2. The scattering lengths $a_0^{(0)}$, $a_0^{(2)}$}

\noindent
The results reported under the headings UFD, CFD
 in Table~1 are what is found by fitting experimental data
on  partial wave amplitudes. However, 
for the scattering lengths  $a_0^{(0)}$ and $a_0^{(2)}$ one can improve the results 
using the Olsson sum rule. One takes the value of the combination 
$2a_0^{(0)}-5a_0^{(2)}$ from the integral in (6.1), a value that is practically independent of 
that obtained fitting data, and thus sets the constraints
$$\eqalign{
a_0^{(0)}=&\,0.223\pm0.010\quad\hbox{[CDF]}\cr
a_0^{(2)}=&\,-0.0451\pm0.0088\quad\hbox{[CDF]}\cr
2a_0^{(0)}-5a_0^{(2)}=&\,0.667\pm0.018\quad\hbox{[Olsson sum rule]}\cr
}
\equn{(6.5)}$$
(in units of $M_\pi$).

\topinsert{
\bigskip
\setbox0=\vbox{\hsize9truecm 
\setbox2=\vbox{{\psfig{figure=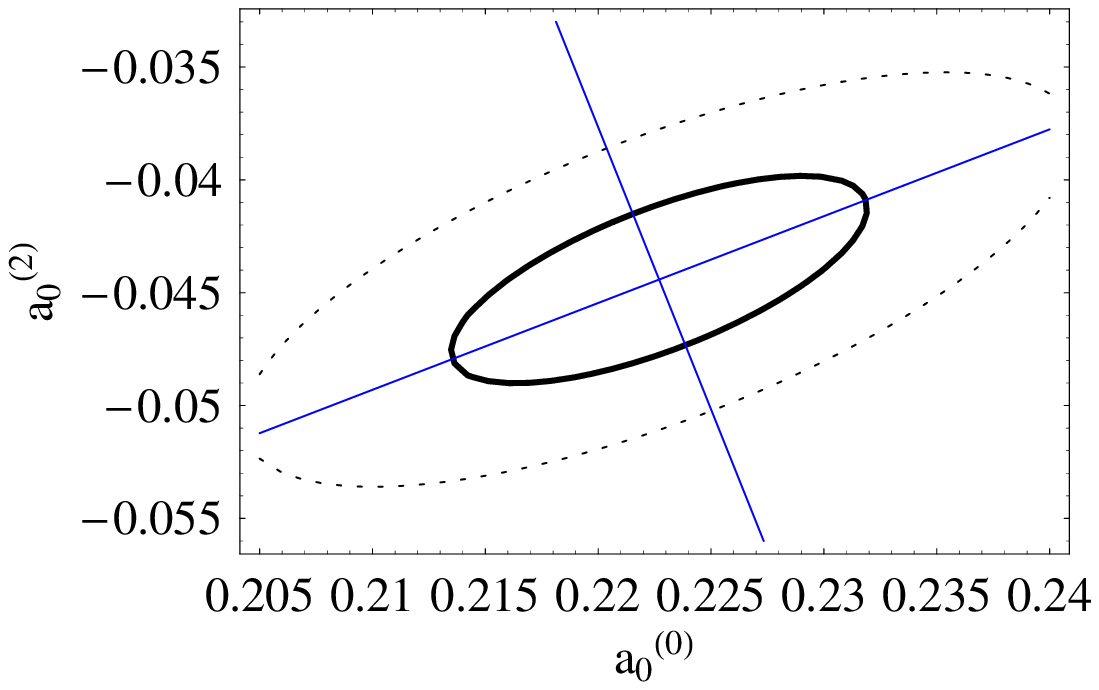,width=9.truecm,angle=0}}}
\centerline{\tightboxit{\box2}}}
\setbox6=\vbox{\hsize 15truecm\captiontype\figurasc{Figure 6.1. }{The 
ellipses  in the $a_0^{(0)} - a_0^{(2)}$ 
plane corresponding to  1-sigma (thick continuous line) and 2-sigma (broken line).}\hb}
\centerline{\box0}
\bigskip
\centerline{\box6}
}\endinsert

We can then fit the  $a_0^{(0)}$, $a_0^{(2)}$ minimizing (6.5).
The resulting errors are strongly correlated (as is well known); 
the corresponding ellipse is shown in \fig~6.1.
We can uncorrelate the errors by using two new variables, 
$x$, $y$ with
$$\eqalign{
a_0^{(0)}=&0.223+0.129\,x+0.335\,y\cr
a_0^{(2)}=&-0.0444-0.335\,x+0.129\,y;\cr
x=&\,0\pm0.0087,\quad y=0\pm0.027.
}
 \equn{(6.6)}$$ 
This gives the central values, and errors, reported under the heading ``Best values"
in Table~1, 
$$a_0^{(0)}=0.223\pm0.009,\quad a_0^{(2)}=-0.0444\pm0.0045.
 \equn{(6.7)}$$
This represents a reasonable improvement on the errors  we had before for 
$a_0^{(2)}$.
We consider (6.7) to be our best result for these scattering lengths.

The S-waves scattering lengths can also be compared with  
other experimental information, not used in our fits, that give directly the combination
 $a_0^{(0)}-a_0^{(2)}$. Indeed, 
the value found in Table~1 for $a_0^{(0)}-a_0^{(2)}$  agrees very well with 
the following independent experimental determinations: from pionic atoms,\ref{16} that give
$$a_0^{(0)}-a_0^{(2)}=0.280\pm0.013\;({\rm St.})\pm0.008\;({\rm Syst.})\;M^{-1}_\pi$$
and from $K_{3\pi}$ decays that imply\ref{17}
$$a_0^{(0)}-a_0^{(2)}=0.268\pm0.010\;({\rm St.})\pm0.013\;({\rm Syst.})\;M^{-1}_\pi.$$

\booksection{7. Conclusions}

\noindent
In the previous Sections we have given a representation of the 
$\pi\pi$ scattering amplitudes obtained fitting experimental data below 
1.42~\gev, supplemented by standard Regge formulas above this energy, what we have called the UFD~Set. 
We have shown that this UFD~Set satisfies very well 
 forward dispersion relations and Roy equations, as well as crossing sum rules.
Then, we have improved the central values of our fits 
requiring, besides fit to data, verification of FDR, Roy equations and sum rules, 
getting what we have called CFD~Set. 
The central values in this  CFD~Set lie well inside those of the UFD~Set, except 
for the D2 wave. FDR are now very well satisfied, while the  
verification of Roy equations is spectacular.

\midinsert{
\setbox0=\vbox{\hsize8truecm 
\setbox1=\vbox{{\psfig{figure=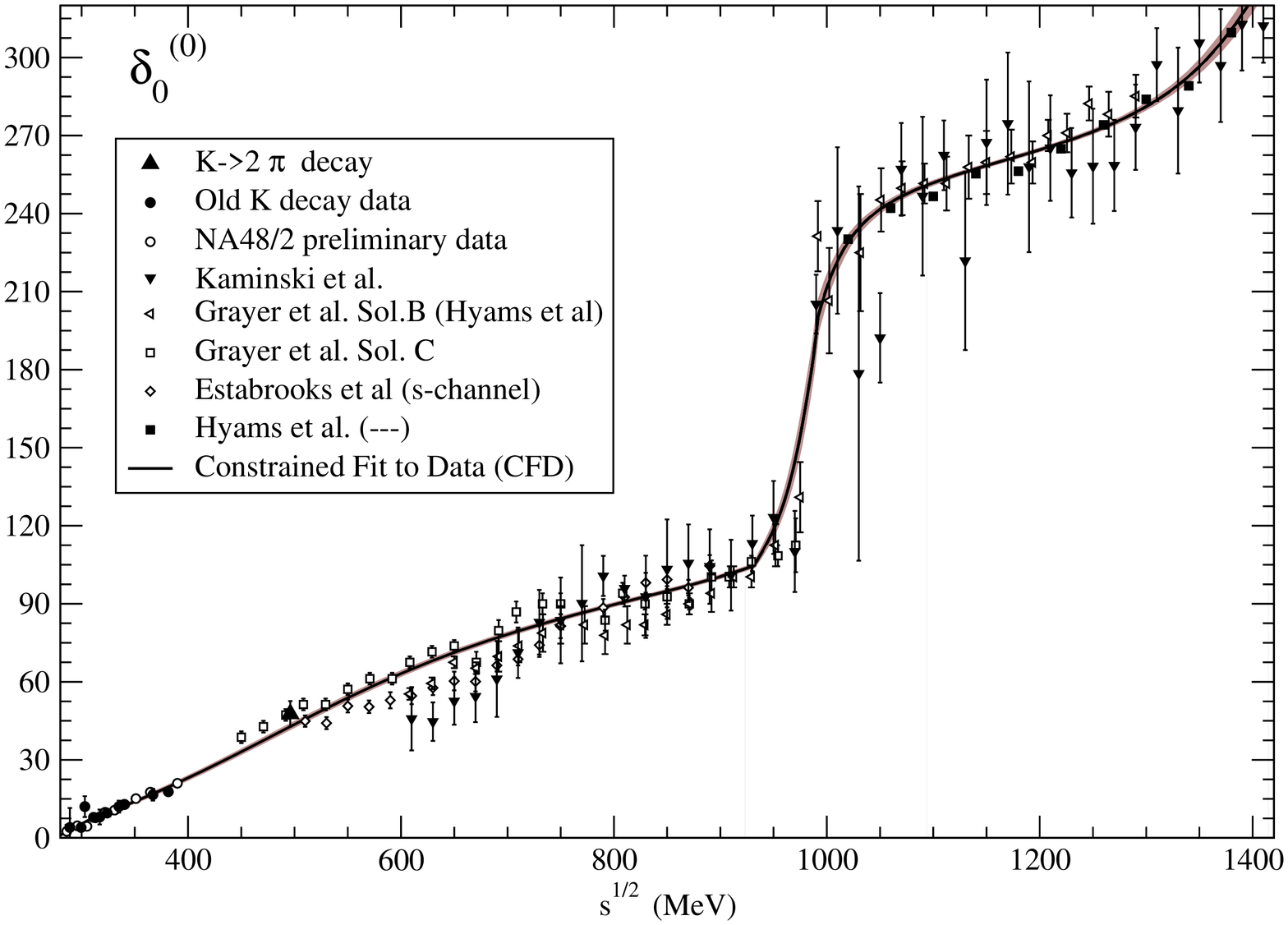,width=14truecm,angle=-90}}} 
\centerline{\tightboxit{\box1}}}
\setbox6=\vbox{\hsize 15truecm\captiontype\figurasc{Figure 7.1{\sc a}. }{S0 wave phase shift
(CFD~Set). 
Some  data from refs.~4, 6 are also shown. Notice the hump around 700~\mev.}\hb}
\centerline{{\box0}}
\bigskip
\centerline{\box6}
}\endinsert 

Altogether, we have confirmed the findings of PY05, KPY06 and 
(for the S0 wave) of GMPY07:
 but we have now errors {\sl much smaller} than 
in PY05.  At present, the low energy P wave is known at
the limit of validity of our  formalism (a limit
given by isospin breaking effects,\fnote{Some isospin breaking effects will be discussed
below.} that our analysis neglects);  while the
S0 and D0 wave are near this same limit.
 The S0, P and D0 phase shifts, with the values of the  CFD~Set, 
are shown in \figs~7.1{\sc a, b, c}.

\topinsert{
\bigskip
\setbox0=\vbox{\hsize8truecm
\setbox2=\vbox{{\psfig{figure=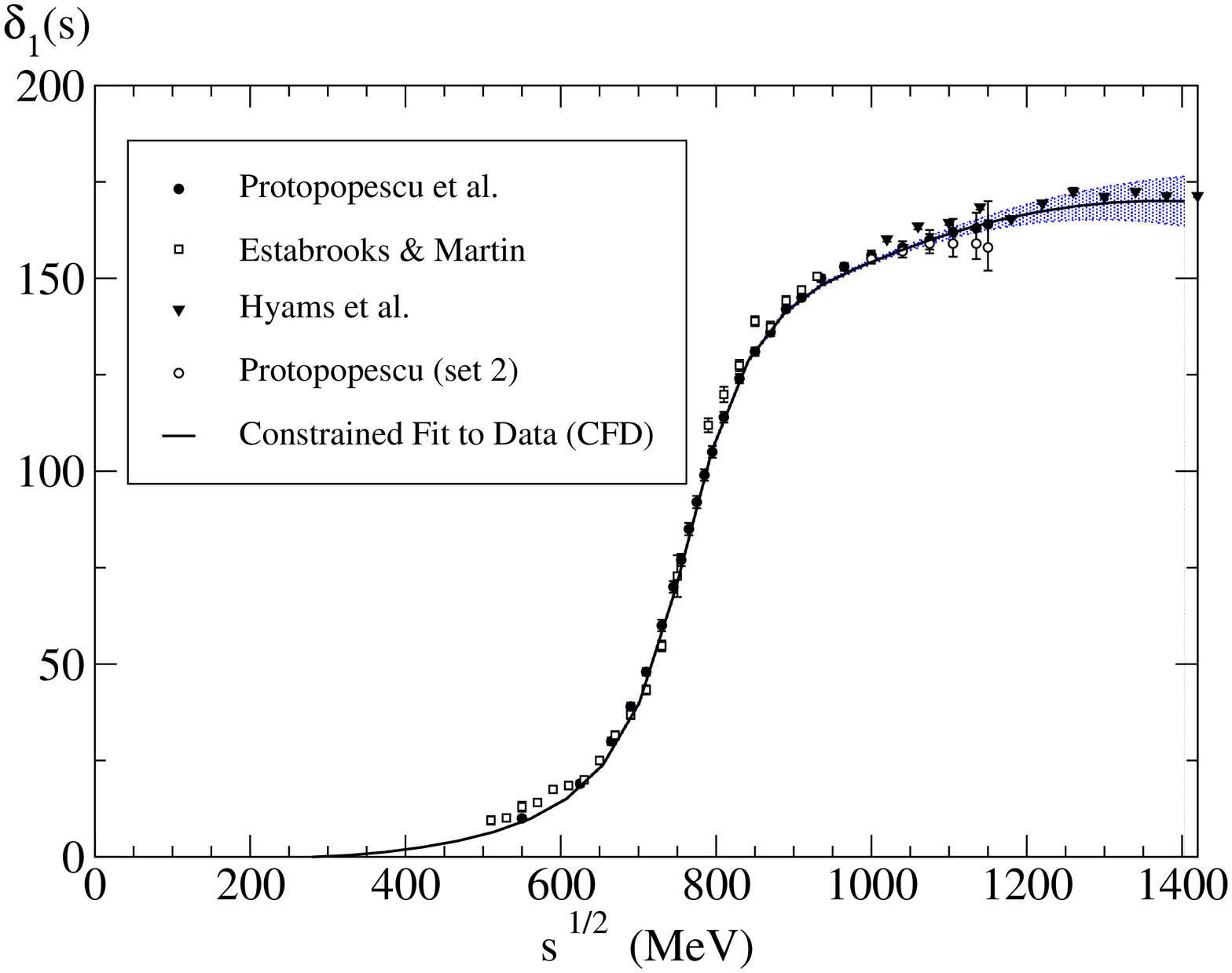,width=12.5truecm,angle=-90}}}
\centerline{\tightboxit{\box2}}}
\setbox6=\vbox{\hsize 15truecm\captiontype\figurasc{Figure 7.1{\sc b}.}
{P wave  phase shift (CFD~Set);
errors below 1~\gev\ are as the thickness of the line.  Some  data from
refs.~4, 6 are also shown. Note,
however, that $\delta_1$ is  obtained  fitting the $\pi\pi$ scattering
data only {\sl above} 
$\bar{K}K$ threshold; below it, it is obtained fitting the pion form factor
(cf.~ref.~8).}\hb}
\centerline{{\box0}}
\bigskip
\centerline{\box6}
}\endinsert

\midinsert{
\medskip
\setbox0=\vbox{\hsize8truecm 
\setbox3=\vbox{{\psfig{figure=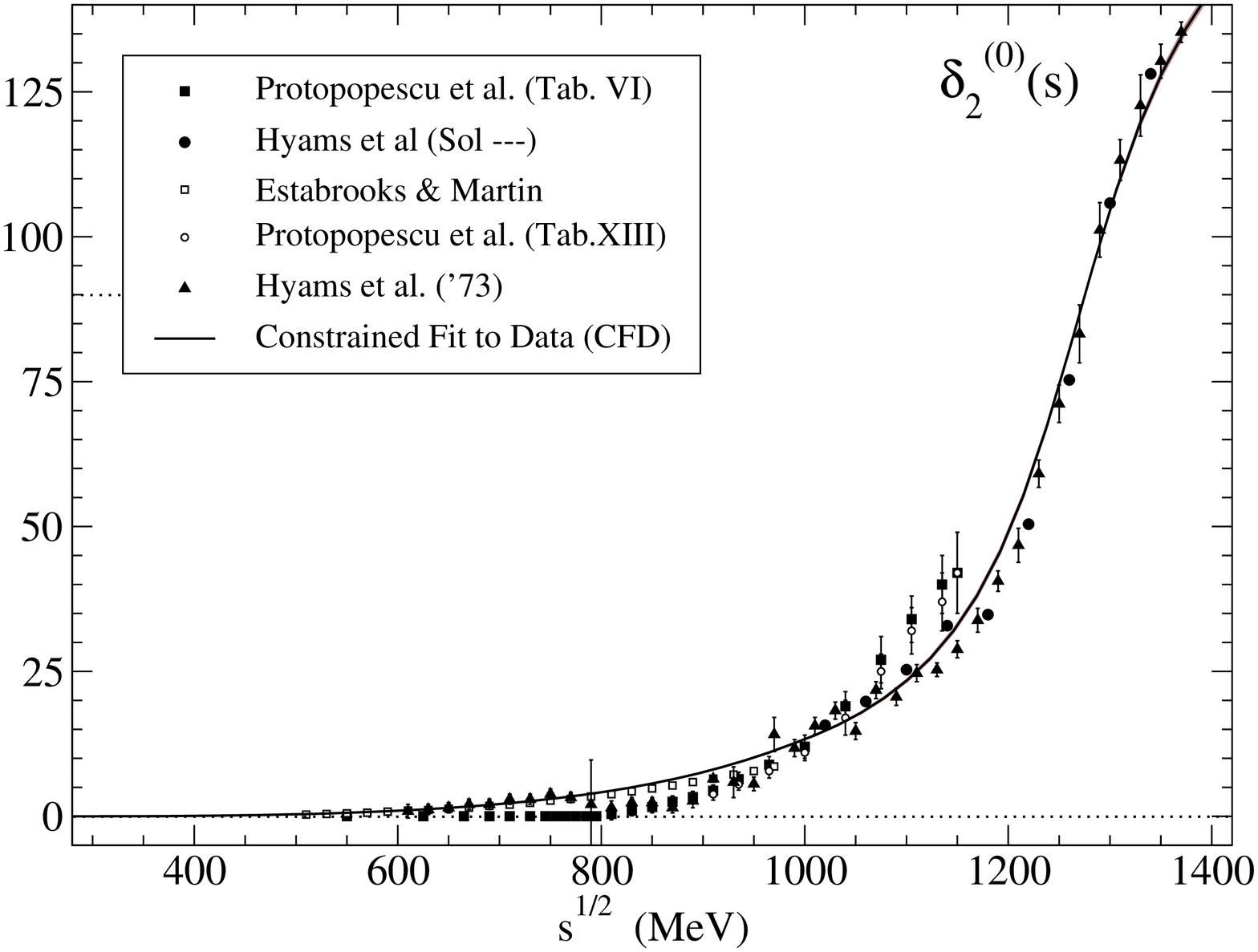,width=12truecm,angle=-90}}}
\centerline{\tightboxit{\box3}}}
\setbox6=\vbox{\hsize 15truecm\captiontype\figurasc{Figure 7.1{\sc c}.}
{D0  wave phase shift (CFD~Set);
errors as the thickness of the line.  Some  data from refs.~4, 6 are also
shown.}\hb}
\centerline{{\box0}}
\bigskip
\centerline{\box6}
}\endinsert

In connection with the S0 wave, there is some  interest attached to  the location of the so-called
``$\sigma$ pole".  This has been discussed in GMPY07; we there found, with the parameters
of the UFD Set,
$$M_\sigma=496\pm6\,({\rm St.)}\pm11\,({\rm Sys.})\;\mev,\quad
 \gammav_\sigma/2=258\pm8\,({\rm St.)}\pm2\,({\rm Sys.})\;\mev.
$$
For the CFD Set we now have
$$M_\sigma=473\pm6\,({\rm St.)}\pm11\,({\rm Sys.})\;\mev,\quad
 \gammav_\sigma/2=257\pm5\,({\rm St.)}\pm2\,({\rm Sys.})\;\mev.
$$
Here the first error is the statistical one, due to the errors in the parameters of the S0 wave, 
and the second the  error induced by the extrapolation, as estimated in GMPY07.
We gave the numbers not
taking into account  isospin breaking; if we took it into account (see 
below for a discussion of this), the central values would change
a little, to 
$M_\sigma=478$ and $\gammav_\sigma/2=242$ for the CFD Set.

Nevertheless, we must remark that a really precise determination of the 
location of this resonance requires the use of the Roy equations. 
We will present this in a future paper.

 The S2 wave is not found with a precision similar to that of the S0 or P waves. 
This is, of course, due to the lack of accuracy of the experimental data on  states
with isospin 
two (but, on the other hand, the corresponding scattering length,
 $a_0^{(2)}$,
is found with good accuracy).
 The lack of
experimental accuracy is what prevents more precise values for the D2 wave, and something similar 
happens for the F wave.
 
By comparison with an analysis similar to ours,  our 
values for S0, D waves are a factor of three or four more precise than those in ref.~14.
Our D2 wave, uncertain as it is, is still 
much more reliable than what is given in ref.~14: where a  parametrization
for this wave is given that  is incompatible with 
 their own findings at low energy (the scattering
length),
 and with
Regge theory and experimental data at
high energy. 
Finally,  at high energy (in the Regge region),
 our amplitudes fit well the existing data, something that
the  amplitudes in ref.~14 fail to do by a factor of two.

We have also used these scattering amplitudes to evaluate low energy parameters 
for P, D0, D2, and F waves, and other observables, 
clearly improving on previous work.
These parameters may then be used to test chiral
 perturbation theory to one and two loops,
or to find quantities relevant for 
CP 
violating kaon decays. 
In particular, for the S-wave scattering lengths we find the very accurate determinations
$$a_0^{(0)}=0.223\pm0.009\,M^{-1}_\pi,\quad a_0^{(2)}=-0.0444\pm0.0045\,M^{-1}_\pi.
\equn{(7.1)}$$

This can be compared with what is found in ref.~18, using chiral perturbation theory with Roy equations
and experimental data input,
$$a_0^{(0)}=0.220\pm0.005\;M^{-1}_\pi,\quad
 a_0^{(2)}=-0.0444\pm0.0010\;M^{-1}_\pi.
\equn{(7.2)}$$
 that is, a remarkable agreement at very low energies.
However,  at higher energies our S0 wave phase shift deviates somewhat from that obtained 
in ref.~18. This deviation is due to the  hump we find in the 400 to 900 MeV
region, that makes our phase shifts  larger than those of ref.~18.
[It is to be noted, on the other hand, that this hump is also generated naturally 
within the framework of unitarized chiral perturbation theory, 
ref.~19, when fitting all existing scattering data (using a large systematic error to
cover all phase shift  sets)]. The S2 wave also devites slightly (but significantly) 
from that of ref.~18 above 500~\mev.

One  can ask the question whether it would be possible to improve on {\sl our} 
precision. The answer is, no in the sense that our amplitudes agree, within errors, 
with theoretical requirements and with data.\fnote{One may  think
 that imposing chiral perturbation
theory could lead to  decreasing the errors of  the 
$\pi\pi$ scattering amplitudes. 
However, this matter is complicated and 
 is left for a future publication: at least the analysis can be 
now made from the well grounded set of pion-pion amplitudes given 
by our fits.}  Moreover, for some waves (notably, S0,
P and D0, shown in \figsŒ~7.1) our precision is at the limit  of the estimated
 corrections due to breaking of isospin invariance. 
For other waves, known less precisely, 
a sizable improvement would require substantially improved experimental data; 
certainly for the  S2 wave at low energy, and for all the waves above $\sim1\,\gev$.
This is particularly true for 
 the {\sl inelasticities}, very poorly 
determined from experiment, except for the S0 wave where
 the lucky coincidence that it is mostly given 
by $\bar{K}K$ states and the existence of data on 
the process $\pi\pi\to\bar{K}K$ saves the day.
It may be assumed that, perhaps, imposing {\sl exact} fulfillment of FDR and Roy equations 
could improve our errors. 
We have found this impossible: increasing the weight of FDR and Roy equations in 
the CFD, Eq.~(5.4), results on amplitudes that indeed satisfy better 
(but not that much better)
FDR and Roy equations, but that deviate from experiment by intolerable amounts.
This is not surprising; there are important features that fail to be resolved 
by existing experimental analyses. For example, we have the matter of the 
inelasticity of the D0 wave near 1~\gev, or the lack of information on the 
$\rho(1450)$ resonance, clearly seen in $e^+e^-$ annihilations but for which the 
Particle Data Tables refrain from giving the $2\pi$ branching ratio 
--and which is absent from our analysis. 
This could give a sizable contribution above 1380~\mev: the absence of the 
$\rho(1450)$ resonance in our P wave is likely responsible for
the structure found in $\deltav_{0+}$, $\deltav_1$ above $\sim1\,\gev$; 
cf.~\figs~4.2,~5.2. Unfortunately, taking this $\rho(1450)$ resonance into account
 {\sl correctly} 
requires a multi-channel calculation, which, even if possible, 
lies outside the scope of the present article.

As for the S0 wave, 
  the contribution of 
$4\pi$ states, dominant above $\sim1350\,\mev$ is not 
properly taken into account.
Likewise, our Regge formulas only give the imaginary parts of the scattering 
amplitudes {\sl in the mean}, in the energy region
$1420\,\mev\leq s^{1/2}\lsim1800\mev$; a region where neither
 phase shift analyses nor Regge fits 
give precise results: our scattering amplitudes are not well determined in the 
region from 1350 to 1800~\mev.
Until much better experimental data, in particular on phase shifts and
 (above all) inelasticities 
above $\bar{K}K$ thresholds are available,
 the Set~CFD of pion-pion amplitudes will remain the best that one can
do, {\sl from experiment}.\fnote{That the poor 
information above $\bar{K}K$ threshold is responsible 
for the limitations in the fulfillment of dispersion relations is seen if we 
realize that FDR are satisfied to an average $\bar{d}^2_{\rm FDR}=0.49$ below 932~\mev,
and Roy equations are satisfied to an average $\bar{d}^2_{\rm Roy}=0.17$ 
below $\bar{K}K$ threshold.}

\topinsert{
\setbox0=\vbox{\hsize8truecm 
\setbox1=\vbox{{\psfig{figure=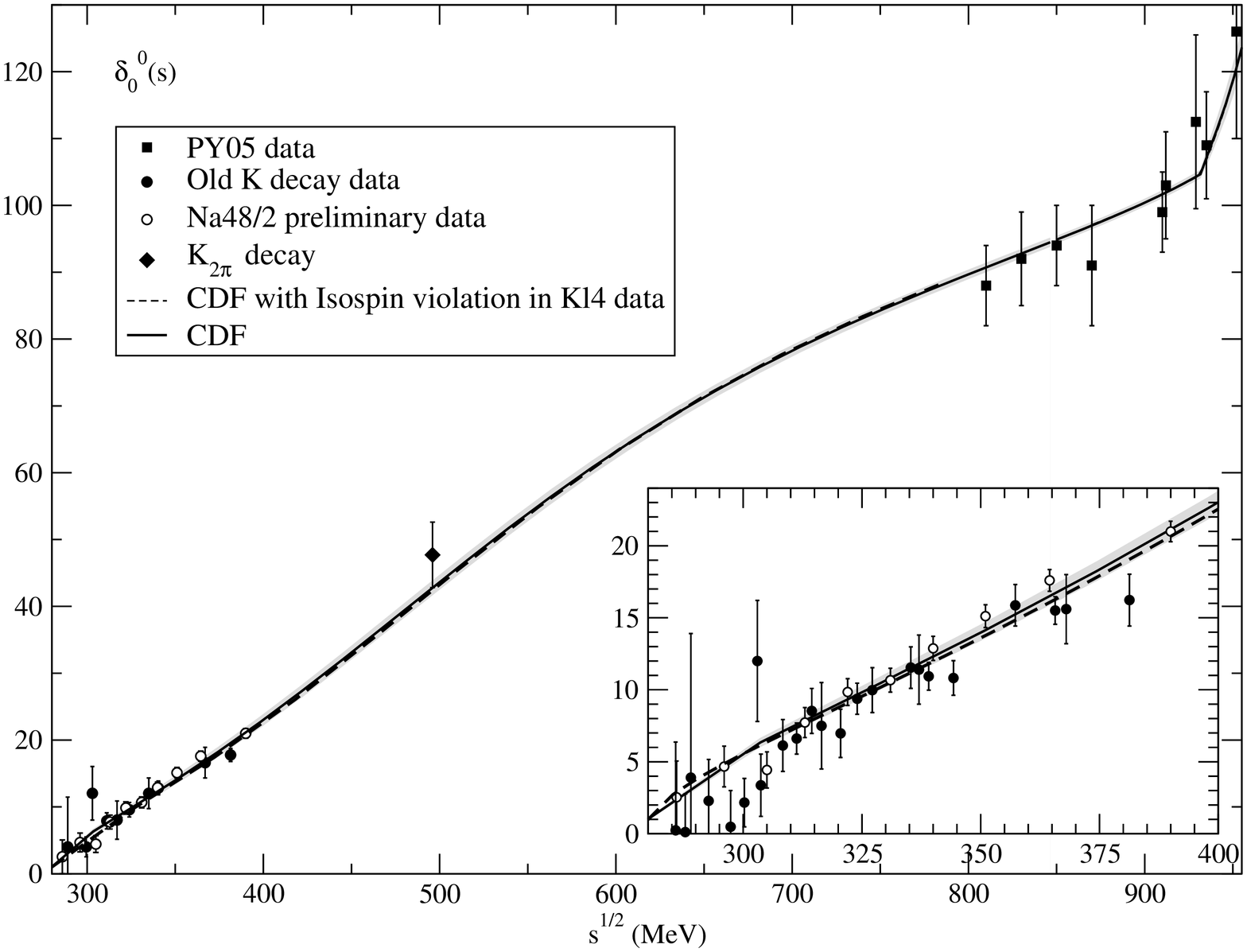,width=15.8truecm,angle=-90}}} 
\centerline{\tightboxit{\box1}}}
\setbox6=\vbox{\hsize 15truecm\captiontype\figurasc{Figure 7.2. }{S0 wave phase shift
(CFD~Set): continuous line, neglecting isospin corrections; 
broken line, with isospin corrections taken into account.
 Notice that the hump around 700~\mev\ 
is basically identical for both. 
Some  data from refs.~4, 6 are also shown.}\hb}
\centerline{{\box0}}
\bigskip
\centerline{\box6}
}\endinsert

Next, we say a few words with respect to isospin breaking corrections,
 neglected in our analysis up till 
now. 
 In most cases, our errors are sufficiently 
large to cover the estimated 
values of such effects; but there are two exceptions.
For P wave, our errors {\sl include} isospin breaking corrections. This is possible 
because we can evaluate this wave either from the pion form factor in 
$e^++e^-\to\pi^++\pi^-$ or in $\tau\to\nu+\pi^++\pi^0$: 
see ref.~8 for the details. And a special case is the S0 wave. 
Here a recent calculation has been made\ref{20} 
in which account is taken, for Ke4 decays, of the fact that in the real world 
  isospin is broken. 
According to ref.~20, this is done by subtracting, from the {\sl experimental} phase shift, as 
given in refs.~4, the correction
$$\deltav\delta_0^{(0)}=\dfrac{1}{32\pi F^2_\pi}\Bigg\{\left[ 4(M^2_\pi-M^2_{\pi^0}) +s\right]
\beta(s)+
(s-M^2_{\pi^0})\left[1+\tfrac{3}{2}r\right]\beta_0(s)\Bigg\}
\equn{(7.3)}$$
where $F_\pi\simeq 92.6\,\mev$ is the pion decay constant, $M_{\pi^0}$ is the mass of the neutral pion, 
$\beta(s)=\sqrt{1-4M^2_\pi/s}$, $\beta_0(s)=\sqrt{1-4M^2_{\pi^0}/s}$ and 
$r=(m_d-m_u)/(m_s-m_d/2-m_u/2)$. 

If we take the  results of this calculation at face value, we can repeat our fits taking 
it into account.  
For the UFD Set, only the low energy S0 wave is affected, and it is so very little: by 
less than $1\degrees$ above 400~\mev\ [which is not surprising, since, in fact, 
the correction (7.3) is actually somewhat smaller than the experimental errors over most of the range]. 
Only the scattering length moves beyond one standard deviation, to 
$a_0^{(0)}=0.210\pm0.010$. For the CFD~Set, the changes induced by incorporation 
of isospin breaking corrections in the waves other than the
S0 wave  are almost negligible. The S0 wave moves closer to the one given in the present paper, 
from which it is almost indistinguishable (\fig~7.2);
and the scattering length becomes  compatible with what we found neglecting 
isospin breaking: we now get
$a_0^{(0)}=0.213\pm0.011$. Indeed, 
including isospin breaking corrections does not much affect our results.
The corresponding values of the parameters for
 the S0 wave, the only ones that change appreciably,
 are given in Appendices~A and B; a comparison 
between what one finds with/without isospin corrections for the S0 phase shift can be seen, 
with difficulty (because they are so close one to the other)  in \fig~7.2.

We finish with a comment regarding the relative merits of Sets~UFD and CFD. 
It is clear that the central values of the CFD~Set should be considered 
as the preferred ones since they incorporate fulfillment of 
analyticity requirements:   
analyticity of the individual waves themselves (for the low 
energy region), as well as analyticity and crossing symmetry 
 of the scattering amplitudes in the form of forward 
dispersion relations, sum rules, and Roy equations. 
However, the price to pay for the last two requirements is that all waves are now 
correlated: in this sense, the UFD~Set is more robust. 
Since both solutions fit data, and are very similar, 
one can use one or the other almost interchangeably, 
except for S2 and, above all, D2 waves, for which the CFD~Set is clearly superior.

\vfill\eject
\booksection{Appendix A: fits to data up to $1.42\;\gev$ (UFD Set)}

\noindent
In this Appendix we collect the best values for the parametrizations of the various
partial waves obtained by fitting {\sl experimental data}, with the procedures defined in PY05 and
KPY06; see also \sect~2 here.
\booksubsection{A.1. The S0 wave}
\noindent{\sl The region $s^{1/2}\leq932\,\gev$}.\quad
We take $s_0=4m^2_K$, and take the Adler zero at $s=\tfrac{1}{2}z_0^2$, with 
$z_0$ fixed at $M_\pi$. 
We find
$$\eqalign{
\cot\delta_0^{(0)}(s)=&\,\dfrac{s^{1/2}}{2k}\,\dfrac{M_{\pi}^2}{s-\tfrac{1}{2}z_0^2}\,
\left\{\dfrac{z_0^2}{M_\pi\sqrt{s}}+{B}_0+
{B}_1\dfrac{\sqrt{s}-\sqrt{s_0-s}}{\sqrt{s}+\sqrt{s_0-s}}+
{B}_2\left[\dfrac{\sqrt{s}-\sqrt{s_0-s}}{\sqrt{s}+\sqrt{s_0-s}}\right]^2\right\},
\quad z_0\equiv M_\pi;\cr
{B}_0=&\,4.3\pm0.3,\quad {B}_1=-26.7\pm0.6,\quad
{B}_2=-14.1\pm1.4;\cr}
\equn{(A.1.1a)}$$

This corresponds to neglect of isospin breaking. If isospin breaking is taken into account as in ref.~20, 
we find instead
$$\matrix{
B_0=3.80\pm0.34,&\quad B_1=-27.1\pm0.8,&\quad B_2=-8.3\pm1.8;&\cr
a_0^{(0)}=0.211\pm0.010\;M_{\pi}^{-1},&\; b_0^{(0)}=0.278\pm0.010\;M_{\pi}^{-3};
\cr}
\equn{(A.1.1b)}$$ 
  
\medskip
 \noindent{\sl The  S0 wave between $932\,\mev$ and $1420\,\mev$}.\quad
We here use the K-matrix fit of ref.~2.\fnote{A polynomial fit, giving results 
very similar to the K-matrix fit, can also be given. It 
may be found in Appendix~B of KPY06.}$$
\tan\delta_0^{(0)}(s)=\cases{\dfrac{k_1|k_2|\det {\bf K}+k_1K_{11}}{1+|k_2|K_{22}},
\quad s\leq 4m^2_K,\cr
\eqalign{&\,\dfrac{1}{2k_1[K_{11}+k_2^2K_{22}\det {\bf K}]}\Bigg\{
k^2_1K^2_{11}-k^2_2K^2_{22}+k_1^2k_2^2(\det{\bf K})^2-1\cr
+&\,
\sqrt{(k^2_1K^2_{11}+k^2_2K^2_{22}+k_1^2k_2^2(\det{\bf K})^2+1)^2-
4k_1^2k_2^2K^4_{12}}\;\Bigg\},\quad s\geq 4m^2_K} 
\cr}
\equn{(A.1.3a)}$$
and
$$\eta_0^{(0)}(s)=\sqrt{\dfrac{(1+k_1k_2\det{\bf K})^2+(k_1K_{11}-k_2K_{22})^2}
{(1-k_1k_2\det{\bf K})^2+(k_1K_{11}+k_2K_{22})^2}},\quad s\geq 4m^2_K.
\equn{(A.1.3b)}$$
Here $k_1=\tfrac{1}{2}\sqrt{s-4M^2_\pi}$,  $k_2=\tfrac{1}{2}\sqrt{s-4m^2_K}$
and the K-matrix elements are
$$K_{ij}(s)=\dfrac{\mu\alpha_i \alpha_j}{M_1^2-s}+
\dfrac{\mu\beta_i\beta_j}{M_2^2-s}+\dfrac{1}{\mu}\gamma_{ij}.
$$
$\mu$ is a mass scale, that we take $\mu=1\,\gev$.
The powers of $\mu$ have been arranged so that the $\alpha_i$, $\beta_i$, $\gamma_{ij}$ are
dimensionless;  they are also assumed to be 
 constant.
The pole at $M_1^2$ simulates the left hand cut of $\bf K$, and 
the pole at $M_2^2$ 
is connected with the phase shift crossing $270\degrees$ around 1.3~\gev;
both poles are necessary to get a good fit.
The values of the parameters
are
$$\eqalign{\alpha_1=&\,0.843\pm0.017,\quad \alpha_2=0.20\pm0.06,\quad \beta_1=1.02\pm0.02;\quad
\beta_2=1.33\pm0.013,\cr 
\gamma_{11}=&\,3.10\pm0.11,\quad \gamma_{12}=1.82\pm0.05,\quad \gamma_{22}=-7.00\pm0.04;\cr
M_1=&\,0.888\pm0.004\;\gev,\quad M_2=1.327\pm0.004\;\gev;
\cr}
\equn{(A.1.3c)}$$
 $M_1$   lies near the beginning of the left hand cut for
 $\bar{K}K\to\pi\pi$ scattering, located at 0.952~\gev.
The parameters in (A.1.3c) are strongly correlated. In fact, we have verified that there exists 
a wide set of minima, with very different values of the parameters. 
 Nevertheless, 
the corresponding values of $\delta_0^{(0)}$ and $\eta_0^{(0)}$  vary
very little in all
these minima, so that (A.1.3c) can be considered a faithful  
representation of the S0 wave for $\pi\pi$ scattering. 
The representations of the S0 wave are matched exactly at 932~\mev, where one 
has $$\delta_0^{(0)}((932\;\mev)^2)=104.9\pm0.5\degrees.$$

\booksubsection{A.2. Parametrization of the S2 wave}
\noindent{\sl The region $s^{1/2}\leq992\,\gev$}.\quad
For isospin 2, there is no low energy resonance, but this wave
 presents the feature that a zero 
is expected (and, indeed, confirmed by the fits). It is 
 related to the so-called Adler zeros; 
 to lowest order in chiral perturbation theory, one has the zero at  
 $s=2z_2^2$, with $z_2=M_\pi$. 
We note that, unlike the corresponding zero for the S0 wave, 
$2z_2^2$ is inside the region where the conformal expansion is expected to converge well.
We here fix $z_2=M_\pi$ and
write
$$\cot\delta_0^{(2)}(s)=\dfrac{s^{1/2}}{2k}\,\dfrac{M_{\pi}^2}{s-2z_2^2}\,
 \left\{B_0+B_1\dfrac{\sqrt{s}-\sqrt{{s_l}-s}}{\sqrt{s}+\sqrt{{s_l}-s}}\right\},
\quad z_2\equiv M_\pi;\quad {s_l}^{1/2}=1.05\;\gev.
\equn{(A.2.1a)}$$
Then we get  
$$\eqalign{
B_0=&\,-80.4\pm2.8,\quad B_1=-73.6\pm10.5.\cr
}
\equn{(A.2.1b)}$$

\medskip
\noindent{\sl The  S2 wave between $932\,\mev$ and $1420\,\mev$}.\quad
We require junction with the 
low energy phase shift, and its derivative, at 932~\mev, neglect inelasticity 
below $1.45\,\gev$, and   write
$$\eqalign{
\cot\delta_0^{(2)}(s)=&\,\dfrac{s^{1/2}}{2k}\,\dfrac{M_{\pi}^2}{s-2M^2_\pi}\,
\left\{B_{h0}+B_{h1}\left[w_h(s)-w_h(s_M)\right]+
B_{h2}\left[w_h(s)-w_h(s_M)\right]^2\right\},\cr s^{1/2}\geq &\,932\;\mev;\cr
 B_{h0}=&\,B_0+B_1w_l(s_M),\quad B_{h1}=
B_1\left.\dfrac{\partial w_l(s)}{\partial w_h(s)}\right|_{s=s_M};\cr
}
\equn{(A.2.2a)}$$
$$\eqalign{w_l(s)=\dfrac{\sqrt{s}-\sqrt{s_l-s}}{\sqrt{s}+\sqrt{s_l-s}};\quad
s_l=(1050\;\mev)^2,\cr
w_h(s)=\dfrac{\sqrt{s}-\sqrt{s_h-s}}{\sqrt{s}+\sqrt{s_h-s}};\quad s_h=(1450\;\mev)^2.\cr
}
$$
$B_{h2}$  is a free parameter. We get 
$$B_{h2}=112\pm38.
\equn{(A.2.2b)}$$
 
The inelasticity  is described by the 
empirical fit
$$\eta_0^{(2)}(s)=1-\epsilon(1-{s_l}/s)^{3/2},\quad 
\epsilon=0.17\pm0.12\quad({s_l}^{1/2}=1.05\;\gev).
\equn{(A.2.2c)}$$

\booksubsection{A.3. The P  wave}
\noindent{\sl The region $s^{1/2}\leq2m_K$}.\quad
 We have
$$\cot\delta_1(s)=\dfrac{s^{1/2}}{2k^3}
(M^2_\rho-s)\left\{\dfrac{2M^3_\pi}{M^2_\rho
\sqrt{s}}+B_0+B_1\dfrac{\sqrt{s}-\sqrt{s_0-s}}{\sqrt{s}+\sqrt{s_0-s}}
\right\};\quad s_0^{1/2}=1.05\;\gev.
\equn{(A.3.1a)}$$
The best result is 
\smallskip
$$\eqalign{
B_0=&\,1.055\pm0.011,\quad B_1=0.15\pm0.05,\quad M_{\rho}=773.6\pm0.9\,{\rm MeV}.
\cr }
\equn{(A.3.1b)}$$

\medskip
\noindent\noindent{\sl The  P wave between $2m_K$ and $1420\,\mev$}.\quad
We use a purely phenomenological parametrization:
$$\eqalign{\delta_1(s)=&\,\lambda_0+\lambda_1(\sqrt{s/4m^2_K}-1)+
\lambda_2(\sqrt{s/4m^2_K}-1)^2,\cr 
\eta_1(s)=&\,1-\epsilon_1\sqrt{1-4m^2_K/{s}}-\epsilon_2(1-4m^2_K/{s});
 \quad  s> 4m_K^2.\cr
}
\equn{(A.3.2a)}$$ 
The phase at the low energy edge,
$\delta_1((0.992\,\gev)^2)=153.63\pm0.55\degrees$, 
is obtained from the fit at low energy above; this fixes $\lambda_0$.
The rest of the parameters follow from the fit at intermediate energy. We have,
$$\eqalign{\lambda_0=&\,
2.681\pm0.010,\quad\lambda_1=1.57\pm0.18,\quad \lambda_2=-1.96\pm0.49;\cr
\epsilon_1=&\,0.10\pm0.06,\quad \epsilon_2=0.11\pm0.11.\cr
}
\equn{(A.3.2b)}$$

\booksubsection{A.4. Parametrization of the D0 wave}
\noindent{\sl The region $s^{1/2}\leq2m_K$}.\quad
To 
take into account the analyticity structure, we
fit with different expressions for energies below and above $\bar{K}K$ threshold, 
requiring however exact matching at $s=4m^2_K$.
Below $\bar{K}K$ threshold we take into account the existence of nonnegligible 
inelasticity above 1.05~\gev, which is near the $\rho\pi\pi$ 
threshold, by choosing a conformal variable $w$ appropriate to a plane cut 
for $s>(1.05\,\gev)^2$. 
So we write
$$\eqalign{
\cot\delta_2^{(0)}(s)=&\,\dfrac{s^{1/2}}{2k^5}\left(M^2_{f_2}-s\right)M^2_\pi
\Big\{B_0+B_1w\Big\},\quad  s< 4m_K^2;\cr
w=&\,\dfrac{\sqrt{s}-\sqrt{s_0-s}}{\sqrt{s}+\sqrt{s_0-s}},\quad
s_0^{1/2}=1.05\;\gev.\cr
}
\equn{(A.4.1a)}$$
The mass of the $f_2$ we fix at $M_{f_2}=1275.4\,\mev$; no error is taken for 
this quantity, since it is
negligibly small (1.2~\mev) when compared with the other errors. We find 
 the values of the parameters
$$B_{0}=12.47\pm0.12;\quad B_{1}=10.12\pm0.16.
\equn{(A.4.1b)}$$

\medskip
\noindent\noindent{\sl The  D0 wave between $2m_K$ and $1420\,\mev$}.\quad
Above  $\bar{K}K$ threshold we use the following formula for the phase shift:
$$\eqalign{
\cot\delta_2^{(0)}(s)=&\,\dfrac{s^{1/2}}{2k^5}\left(M^2_{f_2}-s\right)M^2_\pi
\Big\{B_{h0}+B_{h1}w\Big\},\quad  s> 4m_K^2;\cr
w=&\,\dfrac{\sqrt{s}-\sqrt{{s_h}-s}}{\sqrt{s}+\sqrt{s_h-s}};\quad
s_h^{1/2}=1.45\;\gev.\cr
}
\equn{(A.4.2a)}$$
This neglects inelasticity below 1.45~\gev, which is approximately the $\rho\rho$ threshold; 
inelasticity will be added by hand, see below.
We require exact matching with the low energy expression, 
which yields the value of $B_{h0}$. $B_{h1}$ follows from the fit at intermediate energy. 
We find
$$B_{h0}=18.77\pm0.16;\quad B_{h1}=43.7\pm1.8.
\equn{(A.4.2b)}$$ 

For the inelasticity  we  write, as discussed in the main text,
$$\eqalign{
\eta_2^{(0)}(s)=&\,\cases{1,\qquad  s< 4m_K^2,\phantom{\Big|}
\cr
1-\epsilon\,\left(1-\dfrac{4m^2_K}{s}\right)^{5/2}
\left(1-\dfrac{4m^2_K}{M^2_{f_2}}\right)^{-5/2}
\left\{1+r\left[1-\dfrac{k_2(s)}{k_2(M^2_{f_2})}\right]\right\},\cr
 s> 4m_K^2;\quad
 k_2=\sqrt{{s}/{4}-m^2_K}.\cr} \cr
\epsilon=&0.284\pm0.030;\quad r=2.54\pm0.31.\cr
}
\equn{(A.4.2c)}$$
\booksubsection{A.5. Parametrization of the D2 wave}
\noindent
For isospin equal 2, there are no resonances in the D wave. 
If we want a parametrization that 
applies down to threshold, we must incorporate the  
zero of the corresponding phase shift.  
We write
$$\cot\delta_2^{(2)}(s)=
\dfrac{s^{1/2}}{2k^5}\,\Big\{B_0+B_1 w(s)+B_2 w(s)^2\Big\}\,
\dfrac{{M_\pi}^4 s}{4({M_\pi}^2+\deltav^2)-s},\quad
s^{1/2}\leq1.05\;\gev,
\equn{(A.5.1a)}$$
with $\deltav$ a free parameter and
$$w(s)=\dfrac{\sqrt{s}-\sqrt{s_0-s}}{\sqrt{s}+\sqrt{s_0-s}},\quad
 s_0^{1/2}=1450\,\mev.$$ 
 Moreover, we impose  the 
value for the scattering length 
that follows from the Froissart--Gribov representation.
We  find
$$B_0=(2.4\pm0.5)\times10^3,\quad B_1=(7.8\pm1.0)\times10^3,\quad
 B_2=(23.7\pm4.2)\times10^3,\quad
\deltav=196\pm25\,\mev.
\equn{(A.5.1b)}$$ 
For the inelasticity, 
 above $1.05\,\gev$,
$$\eta_2^{(2)}(s)=1-\epsilon (1-\hat{s}/s)^3,\quad \hat{s}^{1/2}=1.05\;\gev,\quad
\epsilon=0.2\pm0.2;
\equn{(A.5.2c)}$$
this is negligible up to $1.25\,\gev$.

\booksubsection{A.6. The F wave}
\noindent For the   F wave below $s^{1/2}=1.42\,\gev$ we 
 fit the experimental phase shifts  plus the scattering length as given 
by the Froissart--Gribov representation. 
We have 

$$\eqalign{
\cot\delta_3(s)=&\,\dfrac{s^{1/2}}{2k^7}\,M^6_\pi\,
\left\{\dfrac{2\lambda
M_\pi}{\sqrt{s}}+B_0+B_1\dfrac{\sqrt{s}-\sqrt{s_0-s}}{\sqrt{s}+\sqrt{s_0-s}}\right\},\quad
s_0^{1/2}=1.45\;\gev;
\cr
 B_0=&\,(1.09\pm0.03)\times 10^5,\quad
B_1=(1.41\pm0.04)\times 10^5,\quad \lambda=0.051\times 10^5.
\cr}
\equn{(A.6.1)}$$
We neglect the inelasticity of the F wave below 1.45~\gev.
The contribution of the F wave to all our sum rules is very small 
(but not always negligible); the interest 
of calculating it lies in that it provides a test (by its very smallness)
 of the convergence of the partial wave expansions.

\booksubsection{A.7. The G waves}
\noindent 
For the G0 wave, we take its imaginary part to be given by
 $$\eqalign{
\imag \hat{f}_4^{(0)}(s)=&\,\left(\dfrac{k(s)}{k(M^2_{f_4})}\right)^{18}
{\rm BR}\dfrac{M^2_{f_4}\gammav^2\ee^{2c(1-s/M^2_{f_4})^2}}{(s-M^2_{f_4})^2+M^2_{f_4}
\gammav^2[k(s)/k(M^2_{f_4})]^{18}};\cr
{\rm BR}=&\,0.17\pm0.02,\quad
 M_{f_4}=2025\pm8\;\mev,\quad \gammav=194\pm13\;\mev;\quad c=9.23\pm0.46.\cr}
\equn{(A.7.1)}$$
For the wave G2, we can write, neglecting its eventual inelasticity,
$$\cot\delta_4^{(2)}(s)=\dfrac{s^{1/2} {M_\pi}^8}{2k^9}\,B,\quad 
B=(-9.1\pm3.3)\times 10^6;\quad
s^{1/2}\geq 1\;\gev.
\equn{(A.7.2)}$$

It should be noted that  the expressions  for 
the G0, G2 waves, are little more than order of magnitude estimates. 
Moreover, at low energies the expression for G2 certainly fails; 
below 1~\gev, an expression in terms of the scattering length 
approximation, with
$$ a_4^{(2)}=(4.5\pm0.2)\times10^{-6}\,M_{\pi}^{-9},$$
  is more appropriate.

\booksection{Appendix B: fits up to $1.42\;\gev$,
 improved with dispersion relations (CFD~Set)}

\noindent
In this Appendix we collect the best values for the parametrizations of the various
partial waves, after improving with the help of dispersion 
relations: forward dispersion relations and Roy equations. In addition, 
we required verification (within errors) 
of the two crossing sum rules in \sect~4.

All the formulas are as before improvement, i.e., as in Appendix~A; only the 
central values  of the parameters change.
We give in some detail only the  
 S0 and S2 waves, because now we allow variation of the location of the 
Adler zeros.
For the other waves, all the formulas are exactly as in Appendix~A.

\booksubsection{B.1. The S0 wave}

\noindent{\sl The region $s^{1/2}\leq932\,\gev$}.\quad
We take $s_0=4m^2_K$, and impose the Adler zero at $s=\tfrac{1}{2}z_0^2$, with 
$z_0$ free. 
We find
$$\eqalign{
\cot\delta_0^{(0)}(s)=&\,\dfrac{s^{1/2}}{2k}\,\dfrac{M_{\pi}^2}{s-\tfrac{1}{2}z_0^2}\,
\left\{\dfrac{z_0^2}{M_\pi\sqrt{s}}+{B}_0+
{B}_1\dfrac{\sqrt{s}-\sqrt{s_0-s}}{\sqrt{s}+\sqrt{s_0-s}}+
{B}_2\left[\dfrac{\sqrt{s}-\sqrt{s_0-s}}{\sqrt{s}+\sqrt{s_0-s}}\right]^2\right\};
\cr
{B}_0=&\,4.41\pm0.30,\quad {B}_1=-26.25\pm0.60,\quad
{B}_2=-15.8\pm1.4;\quad z_0=166.1\pm4.2\, {\rm MeV}.
\cr    }
\equn{(B.1.1a)}$$ 

This corresponds to neglect of isospin breaking. 
If isospin breaking is taken into account as in ref.~20, 
we find instead
$$\matrix{
B_0=3.93\pm0.34,&\quad B_1=-26.84\pm0.78,&\quad B_2=-10.5\pm1.8;\quad z_0=150.2\pm4.2\, {\rm MeV}.
\cr}
\equn{(B.1.1b)}$$

\medskip
 \noindent{\sl The  S0 wave between $932\,\mev$ and $1420\,\mev$}.\quad
The K-matrix parameters  are almost unchanged; we have now
$$\eqalign{\alpha_1=&\,0.843\pm0.017,\quad \alpha_2=0.20\pm0.06,\quad \beta_1=1.02\pm0.02;\quad
\beta_2=1.33\pm0.013,\cr 
\gamma_{11}=&\,3.10\pm0.11,\quad \gamma_{12}=1.81\pm0.05,\quad \gamma_{22}=-7.00\pm0.04;\cr
M_1=&\,0.888\pm0.004\;\gev,\quad M_2=1.327\pm0.004\;\gev;
\cr}
\equn{(B.1.2)}$$

\booksubsection{B.2. The S2 wave}

\noindent{\sl The region $s^{1/2}\leq992\,\mev$}.\quad
We here leave $z_2$ free. Then, 
$$\cot\delta_0^{(2)}(s)=\dfrac{s^{1/2}}{2k}\,\dfrac{M_{\pi}^2}{s-2z_2^2}\,
\left\{B_0+B_1\dfrac{\sqrt{s}-\sqrt{s_l-s}}{\sqrt{s}+\sqrt{s_l-s}}\right\},
\quad s_l^{1/2}=1.05\;\gev;
\equn{(B.2.1a)}$$
  
$$\eqalign{
B_0=&\,-80.2\pm2.8,\quad B_1=-69.4\pm10.5;\quad z_2=145.0\pm3.6\,{\rm MeV}.\cr
}
\equn{(B.2.1b)}$$

\medskip
\noindent{\sl The  S2 wave between $932\,\mev$ and $1420\,\mev$}.\quad
We   write

$$\eqalign{
\cot\delta_0^{(2)}(s)=&\,\dfrac{s^{1/2}}{2k}\,\dfrac{M_{\pi}^2}{s-2z_2^2}\,
\left\{B_{h0}+B_{h1}\left[w_h(s)-w_h(s_M)\right]+
B_{h2}\left[w_h(s)-w_h(s_M)\right]^2\right\},\cr s^{1/2}\geq &\,932\;\mev;\cr
 B_{h0}=&\,B_0+B_1w_l(s_M),\quad B_{h1}=
B_1\left.\dfrac{\partial w_l(s)}{\partial w_h(s)}\right|_{s=s_M};\cr
}
\equn{(B.2.2a)}$$
$z_2$ is given in (B.2.1a);

$$\eqalign{w_l(s)=\dfrac{\sqrt{s}-\sqrt{s_l-s}}{\sqrt{s}+\sqrt{s_l-s}};\quad
s_l=(1050\;\mev)^2,\cr
w_h(s)=\dfrac{\sqrt{s}-\sqrt{s_h-s}}{\sqrt{s}+\sqrt{s_h-s}};\quad s_h=(1450\;\mev)^2.\cr
}
$$
 We get 
$$B_{h2}=120\pm38.
\equn{(B.2.2b)}$$

The inelasticity  is now
$$\eta_0^{(2)}(s)=1-\epsilon(1-\hat{s}/s)^{3/2},\quad 
\epsilon=0.18\pm0.12\quad(\hat{s}^{1/2}=1.05\;\gev).
\equn{(B.2.2c)}$$

\booksubsection{B.3. The P  wave}

\noindent{\sl The region $s^{1/2}\leq2m_K\,\gev$}.\quad
 We have now
\medskip
$$\eqalign{
B_0=&\,1.052\pm0.011,\quad B_1=0.17\pm0.05,\quad M_{\rho}=773.6\pm0.9\;{\rm MeV}.
\cr }
\equn{(B.3.1b)}$$

\medskip
\noindent\noindent{\sl The  P wave between $2m_K$ and $1420\,\mev$}.\quad
The parameters are now
\medskip
$$\eqalign{\lambda_0=&\,
2.684\pm0.009,\quad\lambda_1=1.50\pm0.18,\quad \lambda_2=-1.97\pm0.49;\cr
\epsilon_1=&\,0.09\pm0.06,\quad \epsilon_2=0.12\pm0.11.\cr
}
\equn{(B.3.2b)}$$

\booksubsection{B.4. Parametrization of the D0 wave}

\noindent
The parameters of this wave do not differ appreciably from those in Appendix~A.

\booksubsection{B.5. Parametrization of the D2 wave}

\noindent
We have now
$$B_0=(3.1\pm0.5)\times10^3,\quad B_1=(7.9\pm1.0)\times10^3,\quad
 B_2=(24.7\pm4.2)\times10^3,\quad
\deltav=205\pm25\,\mev
\equn{(B.5.1)}$$
and 
$$\epsilon=0.15\pm0.2.
\equn{(B.5.2)}$$

\booksubsection{B.6. The F wave}

\noindent
The parameters of this wave do not differ appreciably from those in Appendix~A.

\booksubsection{B.7. The G waves}
\noindent 
We have not varied the parameters of the G waves, which therefore are as in Appendix~A above.

\booksubsection{B.8. Regge parameters}

\noindent
We here give the Regge parameters, obtained with the constrained fits. 
They are to be used with the formulas of \sect~3. 
\medskip\noindent
{\sl Isospin 0}.
\medskip
$$\beta_P=2.54\pm0.04;\quad c_P=0.53\pm1.0\;{\gev}^{-2};\quad \alpha'_P=0.20\pm0.10\;{\gev}^{-2},
\equn{(B.8.1a)}$$

$$\beta_{P'}=0.83\pm0.05;\quad c_{P'}=-0.38\pm0.4\;{\gev}^{-2};\quad \alpha_{P'}(0)=0.54\pm0.02;
\quad \alpha'_{P'}=0.90\;{\gev}^{-2}.
\equn{(B.8.1b)}$$
\medskip\noindent
{\sl Isospin 1}.
\medskip
$$\eqalign{
\beta_\rho=&1.30\pm0.14;\quad \alpha_\rho(0)=0.46\pm0.02;\quad \alpha'_\rho=0.90 \;{\gev}^{-2};
\quad\alpha''_\rho=-0.3\;{\gev}^{-4};\cr
d_\rho=&2.4\pm0.5;\quad e_\rho=0.0\pm2.5\;{\gev}^{-4}.
\cr}
\equn{(B.8.2)}$$
\medskip\noindent
{\sl Isospin 2}.\quad
$$\beta_2=0.22\pm0.2.
\equn{(B.8.3)}$$
\booksection{Appendix C: The conformal mapping method}

\noindent
In this Appendix we explain a few of the features of the method of conformal mapping 
 expansion. Although it is a standard method for calculation analytic functions with cuts
(for example, it is one of the methods used by computers to evaluate logarithms), 
we hope that devoting a few lines 
to the matter would not be a waste.
 To make the discussion more adapted to our case, we 
will exemplify our discussion with  the S0 wave.

The key point in the method is the remark that the analyticity and unitarity properties of 
a $\pi\pi$ partial wave amplitude,\fnote{To lighten the notation we will, in the present Appendix, 
suppress indices. The rest of the notation is as in the main text.}
 $f(s)$, imply analyticity of the 
{\sl effective range function}, $\psi(s)$, given by (for the S0 wave)
$$\eqalign{
\cot\delta(s)=&{{s^{1/2}}\over{2k}}\,{{M^2_\pi}\over{s-\tfrac{1}{2}z_0^2}}\,
\Big\{\dfrac{z_0^2}{M_\pi\sqrt{s}}+\psi(s)\Big\},\cr
}
\equn{(C.1)}$$
in the full complex $s$ plane cut from $-\infty$ to 0, and from $s_0=4m^2_K$ to $+\infty$; 
we are neglecting here inelasticity below the $\bar{K}K$ threshold. The function $\psi(s)$ 
is so constructed that it does not have the elastic cut. 
To find an expansion that respects this analyticity of $\psi$, we map 
the cut plane into a circle (\fig~C.1), which is accomplished in our case by the 
change of variable ({\sl conformal mapping})
$$s\to w(s)=\dfrac{\sqrt{s}-\sqrt{s_0-s}}{\sqrt{s}+\sqrt{s_0-s}}.
\equn{(C.2)}$$

\midinsert{
\setbox0=\vbox{{\epsfxsize 10.2truecm\epsfbox{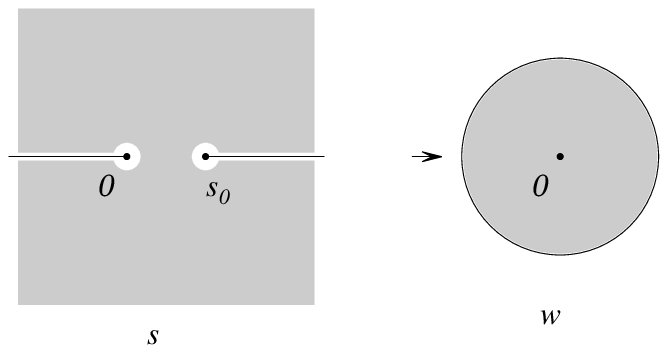}}} 
\setbox6=\vbox{\hsize 6truecm\captiontype\figurasc{Figure C.1. }{ The  
mapping $s\to w$.\hb
\phantom{XX}}\hb
\vskip.1cm} 
\medskip
\centerline{\box0}
\centerline{\box6}
\medskip
}\endinsert

Under the mapping, the left hand cut is mapped into the left half unit circle, and the 
inelastic cut into the right half of the circle; see \fig~C.2. 
The analyticity region (cut plane) is mapped into the interior of the unit circle.
The function $\psi(w)$ is then analytic, in the variable $w$, in the unit disk: 
hence, the analyticity properties of $\psi$ are {\sl }strictly equivalent to the 
convergence of the Taylor expansion,\fnote{Other methods use 
mapping into an ellipse, and expansions in orthogonal
polynomials;  see, e.g., refs.~21.}
$$\psi(w)=B_0+B_1w+B_2w^2+\cdots,
\equn{(C.3)}$$ 
in the unit disk, $|w|<1$.

Reverting to the variable $s$, the expansion (C.3) becomes, for the cotangent of the phase 
shift, the expansion
$$\eqalign{
\cot\delta(s)=&{{s^{1/2}}\over{2k}}\,{{M^2_\pi}\over{s-\tfrac{1}{2}z_0^2}}\,
\Bigg\{\dfrac{z_0^2}{M_\pi\sqrt{s}}+
B_0+B_1\dfrac{\sqrt{s}-\sqrt{s_0-s}}{\sqrt{s}+\sqrt{s_0-s}}+
B_2\left(\dfrac{\sqrt{s}-\sqrt{s_0-s}}{\sqrt{s}+\sqrt{s_0-s}}\right)^2+\cdots\Bigg\}.\cr
}
\equn{(C.4)}$$

\topinsert{
\setbox0=\vbox{\hsize 16truecm{\epsfxsize 10truecm\epsfbox{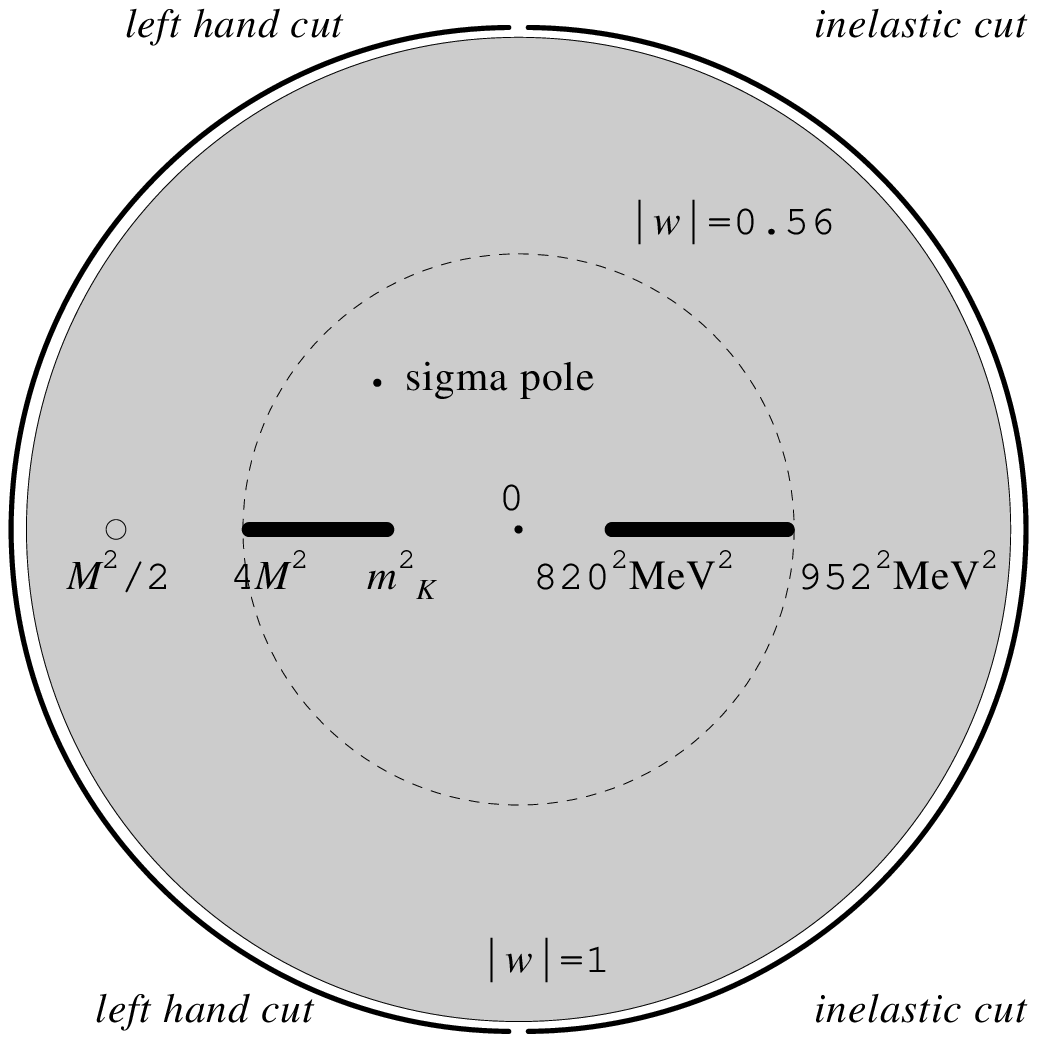}}}
\centerline{\box0}
\setbox6=\vbox{\hsize 12truecm\captiontype\figurasc{Figure C.2. }{
The $w$ disk, 
$|w|<1$.  The dashed line is the line $|w|=0.56$. 
The thick lines are the regions 
where one has reliable experimental data (for the S0 wave). 
Images of the left hand cut and of the inelastic cut are depicted. 
The location of the sigma pole is also shown.}}
\centerline{\box6}
}\endinsert

In our work we only use the expansion in a small region, $|w|\lsim0.56$, which is away from 
the cuts (this will be discussed in more detail later); but it is not difficult to 
prove that (C.4) {\sl also} represents the function $\cot\delta(s)$ on the cuts.
We show this next.

 We write a dispersion relation for the function $\psi^{\rm ex.}(s)$, 
taken to be the exact function:
$$\psi^{\rm ex.}(s)=\dfrac{1}{\pi}\int_{-\infty}^0\dd s'\dfrac{\imag \psi^{\rm ex.}(s')}{s'-s}+
\int_{s_0}^\infty\dd s'\dfrac{\imag \psi^{\rm ex.}(s')}{s'-s};
$$
we neglect eventual subtractions, that play no role here.
It is convenient to rewrite this with a change of integration variable,
$y=s_0/(2s-s_0)$ so that we have
$$\psi^{\rm ex.}(s)=\dfrac{2s_0}{\pi}\int_{-1}^{+1}\dd y\,\dfrac{1}{[2s'(y)-s_0]^2}
\dfrac{\imag \psi^{\rm ex.}(s'(y))}{s'(y)-s}.
\equn{(C.5)}$$
In this new variable, the left hand cut is transformed in the 
interval $[-1,0]$, and the inelastic cut in $[0,+1]$.

Next, it is clear  from the presence of the functions 
$\sqrt{s}$, $\sqrt{s-s_0}$ that $\cot\delta(s)$, or $\psi(s)$, 
have the correct left and right hand cuts.
We  write a dispersion relation for the function expanded to 
order $N$, $\psi^{N}(s)=\sum_0^N B_n w^n$; since it  has  cuts in the same places  as the exact
function, it will read
$$\psi^{N}(s)=\dfrac{2s_0}{\pi}\int_{-1}^{+1}\dd y\,\dfrac{1}{[2s'(y)-s_0]^2}
\dfrac{\imag \psi^{N}(s'(y))}{s'(y)-s}.
\equn{(C.6)}$$
Now, we identify $\psi^{N}(s)$ and $\psi^{\rm ex.}(s)$ at a set of $\nu$ experimental points, 
$s_j$ (we neglect experimental errors), comprised in the interval $[4M^2_\pi, s_0]$: 
$\psi^{N}(s_j)=\psi^{\rm ex.}(s_j)$, $j=1,\dots \nu$. 
This means that the functions {\sl on the cut}, have the same averages with the set of functions
$$\varphi_j(y)=\dfrac{1}{[2s'(y)-s_0]^2}\dfrac{1}{s'(y)-s_j}:
\equn{(C.7)}$$
$$\int_{-1}^{+1}\dd y\,\varphi_j(y)
\left\{\imag \psi^{N}(s'(y))-\imag \psi^{\rm ex.}(s'(y))\right\}=0,\quad j=1,\dots,\nu.
\equn{(C.8)}$$
Hence, the functions built with the $\psi^{N}$ converge to the $\psi^{\rm ex.}$, 
corresponding to the exact partial wave
amplitude,  both on the left hand cut and on the right hand (inelastic) cut, in the mean. 
In the limit in which one had an infinite number of experimental points, the function 
$\psi^{\rm ex.}$ would be represented exactly, because the set of functions in (C.7) 
form a complete set in the interval $[-1,+1]$
(for a proof of this in a physical context, cf. ref.~22). In our case we have only a finite
number of 
points (31 for the S0 case) and, moreover, they have experimental 
errors, so the representation on the cuts is
 valid only in the mean, and up to experimental errors.

In our applications, however, this convergence on the cuts is irrelevant, as we are only fitting
experimental data, 
which are located in a region away both from the left hand cut and the inelastic cut: cf. \fig~C.2,
where we represent the experimental data we are fitting in the case of the S0 wave 
(to the left we have the data obtained from $Ke4$ decays, and $K_{2\pi}$ decays,
 and, to the right, those  higher energy $\pi\pi$ scattering data points included in our fits). 
All of them fall inside the circle $|w|<0.56$.

\vfill\eject
\booksection{Acknowledgments}

\noindent
FJY's work was supported in part by the Spanish DGI of the MEC under contract FPA2003-04597.
JRP's research is partially funded by Spanish CICYT contracts
FPA2005-02327, BFM2003-00856 as well as Banco Santander/Complutense
contract PR27/05-13955-BSCH, and is part of the EU integrated
infrastructure initiative HADRONPHYSICS PROJECT,
under contract RII3-CT-2004-506078. We are grateful to R.~Garc\'{\i}a ~Mart\'{\i}n 
for informartion concerning the location of the $\sigma$~pole. Finally, 
R.~Kami\'nski thanks the Universidad Aut\'onoma de Madrid and Universidad Complutense 
de Madrid, 
where part of this research was carried out.

\booksection{References}
\item{1 }{(PY05): Pel\'aez, J. R., and Yndur\'ain,~F.~J., {\sl Phys. Rev.} {\bf D71}, 074016
(2005).}
\item{2 }{(KPY06): Kami\'nski, R., Pel\'aez, J. R., and Yndur\'ain,~F.~J., {\sl Phys. Rev.} 
{\bf D74}, 014001 (2006) 
and (E), {\bf D74}, 079903 (2006).}
\item{3 }{(GMPY07): Garc\'{\i}a ~Mart\'{\i}n,~R., 
Pel\'aez, J. R., and Yndur\'ain, F. J.,
 hep-ph/0701025 (to be published in Phys. Rev. D.)}
\item{4 }{{\sl Kl4 decays}\/: Rosselet, L., et al. {\sl Phys. Rev.} {\bf D15}, 574  (1977); 
Pislak, S.,  et al.  {\sl
Phys. Rev. Lett.}, {\bf 87}, 221801 (2001).
NA48/2 ( CERN/SPS experiment); 
Bloch-Devaux, B., presented at QCD06 in Montpellier
(France), 3-7 July 2006 and
Masetti, L.,  presented at ICHEP06 in Moscow (Russia), 26 July to 2 August 2006.
{\sl $K\to2\pi$ decays}\/: Aloisio, A., et al., {\sl Phys. Letters}, {\bf B538}, 21 
(2002), 
and private communication by C.~Gatti and V.~Cirigliano.}
\item{5 }{Roy, S. M., {\sl Phys. Letters} {\bf 36B}, 353 (1971).}
\item{6 }{Hyams, B., et al., {\sl Nucl. Phys.} {\bf B64}, 134, (1973);
Estabrooks, P., and Martin, A. D., {\sl Nucl. Physics}, {\bf B79}, 301, 
(1974); Grayer, G., et al.,  {\sl Nucl. Phys.}  {\bf
B75}, 189, (1974); Protopopescu, S. D., et al., {\sl Phys Rev.} {\bf D7}, 1279, (1973);
Kami\'nski, R., Lesniak, L, and Rybicki, K.,
 {\sl Z. Phys.} {\bf C74}, 79 (1997) and 
{\sl Eur. Phys. J. direct} {\bf C4}, 4 (2002); 
Hyams, B., et al., {\sl Nucl. Phys.} {\bf B100}, 205, (1975); 
Losty, M.~J., et al.  {\sl Nucl. Phys.}, {\bf B69}, 185 (1974); 
Hoogland, W., et al. 
{\sl Nucl. Phys.}, {\bf B126}, 109 (1977); 
Durusoy,~N.~B., et al., {\sl Phys. Lett.} {\bf B45}, 517 (1973).}
\item{7 }{$\pi\pi\to\bar{K}K$ scattering: 
Wetzel,~W., et al., {\sl Nucl. Phys.} {\bf B115}, 208 (1976);
Cohen, ~D. et al., {\sl Phys. Rev.} {\bf D22}, 2595 (1980); 
Etkin,~E. et al.,  {\sl Phys. Rev.} {\bf D25}, 1786 (1982).}
\item{8 }{de Troc\'oniz, J. F., and Yndur\'ain, F. J., {\sl Phys. Rev.},  {\bf D65},
093001, (2002)  and {\sl Phys. Rev.} {\bf D71}, 073008 (2005).}
\item{9 }{Pel\'aez, J.~R., and Yndur\'ain, F. J., {\sl Phys. Rev.} {\bf D69}, 114001 (2004).
See also Cudell, J. R., et al., {\sl Phys. Letters} 
{\bf B587}, 78 (2004); Pel\'aez, J.~R., in {\sl Proc. Blois. Conf.
on  Elastic and Diffractive Scattering} (hep-ph/0510005). Note, however, that 
the first reference  fits data for $\pi N$ and $NN$, 
but not for $\pi\pi$, and only for energies above $\sim4\,\gev$; while the last 
article contains only {\sl preliminary} results and, indeed, the parameters for exchange of
isospin zero are not well determined.}
\item{10}{Rarita, W., et al., {\sl Phys. Rev.} {\bf 165}, 1615, (1968).}
\item{11}{Froggatt,~C.~D.,
and Petersen,~J.~L., {\sl Nucl. Phys.} {\bf B129}, 89 (1977).}
\item{12}{Pel\'aez, J. R., and Yndur\'ain, F. J., {\sl Phys. Rev.} {\bf D68}, 074005 (2003).}
\item{13}{Kami\'nski, R., Pel\'aez, J. R., and Yndur\'ain,~F.~J., 
{\sl IV International Conference on Quarks and 
Nuclear Physics},  Madrid, June 2006 [{\sl Eur. Phys.J.} {\bf A31}, 479 (2007)].}
\item{14}{Ananthanarayan, B., et al., {\sl Phys. Rep.} {\bf 353}, 207,  (2001).}
\item{15}{Adler, S. L., and  Yndur\'ain, F. J., {\sl Phys. Rev.} {\bf D75}, 116002 (2007).}
\item{16}{Adeva, B., Romero Vidal, A. and V\'azquez Doce, O., {\sl Eur. Phys. J.} {\bf 31}, 
522 (2007).}
\item{17}{Cabibbo, N., and Isidori, G., {\sl JHEP} 0503:021 (2005); 
NA48 Experiment: see e.g. Balev, S.,
arXiv: 0705.4183 v2 (2007).}
\item{18}{Colangelo, G., Gasser, J.,  and Leutwyler, H.,
 {\sl Nucl. Phys.} {\bf B603},  125, (2001).}
\item{19}{Dobado, A., and Pel\'aez, J. R. {\sl Phys. Rev.} {\bf D56}, 3057 (1999); Oller,~J.~A., 
Oset,~E, and  and Pel\'aez, J. R. {\sl Phys. Rev.} {\bf D59}, 07001 and 
(E) {\sl ibid.} {\bf D60}, 099906 (1999) and (E) {\sl ibid}, {\bf D75}, 099903 (2007); 
G\'omez-Nicola,~A., and Pel\'aez, J. R. {\sl Phys. Rev.} {\bf
D65} , 05009 (202).}
\item{20}{Gasser, J., in Proceedings of the International Conference KAON'07, Frascati, 2007.
 arXiv:0710.3048 [hep-ph].  
}
\item{21}{Ciulli, S.,  In {\sl Strong Interactions}, 
Lecture Notes in Physics, Springer-Verlag, New~York, (1973); 
Pi\u{s}ut, J.  {\sl Analytic Extrapolations and 
Determination of Pion-Pion Phase Shifts}, in ``Low Energy Hadron
Interactions", Springer, Berlin (1970).}
\item{22}{Yndur\'ain, F. J., {\sl Ann. Phys (NY)}, {\bf 75}, 171 (1973).}

\bye